\documentclass[11pt]{article}
\usepackage[hmargin=2.5cm,vmargin=2.4cm]{geometry}
\usepackage{url}
\usepackage{amssymb,latexsym,amsmath,amsthm,psfrag,array,cite,comment,bigints,mathrsfs,bbm,setspace}
\allowdisplaybreaks
\usepackage{amsfonts}
\usepackage{array}
\usepackage{csquotes}
\usepackage{graphicx, caption,array}
\usepackage[colorlinks,linkcolor={blue},urlcolor ={blue},citecolor={blue}]{hyperref}
\usepackage{authblk}
\usepackage[longnamesfirst]{natbib}
\usepackage{graphics}
\usepackage{pdfpages}
\usepackage{mathtools}
\usepackage{booktabs,colortbl,tabularx}
\usepackage{algorithm2e}
\usepackage{graphicx,subcaption,float}
\usepackage{adjustbox}
\usepackage{tikz}
\usepackage{longtable}
\DeclareMathOperator*{\argmin}{arg\,min}

\SetAlgoCaptionSeparator{}
\AtBeginEnvironment{align}{
  \setlength{\abovedisplayskip}{2.5pt} 
  \setlength{\belowdisplayskip}{2.5pt} 
}
\AtBeginEnvironment{align*}{
  \setlength{\abovedisplayskip}{2.5pt} 
  \setlength{\belowdisplayskip}{2.5pt} 
}

\AtBeginEnvironment{eqnarray}{
  \setlength{\abovedisplayskip}{2.5pt} 
  \setlength{\belowdisplayskip}{2.5pt} 
}
\AtBeginEnvironment{eqnarray*}{
  \setlength{\abovedisplayskip}{2.5pt} 
  \setlength{\belowdisplayskip}{2.5pt} 
}

\newtheorem{thm}{Theorem}

\newtheorem{lem}[thm]{Lemma}

\theoremstyle{definition}

\newtheorem{rem}{Remark}
\numberwithin{equation}{section}
\newcommand{\E}{\mathrm{E}}
\newcommand{\var}{\mathrm{var}}

\begin{document}

\title{Efficient Estimation and Consistent Class Number Selection for Semiparametric Class-Specific Index Regression
}

\author[1]{Jen-Chieh Teng}
\author[2]{Sheng-Hsin Fan}
\author[1]{Chin-Tsang Chiang}
\author[3]{Alvin Lim}

\affil[1]{Institute of Applied Mathematical Sciences, National Taiwan University}
\affil[2]{Department of Mathematics, National Taiwan University}
\affil[3]{Goizueta Business School, Emory University} 

\date{\today}%
\maketitle
 
\begin{abstract}
The study introduces a parsimonious semiparametric class-specific index regression that captures heterogeneous covariate--response relationships across latent subpopulations. In such a latent class model formulation, subject-specific indices are used as a complete-data representation of the realized latent labels. We propose a unified framework for estimating model parameters and recovering latent class memberships. For the class-specific index mean model, estimation and latent class recovery are conducted via a pseudo-least-squares criterion with a separation penalty, and a semiparametric information criterion is proposed for consistently estimating the number of classes. Exploiting the response distribution, we develop a Bayes classification rule that maximizes the asymptotic probability of correct assignment and a pseudo-maximum likelihood criterion that yields class index coefficient estimators with asymptotic semiparametric efficiency. This refinement step characterizes latent class membership probabilities via sufficient dimension reduction of the covariate space. Simulations and applications to marketing and biomedical datasets demonstrate superior finite-sample performance of the proposed methodology in parameter estimation, latent class recovery, and class number selection.
\end{abstract} 
\begin{spacing}{1.9}
\section{Introduction} \label{sec:introduction}
 
Across many scientific disciplines, the characteristics distinguishing subpopulations are often missing, unmeasured, or unobservable. This lack of observability complicates statistical analysis and motivates methods that explicitly account for heterogeneity in covariate--response relationships. Such frameworks are known as finite mixtures of regressions in statistics \citep{mclachlan2000finite}, mixtures of experts in machine learning \citep{jacobs1991adaptive}, switching regression in econometrics \citep{quandt1972new}, and clusterwise regression in marketing \citep{desarbo1988maximum,wedel2000market}. Compared with existing parametric models in the literature, semiparametric class-specific index regression provides a flexible alternative, reducing the risk of misspecifying the parametric response or distribution function and avoiding the need to predefine class-specific central subspaces. 

Analyzing 2018 transaction-level data from a North American supermarket chain, we observed distinct subgroups in purchase amounts in the second half of the year (H2) and an association with covariates measured in the first half (H1) that is invariant across customer segments. In another application, data from the AIDS Clinical Trials Group (ACTG) Study 175 reveal heterogeneity in treatment effects across patient subgroups. In particular, CD4 cell counts measured at $20\pm5$ weeks exhibit substantial heterogeneity suggestive of latent subgroups, and treatment effects---zidovudine monotherapy versus alternative therapies---on CD4 cell counts vary across these subgroups. The estimated response function from the class-specific index-mean model exhibits a non-monotonic pattern. In contrast to conventional class-specific linear, generalized linear, and quasi-likelihood models, our semiparametric class-specific index regression flexibly captures complex functional relationships without imposing restrictive parametric structural or distributional assumptions, providing a robust framework for analyzing data with heterogeneous effects.

Building on the subject-specific formulation of the proposed semiparametric class-specific index mean model, we develop a unified estimation and latent class recovery framework for uncovering latent structure without prespecified class labels. The procedure employs a separation-penalty estimation strategy, implemented via a computationally efficient alternating direction method of multipliers \citep[ADMM;][]{boyd2011distributed}, with a well-designed initialization scheme that ensures rapid and stable convergence, to obtain estimators of subject index coefficients and latent class memberships. Furthermore, we propose a semiparametric information criterion that consistently estimates the number of classes without relying on the response distribution, complementing the estimation and latent class recovery procedure. Under a general characterization of the response distribution, we derive a Bayes classification rule, incorporating a sufficient dimension reduction \citep[SDR;][]{li1991sliced} representation of latent class membership probabilities, to achieve asymptotic optimality by maximizing the probability of correct assignment. We also propose a pseudo-maximum likelihood criterion to improve the efficiency of the separation-penalty estimators of the model parameters. The asymptotic variance matrix of the resulting pseudo-maximum likelihood estimators attains the semiparametric efficiency bound. As noted by \citet{teng2025unsupervised}, their pseudo-maximum likelihood method, developed under a semiparametric latent class index distribution model, is more sensitive to extreme values in the estimated density and requires more restrictive bandwidth conditions for continuous responses. Our method avoids these limitations under the proposed model formulation. Finally,
by modifying the distribution estimator underlying the information criteria of \citet{teng2025unsupervised}, we construct alternative criteria for selecting the number of classes.

In estimating class-specific covariate--response relationships, subject-specific formulations have emerged as a standard approach, evolving from the early combinatorial optimization technique \citep{spath1981correction} to contemporary regularization methods. The latter imposes concave pairwise fusion penalties \citep{ma2017concave,ma2020exploration} or adaptive $\ell_1$ penalties \citep{chen2021identifying}. Recent innovations, including multidirectional separation penalties \citep{tang2021individualized} and center-augmented regularization \citep{he2023center}, aim to reduce computational complexity. Nonetheless, these approaches remain computationally burdensome and susceptible to inaccurate latent class recovery. Recent work by \citet{teng2025unsupervised} partially addresses these challenges within a general semiparametric modeling framework. Building on their computational strategy, we further improve computational efficiency by reducing the number of ADMM loops required from two to one.

Alternatively, mixture regression models explicitly parameterize latent class membership. Early approaches assumed covariate-independent class membership probabilities \citep{quandt1972new,kiefer1978discrete}, but were later extended to accommodate covariate-dependent structures, including hierarchical mixture-of-experts models \citep{jacobs1991adaptive,jordan1994hierarchical} and conditional mixture models \citep{desarbo1988maximum}. Although robust procedures for testing the presence of latent subpopulations have been developed \citep{peng1996bayesian,shen2015inference}, this class of models remains highly sensitive to misspecification of latent class membership models. Our target remains a latent-class model. Rather than estimating the model through a marginal mixture likelihood, we introduce the subject-specific indices as a complete-data representation of the realized latent labels. The separation-penalty step estimates this realized-label representation, while the SDR-based class membership model is used in the distributional refinement step.
In the proposed class-specific index density model with SDR-based class membership probabilities, direct marginal mixture-likelihood estimation is difficult because the unknown functions in these semiparametric settings create severe identifiability and computational complications.

The remainder of the article is organized as follows. Section \ref{sec:sec2} formalizes the semiparametric class-specific index regression and its subject-specific formulation. Section \ref{sec:sec3} introduces a unified estimation and latent class recovery procedure and a novel criterion for selecting the number of classes. Section \ref{sec:sec4} refines the separation-penalty parameter estimator under a class-specific index density model. Section \ref{sec:sec5} assesses the method's finite-sample performance through simulation studies. Section \ref{sec:sec6} illustrates its practical utility in personalized marketing and patient stratification. Finally, Section \ref{sec:sec7} concludes with key findings and directions for future research. Appendix \ref{subsec:notation} summarizes the notation used in Sections \ref{sec:sec2}--\ref{sec:sec4}.

\section{Semiparametric Class-Specific Index Regression}
\label{sec:sec2}

Let $Y\in\mathcal{Y} $ denote a quantitative response variable, $X\in \mathcal{X}$ a $p \times 1$ covariate vector, $k$ the number of latent classes, and $C\in\mathcal{C}=\{ 1, \dots, k\}$ a latent class variable. We consider a semiparametric class-specific index model (CSIM) for the conditional mean:
\begin{align}
\E[Y\mid X=x, C = c] = m\big(\gamma_{c}^{\top}z\big),~ (x,c) \in \mathcal{X}\times \mathcal{C}, \label{fn:CIF_1}
\end{align}
where $\gamma_{c} = (\gamma_{c0}, \gamma_{c1}, \dots, \gamma_{cp})^{\top}$ is the index coefficient vector for Class $c$, $z$ is a realization of the augmented covariate vector $Z = (1, X^{\top})^{\top}$, and $m(\cdot)$ is an unknown response function common across classes.
This latent class model is identifiable under the following conditions. First, $\gamma_{11}\neq 0$, corresponding to a continuous covariate that satisfies the continuity condition of \citet{ichimura1993semiparametric}, and $(\gamma_{10}, \gamma_{11})$ is normalized to $(0,1)$. Second, the separation condition $m \big(\gamma_{c}^{\top}z\big) \neq m \big(\gamma_{\ell}^{\top}z\big)$ holds for all $c \neq \ell$, $c, \ell \in \mathcal{C}$, and almost every $x\in\mathcal{X}$. Under these assumptions, the CSIM (\ref{fn:CIF_1}) is identifiable up to a permutation of the class labels. Notably, the separation condition rules out periodic response functions over the covariate space in the CSIM. Since our approach is based on a subject-specific formulation of the CSIM rather than the mixture representation
$\E[Y\mid X=x]=\sum_{c=1}^{k}P(C=c\mid X=x)\,m(\gamma_c^{\top}z)$,
it avoids the associated identifiability issues. 

The proposed model encompasses, as special cases, parametric mean structures arising in generalized linear models \citep{nelder1972generalized} and quasi-likelihood models \citep{wedderburn1974quasi}.
We decompose $\gamma_{c}$ into a class-specific index coefficient vector $\gamma_{(1)c}$ and a class-invariant index coefficient vector $\gamma_{(2)}$, aligned with the covariate subsets $Z_{(1)}$ and $Z_{(2)}$, so that the class-specific index can be written as $\gamma_{c}^{\top} Z=\gamma_{(1)c}^{\top} Z_{(1)} + \gamma_{(2)}^{\top}Z_{(2)}$ for each $c \in \mathcal{C}$.
Under this parameterization, the CSIM (\ref{fn:CIF_1}) is given by
\begin{align}
\E[Y\mid X=x, C = c]= m\big(\gamma_{(1)c}^{\top} z_{(1)} + \gamma_{(2)}^{\top} z_{(2)}\big),~ (x,c)\in \mathcal{X}\times\mathcal{C}.\label{fn:CSIM_2}
\end{align}

Given observations $\{(X_{i}, Y_{i})\}^{n}_{i=1}$, we adopt a subject-specific formulation of the CSIM (\ref{fn:CIF_1}) within a pseudo-estimation framework:
\begin{align}
E[Y_i \mid X_{i}, \beta_i]= m\big(\beta_{i}^{\top} Z_i\big),~ i = 1, \dots, n,\label{fn:SIM}
\end{align}
where $\beta_i \in \{ \gamma_{1}, \dots, \gamma_{k}\}$ and $\beta_{i}=\gamma_{c}$ if and only if $C_{i}=c$. Although the population model is a latent class model, the subject index coefficients $\beta_i= \gamma_{C_i}$, $i=1,\dots,n$, are used as a complete-data representation of the realized latent class labels.
This subject-specific index model (SSIM) enables direct estimation of the response function $m(\cdot)$ via the subject index coefficient vectors $\{\beta_{i}\}^{n}_{i=1}$, without requiring the latent class memberships to be known in advance.
For each observation $i$, the subject index coefficient vector is partitioned as $\beta_{i} = \big(\beta^{\top}_{(1)i}, \beta^{\top}_{(2)}\big)^{\top}$, where $\beta_{(1)i} \in \{\gamma_{(1)1}, \dots, \gamma_{(1)k}\}$ and $\beta_{(2)} = \gamma_{(2)}$.

Specifying the distribution for $Y$---for example, within an exponential or location family---enables the estimation procedure to fully exploit this information, thereby improving the efficiency of parameter estimators.
In particular, consider the SSIM
\vspace{-0.05in}
\begin{align}
Y_{i}= m\big(\beta^{\top}_{i} Z_{i}\big) + U_{i}~, i=1,\dots,n, \label{fn:SSIM_1}
\end{align}
where $U_{i}$ is independent of $(X_{i},\beta_{i})$ and has an unspecified probability density function $f(\cdot)$. Its corresponding CSIM representation is
\begin{align}
Y= m\big(\gamma_{C}^{\top} Z\big) + U,~ C\in\mathcal{C}. \label{fn:CSIM_1}
\end{align}
Therefore, the conditional probability density function $f(y \mid x,c)$ of $Y$ given $(X,C)$ is
\begin{align}
f(y \mid x,c)=f\big(y-m\big(\gamma_{c}^{\top} z\big) \big),~ (y,x,c)\in\mathcal{Y}\times\mathcal{X}\times\mathcal{C}. \label{fn:CSIM_pdf}
\end{align}
In many applications, including bioinformatics and marketing, responses are nonnegative (e.g., biomarker concentrations or purchase amounts) and are often analyzed on the log 
scale. The CSIM (\ref{fn:CSIM_1}) accommodates such log-linear specifications as special cases.
\vspace{-0.1in}
\begin{rem}
\label{Remark1}
When $Y$ follows a distribution in the exponential family, the conditional probability density function of $Y$ given $(X,C)$ can be written as
\begin{align}
f(y \mid x,c)=\exp\bigg(\frac{y\,\theta\big(m\big(\gamma_{c}^{\top} z\big)\big)-b\big(\theta\big(m\big(\gamma_{c}^{\top} z\big)\big)\big) }{\phi}+c(y,\phi)\bigg),~ (y,x,c)\in\mathcal{Y}\times\mathcal{X}\times\mathcal{C}, \label{fn:GLM_pdf}
\end{align}
where $\theta(\cdot)$, $b(\cdot)$, and $c(\cdot,\cdot)$ are known functions determined with the chosen exponential family, and $\phi$ is a scale parameter.
An estimation procedure can be constructed in close analogy to that under the class-specific index probability density model (\ref{fn:CSIM_pdf}), while the corresponding theoretical analysis proceeds similarly but is simplified by the absence of any need to estimate the error density $f(\cdot)$.
\end{rem}

\vspace{-0.25in}
\section{Estimation and Class Number Selection}
\label{sec:sec3}

Let $\theta_{*}$ denote the unknown components of a generic parameter vector $\theta$. Corresponding to the alternative expression (\ref{fn:CSIM_2}) of the CSIM (\ref{fn:CIF_1}) and the SSIM (\ref{fn:SIM}),
let $\gamma=\big(\gamma_{(1)}^{\top},\gamma_{(2)}^{\top}\big)$ and $\beta=\big(\beta_{(1)}^{\top},\beta_{(2)}^{\top}\big)^{\top}$, where
$\gamma_{(1)}=\big(\gamma_{(1)1}^{\top}, \dots, \gamma_{(1)k}^{\top}\big)^{\top}$ and $\beta_{(1)}=\big(\beta_{(1)1}^{\top}, \dots, \beta_{(1)n}^{\top}\big)^{\top}$. 
The true parameters are denoted by $\gamma^{\text{o}}$ and $\beta^{\text{o}}$. For the estimators defined below, we adopt the convention that a ratio of the form $0/0$ is assigned the value zero.

\subsection{Pseudo-Least-Squares Estimation with a Separation Penalty}
\label{sec:sec3.1}

Given a coefficient vector $\beta$, we estimate the response function $m(v)$ using the kernel estimator
\begin{align}
\widehat{m}_h(v; \beta) = \frac{\sum_{i=1}^{n} Y_i  K_{h}\big(\beta_i^{\top}Z_i - v\big)}{\sum_{i=1}^{n} K_{h}\big(\beta_i^{\top}Z_i - v\big)}, \label{fn:m_est}
\end{align}
where $K_{h}(v) = K(v/h)/h$ is a scaled kernel with bandwidth $h>0$. The kernel $K(\cdot)$ is symmetric and compactly supported, has a constant fourth derivative, and satisfies $\int K^{(2)}(u)du=0$. The biweight kernel $K(u)=15/16(1-u^{2})^2 I(|u|\leq 1)$ is used in the numerical studies.
For fixed $h$ and $\lambda$, the separation-penalty (SP) estimator $\big(\widehat{\beta},\widehat{\gamma}_{(1)}\big)$ of $(\beta^{\text{o}},\gamma_{(1)}^{\text{o}})$ is defined as the minimizer of a pseudo-least-squares criterion augmented with an $\ell_{1}$ separation penalty:
\begin{align}
\textsc{pss}_{\textsc{sp}}\big(\beta,\gamma_{(1)}\big) 
= \frac{1}{2} \sum_{i=1}^n \big(Y_{i}-\widehat{m}^{-i}_{h}\big(\beta_{i}^{\top}Z_i; \beta\big)\big)^2 + \lambda\sum^{n}_{i=1}\min_{1\leq c\leq k} \big\|\beta_{(1)i} -\gamma_{(1)c}\big\|_{1},\label{3.1.1}
\end{align}
where $\widehat{m}^{-i}_{h}\big(v; \beta\big)$ denotes the leave-one-out version of $\widehat{m}_h(v; \beta)$, $i = 1, \dots, n$, $\lambda > 0$ controls class separation, and $\|\cdot\|_{1}$ denotes the $\ell_{1}$-norm. 
Larger values of $\lambda$ yield more homogeneous classes, whereas smaller values allow for greater within-class variability. 
The estimator of $\gamma^{\text{o}}$ is given by $\widehat{\gamma}=\big(\widehat{\gamma}^{\top}_{(1)},\widehat{\beta}^{\,\top}_{(2)}\big)^{\top}$, since $\gamma_{(2)}^{\text{o}}=\beta_{(2)}^{\text{o}}$.
If $\beta_{i}$ contains no subject-invariant component, the penalty is applied to $\big(\beta_{i},\gamma_{c}\big)$ rather than  $\big(\beta_{(1)i},\gamma_{(1)c}\big)$
 for all $i=1,\dots, n$ and $c \in \mathcal{C}$. The local theoretical analysis restricts the criterion to the truth-centered parameter region specified in assumption {A4}. We do not, however, assume that the initial values generated by the initialization system are known to belong to this region. The unrestricted SP estimator is further evaluated in Section \ref{sec:sec5}.
 
Let $\mathcal{G}^{\text{o}} = \{\mathcal{G}_1^{\text{o}}, \dots, \mathcal{G}_k^{\text{o}}\}$, where $\mathcal{G}_c^{\text{o}} = \{i : C_i = c\}$, denote the realized latent partition. Based on the SP estimator $\big(\widehat{\beta},\widehat{\gamma}_{(1)}\big)$, define the estimated latent class partition $\widehat{\mathcal{G}} = (\widehat{\mathcal{G}}_1, \dots, \widehat{\mathcal{G}}_{k})$ of $\mathcal{G}^{\text{o}}$ by assigning observation $i$ to Class $c$ whenever 
\vspace{-0.05in}
\begin{align*}
c \in \argmin_{1 \leq \ell \leq k} \| \widehat{\beta}_{(1)i} - \widehat{\gamma}_{(1)\ell} \|_1. 
\end{align*}
Conditional on the realized labels, $\beta_i^{\text{o}} = \gamma_c^{\text{o}}$ for $i \in \mathcal{G}_c^{\text{o}}$ and $c\in\mathcal{C}$. We define the objective function
\begin{align}
\textsc{pss}\big(\gamma,\mathcal{G}\big)=  \frac{1}{2}\sum_{i=1}^n \bigg(Y_i -\sum^{k}_{c=1}I(i\in \mathcal{G}_{c}) \widehat{m}^{-i}_h\big(\gamma^{\top}_{c}Z_i; \mathcal{G}, \gamma\big) \bigg)^2,\label{fn:pss1}
\end{align}
where $\mathcal{G} = \{ \mathcal{G}_1, \dots, \mathcal{G}_k \}$ denotes a partition of the individual index set $\{1, \dots, n\}$, $I(\cdot)$ is the indicator function, and $\widehat{m}_{h}(v; \mathcal{G}, \gamma)$ is an alternative representation of $\widehat{m}_h(v; \beta)$ in (\ref{fn:m_est}), given by
\begin{align}
\widehat{m}_{h}(v; \mathcal{G}, \gamma) =  \frac{\sum_{i=1}^n Y_i\sum^{k}_{c=1}I\big(i\in \mathcal{G}_{c}\big) K_{h}\big(\gamma^{\top}_{c}Z_{i} - v\big)}{\sum_{i=1}^n \sum^{k}_{c=1}I\big(i\in \mathcal{G}_{c}\big) K_{h}\big(\gamma^{\top}_{c}Z_i - v\big)}. \label{fn:m_n_hat_subject}
\end{align} 
The bandwidth is selected as $\hat{h} = n^{3/80} \tilde{h}$ to guarantee the $\sqrt{n}$-consistency of the SP estimator $\widehat{\gamma}$, where $\tilde{h} = O_p(n^{-1/5})$ is the converged value of $\tilde{h}^{(s)}\in\argmin_{h}\textsc{pss}(\widehat{\gamma}^{(s)}, \widehat{\mathcal{G}}^{(s)})$ for a fixed $\lambda$, and $\widehat{\gamma}^{(s)}$ and $\widehat{\mathcal{G}}^{(s)}$ are obtained from the ADMM iterates at iteration $s \ge 0$. The tuning parameter ${\lambda}$ is chosen to minimize $\textsc{pss}(\widehat{\gamma},\widehat{\mathcal{G}}\,)$ over a pre-specified shrinkage grid in Appendix \ref{sec:Algorithm}. In our exploratory comparison, the SP-stage results obtained using the selected rule are broadly competitive with those obtained by fixing the position in the data-dependent sequence, although they do not uniformly yield the best performance.

Let $B$ and $\Gamma$ denote the compact parameter spaces of $\beta_*$ and $\gamma_*$, respectively, and suppose that $\beta^{\text{o}}_*$ and $\gamma^{\text{o}}_*$ lie in their interiors. 
Define the oracle estimator of $\beta^{\text{o}}$ as $\widehat{\beta}^{or} = \sum_{c = 1}^k (I(1 \in \mathcal{G}^{\text{o}}_{c}), \dots, I(n \in \mathcal{G}^{\text{o}}_{c}) )^{\top} \otimes \widehat{\gamma}^{\,or}_{c}$,
where $\widehat{\gamma}^{\,or}$ minimizes the criterion (\ref{fn:pss1}) with $\mathcal{G} = \mathcal{G}^{\text{o}}$, and $\otimes$ denotes the Kronecker product. 
The oracle estimator is the infeasible estimator that knows $\mathcal{G}^{\text{o}}$. The following theorem establishes the asymptotic properties of the oracle estimator $\widehat{\gamma}^{or}_{*}$. 
\begin{thm} 
\label{Thm2.1}
Under assumptions {A1}--{A3}, {A5}, and {A6}, 
\begin{align*}
\widehat{\gamma}^{or}_{*}\stackrel{p}{\longrightarrow} \gamma^{\text{o}}_{*} \text{ and }\sqrt{n}\big(\widehat{\gamma}^{or}_{*} - \gamma^{\text{o}}_{*}\big)\stackrel{d}{\longrightarrow} N(0,V_{2}^{-1}V_{1}V_{2}^{-1}) \text{ as }n \longrightarrow \infty,
\end{align*}
where $V_{1}= \E[((Y-m(\gamma^{\text{o}\top}W))m^{[1]}(W, \gamma^{\text{o}\top}W))^{\otimes 2} \big]$ and $V_{2}= \E[( m^{[1]}\big(W,\gamma^{\text{o}\top}W))^{\otimes 2}]$, with $W=(I(C=1), \dots, I(C=k))^{\top} \otimes Z\in\mathcal{W}$, $m^{[t]}(w, \gamma^{\top}w)$, $t=0,1,2$, defined in Lemma \ref{A:Lemma_2}, and $(\cdot)^{\otimes}$ denoting the Kronecker power.
\begin{proof}
See Appendix \ref{sec:A.3}.
\end{proof}
\end{thm}
\noindent The next theorem establishes the local oracle property of the proposed SP estimator $(\widehat{\beta},\widehat{\gamma}_{(1)})$.
\begin{thm} 
\label{Thm3.1}
Under assumptions {A1}--{A7}, there exists a local minimizer $\big(\widehat{\beta},\widehat{\gamma}_{(1)}\big)$ of the criterion (\ref{3.1.1}) satisfying
\begin{align*}
P\big(\big(\widehat{\beta},\widehat{\gamma}\big) = \big(\widehat{\beta}^{or},\widehat{\gamma}^{\,or}\big)\big) {\longrightarrow} 1 \text{ as }n {\longrightarrow} \infty.
\end{align*}  
\begin{proof}
See Appendix \ref{sec:A.1}.
\end{proof}
\end{thm}
\noindent Because $\big(\widehat{\beta},\widehat{\gamma}\big) = \big(\widehat{\beta}^{or},\widehat{\gamma}^{\,or}\big)$ entails 
$\widehat{\mathcal{G}}=\mathcal{G}^{\text{o}}$ under the label alignment used in assumption {A4}, it follows that
\begin{align}
P\big(\widehat{\mathcal{G}}=\mathcal{G}^{\text{o}}\big)\longrightarrow 1 \text{ as } n\longrightarrow \infty. \label{consistency:grouping}
\end{align}
The exact latent-partition recovery should not be interpreted as a general perfect-classification result for mixtures with fixed nonvanishing overlap. Under such overlap, a positive per-observation Bayes classification error generally precludes recovery of every realized latent label with probability converging to one.
Within the localized problem, assumption {A4} constrains each candidate subject parameter to a neighborhood of its corresponding true center, while assumption {A7} imposes a sufficiently strong separation penalty to ensure the dominance of the oracle configuration over competing local alternatives. Combining Theorems \ref{Thm2.1} and \ref{Thm3.1} yields
\begin{align}
\sqrt{n}\big(\widehat{\gamma}_{*} - \gamma^{\text{o}}_{*}\big)\stackrel{d}{\longrightarrow} N\big(0,V_{2}^{-1}V_{1}V_{2}^{-1}\big) \text{ as }n \longrightarrow \infty. \label{normality:psispe}
\end{align} 
The asymptotic variance matrix of $\widehat{\gamma}_{*}$ can be consistently estimated by $\widehat{V}_{2}^{-1}\widehat{V}_{1}\widehat{V}_{2}^{-1}$, where
$\widehat{V}_{1}$ and $\widehat{V}_{2}$ are constructed using the optimal bandwidth $\tilde{h}$ for the response function estimator, with
 \begin{align*}
\widehat{V}_{1}=&\frac{1}{n} \sum_{i=1}^n \sum^{k}_{c=1}I(i\in \widehat{\mathcal{G}}_{c}) \bigg[\Big( Y_i  - \widehat{m}_{\tilde{h}}\big(\widehat{\gamma}^{\top}_{c}Z_i; \widehat{\mathcal{G}},\widehat{\gamma}\big)  \Big)  \partial_{\gamma_{*}} \widehat{m}_{\tilde{h}}\big(\widehat{\gamma}^{\top}_{c}Z_i; \widehat{\mathcal{G}},\widehat{\gamma}\big)\bigg]^{\otimes 2} \text{ and}\\
\widehat{V}_{2}=& \frac{1}{n} \sum_{i=1}^n \sum^{k}_{c=1}I(i\in \widehat{\mathcal{G}}_{c}) \Big(\partial_{\gamma_{*}} \widehat{m}_{\tilde{h}}\big(\widehat{\gamma}^{\top}_{c}Z_i; \widehat{\mathcal{G}},\widehat{\gamma}\big) \Big)^{\otimes 2}.
\end{align*}

\begin{rem}
\label{Remark2}
In the spirit of quasi-likelihood methods that avoid full distributional specification, the proposed SSIM (\ref{fn:SIM}) extends naturally to accommodate a class-specific working variance matrix of the form $\sigma^{2}(m(\gamma^{\top}_{c}z))$ for each $c\in\mathcal{C}$, where $\sigma^{2}(\cdot)$ may be specified a priori or left unspecified. For estimation, the SP criterion in (\ref{3.1.1}) is modified by replacing the pseudo sum of squares with $\sum_{i=1}^n \omega_{i} \big(Y_{i}-\widehat{m}^{-i}_{h}\big(\beta_{i}^{\top}Z_i; \beta\big)\big)^2$. The weights $\omega_{1},\dots, \omega_{n}$ are initialized at one and subsequently updated to consistent estimators of $\sigma^{-2}(m(\beta^{\top}_{1}Z_{1})), \dots,\sigma^{-2}(m(\beta^{\top}_{n}Z_{n}))$ until a convergence criterion is met.
\end{rem}

\vspace{-0.2in}
\subsection{Selecting the Number of Classes}
\label{sec:sec3.2}
\vspace{-0.1in}
We select the number of classes, $\hat{k}$, by minimizing the semiparametric information criterion (SPIC)
\begin{align}
\textsc{SPIC}_{0}(k) = \frac{2\,\textsc{pss}\big(\widehat{\gamma},\widehat{\mathcal{G}}\,\big)}{\sum_{i=1}^{n} (Y_i - \bar{Y})^{2}}+ \frac{k \log n }{n^{\frac{4}{5}}} ,~ k \geq 1,\label{fn:spic}
\end{align}
where $\bar{Y} =\sum_{i=1}^n Y_i /n$.
For large samples (e.g., $n \gtrsim 600$ in our numerical studies), we augment the penalty in (\ref{fn:spic}) to $k(\log n / n^{4/5} + \log n / n^{9/10})$ to mitigate overestimation of the number of classes. This adjustment accounts for the interaction between the parametric and nonparametric convergence rates, namely $\sup_{v}|\widehat{m}_{\tilde{h}}(v; \widehat{\mathcal{G}},\widehat{\gamma})-m(v)|= o_{p}\big(\sqrt{\log n /n^{4/5}} \,\big)$ and $\widehat{\gamma}_{*}-\gamma^{\text{o}}_{*}=O_{p}(n^{-1/2})$.

Let $k^*$, $C^{*}\in\mathcal{C}^{*} = \{1, \dots, k^{*}\}$, and  
$\gamma^{*\text{o}}_{1}, \dots, \gamma^{*\text{o}}_{k^{*}}$ denote the true values of the number of classes, the latent class variable, and the class index coefficients, respectively, in the CSIM (\ref{fn:CIF_1}). The realized latent partition is denoted by $\mathcal{G}^{*\text{o}} = \{\mathcal{G}^{*\text{o}}_{1}, \dots, \mathcal{G}^{*\text{o}}_{k^{*}}\}$,
where $\mathcal{G}_c^{*\text{o}} = \{i : C^{*}_i = c\}$. We denote $U^{*}$ as $Y - m\big( \gamma^{* \text{o} \top} W^* \big)$, where $W^{*} =(I(C^* = 1), \dots, I(C^* = k^*))^{\top} \otimes Z$.
Assume that $\lim_{n\rightarrow\infty}P\big(\widehat{\mathcal{G}}=\mathcal{G}^{\text{o}}\big)=1$ for all candidate values $k \geq 1$ under consideration. This condition is not necessarily guaranteed by assumptions {A4} and {A7} when $k\neq k^*$. Under this condition, the following theorem establishes the consistency of $\hat{k}$ to $k^*$.
\begin{thm} \label{Thm3.2}
Under the condition that $\lim_{n\rightarrow\infty}P\big(\widehat{\mathcal{G}}=\mathcal{G}^{\text{o}}\big)=1$  for all $k \geq 1$ and assumptions {A1}--{A3}, {A5}, and {A6},
\begin{align*} 
\hat{k} \stackrel{p}{\longrightarrow} k^{*} \text{ as } n\longrightarrow \infty.
\end{align*} 
 \begin{proof}
See Appendix \ref{sec:A.4}.
\end{proof}
 \end{thm}

\vspace{-0.2in}
\section{Optimal Classification and Semiparametric Efficient Estimation} 
\label{sec:sec4}
\vspace{-0.1in}
Under the class-specific index density model (\ref{fn:CSIM_pdf}), we improve the SP estimation by leveraging SDR-based class membership probabilities.

\subsection{Estimation of SDR-based Class Membership Probabilities}
\label{sec:sec4.1}

We characterize the conditional distribution of $C$ given $X$ through SDR of the covariate space,
\begin{align}
P(C=c \mid X=x) = \pi_{c}\big(A^{\top}_{d}x\big),~ (x, c) \in \mathcal{X} \times \mathcal{C}, \label{fn:SDR}
\end{align}
where $A_{d} $ is a $p\times d$ full rank matrix, and $\{\pi_{c}(\nu)\}^{k}_{c=1}$ are the unknown functions satisfying $0\leq \pi_{c}(\nu)\leq 1$ and $\sum^{k}_{c=1}\pi_{c}(\nu)=1$ for all $\nu\in\mathbb{R}^d$.
Under the consistent latent class recovery condition in (\ref{consistency:grouping}) and the identity $I(C_i = c)=I(i\in\mathcal{G}^{\text{o}}_{c})$ for all $i=1,\dots,n$ and $c\in\mathcal{C}$, we estimate $\pi_{c}(\nu)$, given a basis matrix $A_{d}$, by 
\begin{align}
\widehat{\pi}_{c,h_{d}}(\nu;A_{d})=\frac{\sum^{n}_{i=1}I\big(i\in \widehat{\mathcal{G}}_{c}\big)\mathcal{K}_{r,h_{d}}\big(A^{\top}_{d}X_{i}-\nu\big)}{\sum^{n}_{i=1}\mathcal{K}_{r,h_{d}}\big(A^{\top}_{d}X_{i}-\nu\big)},~c\in\mathcal{C},\label{fn:est1_GM}
\end{align}
where $\mathcal{K}_{r,h_{d}}(\nu)=\prod^{d}_{j=1}K_{r,h_{d}}(\nu_{j})$ and $K_{r,h_{d}}(v) = K_{r}(v/h_{d})/h_{d}$ is a kernel function with bandwidth $h_{d}>0$. The function $K_{r}(\cdot)$ is an $r$th-order kernel that is a symmetric, compactly supported, and twice continuously differentiable, satisfying $\int K_{r}^{(2)}(u)du=0$, with $r > \max\{(4+d)/4,(1+d)/2\}$.
Because an $r$th-order kernel need not be nonnegative, the estimator $\widehat{\pi}_{c,h_d}(\nu;A_d)$ may take values outside $[0,1]$. We therefore truncate the estimator to $[0,1]$ and set $\widehat{\pi}_{c,h_d}(\nu;A_d)=1/k$ for all $c\in\mathcal{C}$ whenever all truncated estimates are zero at a given $\nu$. For fixed $d$ and $h_{d}$, the pseudo-least squares estimator $\widehat{A}_{d}$ is defined as the minimizer of the pseudo sum of squares
\begin{align}
\textsc{pss}_{\textsc{cm}}(A_{d},h_{d})=\frac{1}{2}\sum^{n}_{i=1}\sum^{k-1}_{c=1}\Big(I\Big(i\in \widehat{\mathcal{G}}_{c}\Big)- \widehat{\pi}^{-i}_{c,h_{d}}\big(A^{\top}_{d}X_{i}; A_{d}\big)\Big)^{2},  \label{fn:PSS_l}
\end{align}
where $\widehat{\pi}^{-i}_{c,h_{d}}\big(\nu; A_{d}\big)$ denotes the leave-one-out version of $\widehat{\pi}_{c,h_{d}}\big(\nu; A_{d}\big)$, $i = 1, \dots, n$.
Using $\widehat{A}_{d}$ and a bandwidth selector
$\tilde{h}_{d}\in\arg\min_{h_{d}} \textsc{pss}_{\textsc{cm}}(\widehat{A}_{d},h_{d})$, which attains the rate $O_{p}(n^{-1/(2r+d)})$, the structural dimension $d^{\text{o}}$ is estimated by
a minimizer $\widehat{d}$ of the criterion:
\vspace{-0.1in}
\begin{align}
\textsc{pss}_{\textsc{cm}}(d) =\frac{2}{n}\textsc{pss}_{\textsc{cm}}\big(\widehat{A}_{d},\tilde{h}_{d}\big)+\frac{\log n}{2n^{\frac{2r}{2r+d}}}\text{ for }d\geq 1.
 \label{fn:PSS_d}
\end{align}
Based on the selected structural dimension, the resulting estimator of the central subspace (CS) matrix $A^{\text{o}}$ is denoted by $\widehat{A}$. 
In practice, covariate independence of class membership probabilities can be assessed by comparing $\textsc{pss}_{\textsc{cm}}(1)$ with
\begin{align*}
\frac{1}{2n}\sum^{n}_{i=1}\sum^{k-1}_{c=1}\Big(I\big(i\in \widehat{\mathcal{G}}_{c}\big)- \frac{1}{n}\sum^{n}_{j=1}I\big(j\in \widehat{\mathcal{G}}_{c}\big)\Big)^{2}+\frac{\log n}{2n}.
\end{align*}

Let $\text{vec}(\cdot)$ denote the column-stacking operator, $P_{A}$ the projection matrix onto the column space of a generic matrix $A$, and
$\pi^{[t]}_{c}({x},A^{\top}_{d}x)$ the uniformly convergent function of $\partial^{t}_{\text{vec}(A_{d*})}\widehat{\pi}_{c,h_{d}}(A^{\top}_{d}x;A_{d})$ for $t=0, 1, 2$ and $c\in\mathcal{C}$, with $\pi^{[0]}_{c}({x},A^{\text{o} \top}x) = \pi_{c}(A^{\text{o} \top}x)$.
The following theorem establishes the consistency of $\hat{d}$ to $d^{\text{o}}$ and the asymptotic normality of $\widehat{A}_*$.
\begin{thm} 
\label{Thm3.3_1}
Under assumptions {A1}--{A7} and {B1}--{B4}, 
\begin{align*}
\hat{d}\stackrel{p}{\longrightarrow}d^{\text{o}} \text{ and }\sqrt{n}\,\text{vec}\big(P_{\widehat{A}_*}- P_{A_*^{\text{o}}}\big)\stackrel{d}{\longrightarrow} N(0,\Sigma) \text{ as }n \longrightarrow \infty,
\end{align*}
where $\Sigma=\E\big[\big(\text{vec}\big(\big(I_{p}-P_{A_*^{\text{o}}}\big)\big(I_{d^{\text{o}}},V^{\top}_{0}\big)^{\top}\big)\big(A^{\text{o}\top}A^{\text{o}}\big)^{-1}A^{\text{o}\top}
+A^{\text{o}}(A^{\text{o} \top}A^{\text{o}}\big)^{-1}\big(I_{d^{\text{o}}},V^{\top}_{0}\big)\big(I_{p}-P_{A_*^{\text{o}}}\big)\big)^{\otimes 2}\big]$,
with
$\text{vec}(V_{0})=\big(\sum^{k-1}_{c=1}\E\big[\big(\pi^{[1]}_{c}\big({X}, A^{\text{o} \top}X\big)\big)^{\otimes 2}\big]\big)^{-1}
\big(\sum^{k-1}_{c=1}\big(I(C=c)-\pi_{c}({X}, A^{\text{o}\top}X))\pi^{[1]}_{c}({X}, A^{\text{o}\top}X\big)\big)$.
\begin{proof}
See Theorem 2 in \cite{teng2025unsupervised}.
\end{proof}
\end{thm}
\noindent The proof strategy of Theorem \ref{Thm2.1} can be extended to the present setting and used to strengthen existing theoretical developments by relaxing the compact-support condition on $X$ and the lower-bound condition on the probability density of the CS directions. In particular, the SDR estimator $\widehat{A}_*$ is asymptotically equivalent to the infeasible oracle estimator constructed from the true class indicators $I(C_i = c)$ for $i=1,\dots, n$ and $c\in\mathcal{C}$.

\subsection{Bayes Classification Rule and Pseudo-Maximum Likelihood Estimation}
\label{sec:sec4.2}

We estimate the probability density function $f(u)$ in the class-specific index density model (\ref{fn:CSIM_pdf}) using the SP-estimated latent class partition $\widehat{\mathcal{G}}$ and the SP estimator $\widehat{\gamma}$:
\begin{align}
\widehat{f}_{\xi}(u)=\frac{1}{n}\sum^{n}_{i=1}K_{\xi}\bigg(Y_i - \sum^{k}_{c=1}I\big(i \in \widehat{\mathcal{G}}_{c}\big) \widehat{m}^{-i}_{\tilde{h}}\big(\widehat{\gamma}^{\top}_{c}Z_i;\widehat{\mathcal{G}}, \widehat{\gamma}\,\big)- u\bigg). \label{4.2.1}
\end{align} 
The bandwidth $\xi$ is selected as the minimizer of the cross-validation criterion of \cite{bowman1984alternative},
\begin{align}
\textsc{CV}(h) = \frac{1}{n}\sum\limits^{n}_{i=1}\bigg[\int \Big(\widehat{f}^{-i}_{\xi}(u)\Big)^{2}du-2\widehat{f}^{-i}_{\xi}\bigg(Y_i - \sum^{k}_{c=1}I\big(i \in \widehat{\mathcal{G}}_{c}\big) \widehat{m}^{-i}_{\tilde{h}}\big(\widehat{\gamma}^{\top}_{c}Z_i; \widehat{\mathcal{G}}, \widehat{\gamma}\,\big)\bigg)\bigg],\label{4.2.2.0}
\end{align}
where $\widehat{f}^{-i}_{\xi}(u)$ denotes the leave-one out version of $\widehat{f}_{\xi}(u)$.
Using the estimators $\widehat{f}_{\tilde{\xi}}(y-\widehat{m}_{\tilde{h}}(\widehat{\gamma}^{\top}_{c}z;\widehat{\mathcal{G}}, \widehat{\gamma}\,))$ and $\widehat{\pi}_{c,\tilde{h}_{\hat{d}}}(\widehat{A}\,^{\top}X_{i};\widehat{A})$ for $(y,x,c) \in \mathcal{Y} \times \mathcal{X} \times \mathcal{C}$, we construct the Bayes classification rule by assigning a new observation $(X_{0},Y_{0})$ to Class $c$ whenever
\vspace{-0.1in}
\begin{eqnarray}
\hspace{-0.01in}\widehat{f}_{\tilde{\xi}}\big(Y_{0}-\widehat{m}_{\tilde{h}}\big(\widehat{\gamma}^{\top}_{c}Z_{0};\widehat{\mathcal{G}}, \widehat{\gamma}\,\big)\big)\, \widehat{\pi}_{c,\tilde{h}_{\hat{d}}}\big(\widehat{A}^{\top}X_0;\widehat{A}\,\big)=\max_{\{ 1\leq \ell\leq k\}}\widehat{f}_{\tilde{\xi}}\big(Y_{0}-\widehat{m}_{\tilde{h}}\big(\widehat{\gamma}^{\top}_{\ell}Z_{0};\widehat{\mathcal{G}}, \widehat{\gamma}\,\big)\big)\, \widehat{\pi}_{\ell,\tilde{h}_{\hat{d}}}\big(\widehat{A}^{\top}X_{0};\widehat{A}\,\big).~\label{4.2.3}
\end{eqnarray}
The optimality result for Bayes assignment in \citet{teng2025unsupervised} implies that this classifier achieves the maximal asymptotic probability of correct assignment.
In practice, the same classification rule can be used to construct the estimated latent class partition $\widetilde{\mathcal{G}}=\big(\widetilde{\mathcal{G}}_{1},\dots,\widetilde{\mathcal{G}}_{k}\big)$ of $\mathcal{G}^{\text{o}}$
by assigning observation $i$ to $\widetilde{\mathcal{G}}_{c}$ whenever condition (\ref{4.2.3}) holds after replacing $(X_0,Y_0)$ with $(X_i,Y_i)$. The Bayes-rule-based (BR) estimated latent class partition is not guaranteed to achieve consistent latent class recovery, unlike the SP-estimated latent class partition in (\ref{consistency:grouping}). Nevertheless, it exhibits comparable---and in some cases superior---finite-sample performance when the component probability densities exhibit moderate overlap.

Given a parameter vector $\gamma$, we estimate $f(u)$ using the SP-estimated latent class partition $\widehat{\mathcal{G}}$ and the bandwidth selectors $\hat{h}$ and $\hat{\xi} = n^{3/80} \tilde{\xi}$, where $\tilde{\xi} = O_p(n^{-1/5})$. The bandwidth $\hat{\xi}$ is selected to guarantee the $\sqrt{n}$-consistency of the resulting estimator, defined as 
\begin{align}
\widetilde{f}_{\hat{\xi}}(u;\gamma)=\frac{1}{n}\sum^{n}_{i=1}K_{\hat{\xi}}\bigg(Y_i - \sum^{k}_{c=1}I\big(i \in \widehat{\mathcal{G}}_{c}\big) \widehat{m}_{\hat{h}}\big(\gamma^{\top}_{c}Z_i; \widehat{\mathcal{G}}, \gamma\,\big)- u\bigg). \label{4.2.4}
\end{align} 
The pseudo-maximum likelihood (PML) estimator $\widetilde{\gamma}$ of $\gamma^{\text{o}}$ is obtained by maximizing the log-pseudo-likelihood
\begin{align}
p\ell(\gamma)=\sum^{n}_{i=1} \log \bigg(\widetilde{f}_{\hat{\xi}}\bigg(Y_i - \sum^{k}_{c=1}I\big(i \in \widehat{\mathcal{G}}_{c}\big) \widehat{m}_{\hat{h}}\big(\gamma^{\top}_{c}Z_i; \widehat{\mathcal{G}}, \gamma\,\big); \gamma\bigg)\bigg). \label{4.2.5}
\end{align}
This optimization is initialized at the SP estimator $\widehat{\gamma}$. 

The proposed pseudo-maximum likelihood estimator $\widetilde{\gamma}$ is consistent and asymptotically normal.
\begin{thm} 
\label{Thm4.2}
Under assumptions {A1}--{A7} and {C1}--{C4},
\begin{align*}
\widetilde{\gamma}_*\stackrel{p}{\longrightarrow}\gamma^{\text{o}}_*\text{ and } \sqrt{n}\big(\widetilde{\gamma}_*-\gamma^{\text{o}}_*\big)\stackrel{d}{\longrightarrow} N\big(0,V^{-1}\big) \text{ as } n {\longrightarrow} \infty,
\end{align*} 
where $V=\var\big({f^{[1]}(W, Y-m(\gamma^{\text{o}\top}W);\gamma^{\text{o}})}/{f^{[0]}(W, Y-m(\gamma^{\text{o}\top}W);\gamma^{\text{o}})}\big)$, with $f^{[t]}(w,u;\gamma)$, $t=0,1,2$, defined in Lemma \ref{A:Lemma_3}.
\begin{proof}
See Appendix \ref{sec:A.5}.
\end{proof}
\end{thm}
\noindent Under the assumptions of Theorem \ref{Thm4.2}, the consistent latent class recovery condition in (\ref{consistency:grouping}) establishes that the
PML estimator based on $\widehat{\mathcal{G}}$ is asymptotically equivalent to its infeasible oracle counterpart
constructed from $\mathcal{G}^{\text{o}}$. The asymptotic variance attains the semiparametric efficiency bound; see the proof of Theorem 2 in \cite{ma2013efficient} for a related argument. Moreover, the
asymptotic variance matrix of $\widetilde{\gamma}_{*}$ is estimated by $\widetilde{V}^{-1}$, with
\begin{align*}
\widetilde{V}=\frac{1}{n} \sum_{i=1}^n \Bigg( \frac{\partial_{\gamma_*}\widetilde{f}_{\tilde{\xi}}\big(Y_i - \sum^{k}_{c=1}I\big(i \in \widehat{\mathcal{G}}_{c}\big) \widehat{m}_{\tilde{h}}\big(\widetilde{\gamma}^{\top}_{c}Z_i; \widehat{\mathcal{G}}, \widetilde{\gamma}\,\big); \widetilde{\gamma}\big)}{\widetilde{f}_{\tilde{\xi}}\big(Y_i - \sum^{k}_{c=1}I\big(i \in \widehat{\mathcal{G}}_{c}\big) \widehat{m}_{\tilde{h}}\big(\widetilde{\gamma}^{\top}_{c}Z_i; \widehat{\mathcal{G}}, \widetilde{\gamma}\,\big); \widetilde{\gamma}\big)}\Bigg)^{\otimes 2}. 
\end{align*} 

\begin{rem}
\label{Remark3}
Within the semiparametric generalized linear model with an unspecified response function in (\ref{fn:GLM_pdf}), the parameters $(\gamma,\phi)$ can be estimated by 
the maximizer $(\widehat{\gamma},\hat{\phi})$ of
\begin{align*}
p\ell_{0}(\gamma,\phi)=\sum^{n}_{i=1}\bigg(\sum^{k}_{c=1}I\big(i \in \widehat{\mathcal{G}}_{c}\big)\frac{Y_{i}\,\theta\big( \widehat{m}_{\hat{h}}\big(\gamma^{\top}_{c}Z_i; \widehat{\mathcal{G}}, \gamma\,\big)\big)-b\big(\theta\big( \widehat{m}_{\hat{h}}\big(\gamma^{\top}_{c}Z_i; \widehat{\mathcal{G}},\gamma\,\big)\big)\big) }{\phi}\bigg)+c(Y_{i},\phi).
\end{align*}
\end{rem}

\subsection{Modified Criteria for Selecting the Number of Classes}
\label{sec:sec4.3}

As alternatives to $\textsc{SPIC}_{0}$ in (\ref{fn:spic}), we consider two criteria from \citet{teng2025unsupervised}, replacing their bivariate function estimator with the estimator
$\widetilde{F}_{\tilde{\xi}}(u) = \int_{-\infty}^{u} \widetilde{f}_{\tilde{\xi}}(v;\tilde{\gamma}) dv$ of the distribution function $F(u)=\int^{u}_{-\infty}f(v)dv$, to select the number of classes:
\begin{align}
\textsc{SPIC}_1(k) =  \frac{1}{n} \sum_{i=1}^n \int \Big( I(Y_i \leq y) - \widetilde{F}^{-i}_{\tilde{\xi}}( e_{iy} ) \Big)^2 d\widetilde{F}_{Y}(y)+ \frac{k \log n }{2n^{\frac{4}{5}}}  \label{fn:spic1} 
\end{align}
and
\begin{align}
\textsc{SPIC}_2(k) = &~ -\frac{1}{n} \sum_{i=1}^n \int \Big[ I(Y_i \leq y) \log\Big(\widetilde{F}^{-i}_{0 \tilde{\xi}}(e_{iy}) \Big)+ I(Y_i > y) \log \Big(1-\widetilde{F}^{-i}_{1 \tilde{\xi}}(e_{iy})\Big) \Big] d\widetilde{F}_{Y}(y)\nonumber\\
&~ + \frac{k \log n }{n^{\frac{4}{5}}},~ k\geq 1,\label{fn:spic2}
\end{align}
where $\widetilde{F}_{Y}(y)= \sum_{i=1}^n I(Y_{i}\leq y)/n$, $e_{iy} = y - \sum_{c=1}^{k} I(i \in \widehat{\mathcal{G}}_{c})\widehat{m}^{-i}_{\tilde{h}}(\widetilde{\gamma}^{\top}_{c} Z_i; \widetilde{\gamma})$, and $\widetilde{F}^{-i}_{s \tilde{\xi}}(e_{iy}) = \widetilde{F}^{-i}_{\tilde{\xi}}(e_{iy}) + (-1)^s I\big(\widetilde{F}^{-i}_{\tilde{\xi}}(e_{iy}) = s\big)$ for $i = 1, \dots, n$ and $s = 0, 1$.
The penalty term in (\ref{fn:spic1}) and (\ref{fn:spic2}) corresponds to the uniform convergence rate $\sup_{u}|\widetilde{F}_{\tilde{\xi}}(u)-F(u)|=O_{p}(n^{-2/5})$. Under the conditions of Theorem \ref{Thm3.2} and the uniform convergence rate of $\widetilde{F}_{\tilde{\xi}}(u)$ to $F(u)$, we establish the consistency of the minimizers $\hat{k}_1$ of $\textsc{SPIC}_{1}$ and $\hat{k}_2$ of $\textsc{SPIC}_{2}$ for estimating $k^*$, using arguments similar to those in the proof of Theorem 3 in \cite{teng2025unsupervised}.

\begin{rem}
\label{Remark4}
Alternatives to the goodness-of-fit terms in (\ref{fn:spic1}) and (\ref{fn:spic2}) include the log pseudo-likelihood and the log pseudo-marginal likelihood.
Numerical experiments show that the criterion based on the log pseudo-likelihood overestimates the number of classes, even for large sample sizes. Furthermore, the log pseudo-marginal likelihood induces complex dependencies through SDR-based class membership probabilities, precluding a tractable theoretical adjustment for overfitting.
\end{rem}

\section{Monte Carlo Simulations} 
\label{sec:sec5}

The simulation settings were motivated by transaction-level data from a North American supermarket chain and clinical data from ACTG 175. We conducted simulation studies to assess the finite-sample performance of the proposed methods for the two models (\ref{fn:CIF_1}) and  (\ref{fn:CSIM_1}), along with benchmark methods for class-specific linear regression, using sample sizes $n\in\{200, 400, 600\}$ and 500 replications per scenario. The comparisons evaluate the performance of the implemented procedures under linear versus nonlinear regression specifications and normal versus nonnormal error distributions; they are not intended as a comprehensive comparison of latent class regression methods.

\subsection{Simulation Design} \label{sec:simulation_settings}

Data were generated from the CSIM (\ref{fn:CSIM_1}) with the response function $m(v)=(v+5)^{3/5}$ and the number of classes set to $k=2$ or $k=3$. The error term $U$ was independently drawn from one of three distributions: $N(0,\sigma^{2})$, $\text{Logistic}(0,\sqrt{3}\sigma/\pi)$, or $\text{Gamma}(4,\sigma/2)-2\sigma$, with $\sigma=0.15$ for $k=2$ and $\sigma=0.10$ for $k=3$.

The class-invariant index coefficient vector was fixed at $\gamma_{(2)}=(1,-1,1,-1,1)^{\top}/\sqrt{5}$ across all scenarios. The first four components of $Z_{(2)}$ were sampled from a multivariate normal distribution $N_{4}(0_{4}, 0.8 I_4 + 0.2 1_4 1_4^{\top})$, while the fifth component was independently sampled from $\{-0.5,0.5\}$ with equal probability. The class-specific index coefficient vector was set to $\gamma_{(1)}=(0, 2.5)^{\top}$ with $Z_{(1)}=1$ for $k=2$, 
and to $\gamma_{(1)}=(0, 0.1, 0.1, 1.5, 0.3, 1, 3, 0.3, 1)^{\top}$, with $Z_{(1)}$ drawn uniformly from $\{(1, 0, 0)^\top, (1, 1, 0)^\top, (1, 0, 1)^\top\}$, for $k=3$. The latent class membership probabilities were specified as
\begin{align*}
&\pi(1 \mid x)=0.25+\frac{2.5\phi\big(2.8\alpha^{\top}_{1}x\big)}{1+\phi\big(2.8\alpha^{\top}_{1}x\big)}, \text{ with }\alpha_1 =(1,0,0,0,0.29)^{\top}, \text{ for } k=2,\text{ and}\\
&\pi(1 \mid x)=0.15+\frac{1.5\phi\big(0.1\alpha^{\top}_{1}x\big)}{1+\phi\big(0.1\alpha^{\top}_{1}x\big)}\text{ and }\pi(2 \mid x) = 0.3, \text{ with } \alpha_1 = (1,1,35,0,0,0,10)^{\top}, \text{ for } k=3,
\end{align*}
where $\phi(u)$ denotes the standard normal density. 

\subsection{Performance of Estimation, Class Recovery, and Class Number Selection} \label{sec:sec5.1}

Agreement between the realized latent class partition and each estimated latent class partition was quantified using the Rand index (RI; \citealp{rand1971objective}). 
Table \ref{tab:RI} reports the RI values for the SP- and BR-estimated latent class partitions when the number of classes is known. Both estimators perform well across the examined settings; however, when the distributional structure is correctly specified, the BR-estimated latent class partition more accurately recovers the realized latent class partition. Accuracy improves for both estimated latent class partitions as the sample size increases. However, distinct patterns emerge: the performance gap narrows for $k=2$, whereas the advantage of the BR-estimated latent class partition becomes apparent for $k=3$. These patterns persist across the three error distributions considered.

\begin{table}[htbp]
  \centering 
  \caption{Means (standard deviations) of 500 RI values for the SP- and BR-estimated latent class partitions under various model settings.}
      \begin{adjustbox}{max width=1\linewidth}
    \begin{tabular}{cccccccccccccccccc}
    \toprule
            & \multicolumn{8}{c}{$k=2$} & & \multicolumn{8}{c}{$k=3$}\\
      \cmidrule(rl){2-9} \cmidrule(rl){11-18}
                & \multicolumn{2}{c}{normal} &       & \multicolumn{2}{c}{logistic} &       & \multicolumn{2}{c}{Gamma} & & \multicolumn{2}{c}{normal} &       & \multicolumn{2}{c}{logistic} &       & \multicolumn{2}{c}{Gamma} \\
          \cmidrule(rl){2-3} \cmidrule(rl){5-6} \cmidrule(rl){8-9} \cmidrule(rl){11-12} \cmidrule(rl){14-15} \cmidrule(rl){17-18}
            $n$    & SP    & BR    &       & SP    & BR    &       & SP    & BR &       & SP    & BR   &       & SP    & BR    &       & SP    & BR \\
            \midrule
    200   & 0.967 & 0.969 &       & 0.962 & 0.964 &       & 0.964 & 0.966 &       & 0.872 & 0.878 &       & 0.865 & 0.873 &       & 0.867 & 0.871 \\
          & (0.0382) & (0.0383) &   & (0.0308) & (0.0302) &   & (0.0269) & (0.0247) &   & (0.0989) & (0.1040) &   & (0.1014) & (0.1066) &   & (0.1047) & (0.1124) \\
    400   & 0.978 & 0.979 &       & 0.970 & 0.971 &       & 0.973 & 0.974 &       & 0.944 & 0.954 &       & 0.938 & 0.947 &       & 0.943 & 0.953 \\
          & (0.0116) & (0.0108) &   & (0.0126) & (0.0122) &   & (0.0130) & (0.0128) &   & (0.0361) & (0.0302) &   & (0.0410) & (0.0359) &   & (0.0393) & (0.0325) \\
    600   & 0.981 & 0.981 &       & 0.973 & 0.974 &       & 0.974 & 0.975 &       & 0.956 & 0.961 &       & 0.955 & 0.960 &       & 0.957 & 0.964 \\
          & (0.0082) & (0.0081) &   & (0.0099) & (0.0098) &   & (0.0110) & (0.0109) &   & (0.0220) & (0.0189) &   & (0.0189) & (0.0136) &   & (0.0226) & (0.0166) \\
          \bottomrule 
    \end{tabular}%
           \end{adjustbox}
  \label{tab:RI}%
\end{table}%

Estimation performance for a generic parameter vector $\theta^{\mathrm{o}}$ was evaluated using the normalized root squared error (RSE): $\sqrt{(\breve{\theta}-\theta^{\text{o}})^{\top}(\breve{\theta}-\theta^{\text{o}})/|\theta^{\text{o}} |}$, where 
$\breve{\theta}$ denotes the corresponding estimator. 
Table \ref{tab:oracle} compares the RSEs of the SP and PML estimators with those of their respective oracle counterparts when the number of classes is known. Across all simulation settings, the PML estimator and its corresponding oracle estimator achieve lower RSEs than their SP counterparts, respectively, reflecting efficiency gains from leveraging the underlying distributional structure. As the sample size increases, the RSEs of both SP and PML estimators decrease and converge to those of their oracle counterparts. This pattern confirms that the proposed estimators achieve near-oracle efficiency in large samples. Notably, the deviations from the oracle estimators are smaller for $k=2$ than for $k=3$, reflecting the superior latent class recovery accuracy reported in Table \ref{tab:RI}.

\begin{table}[htbp]
  \centering 
  \caption{Means of 500 RSEs for SP and PML estimates of the class index coefficients, along with the corresponding oracles ($\text{ORCL}_{1}$ and $\text{ORCL}_{2}$), under various model settings.}
    \scriptsize
      \begin{adjustbox}{max width=1\linewidth}
    \begin{tabular}{cccccccccccccccc}
    \toprule
              &       & \multicolumn{4}{c}{normal} &       & \multicolumn{4}{c}{logistic} &       & \multicolumn{4}{c}{Gamma}\\ 
          \cmidrule(rl){3-6} \cmidrule(rl){8-11} \cmidrule(rl){13-16}
         & $n$      & SP    & PML   & $\text{ORCL}_{1}$    & $\text{ORCL}_{2}$   &       & SP    & PML   & $\text{ORCL}_{1}$    & $\text{ORCL}_{2}$  &       & SP    & PML   & $\text{ORCL}_{1}$    & $\text{ORCL}_{2}$ \\
      \midrule
    $k=2$    & 200   & 0.226 & 0.197 & 0.201 & 0.174 &       & 0.208 & 0.189 & 0.198 & 0.176 &       & 0.192 & 0.171 & 0.187 & 0.156 \\
          & 400   & 0.125 & 0.110 & 0.122 & 0.108 &       & 0.116 & 0.103 & 0.113 & 0.093 &       & 0.117 & 0.092 & 0.112 & 0.085 \\
          & 600   & 0.087 & 0.077 & 0.087 & 0.076 &       & 0.089 & 0.074 & 0.086 & 0.073 &       & 0.090 & 0.065 & 0.089 & 0.064 \\
        \midrule\midrule
    $k=3$    & 200   & 0.351 & 0.335 & 0.273 & 0.244 &       & 0.343 & 0.333 & 0.277 & 0.251 &       & 0.341 & 0.338 & 0.279 & 0.240 \\
          & 400   & 0.189 & 0.168 & 0.133 & 0.110 &       & 0.178 & 0.159 & 0.138 & 0.107 &       & 0.172 & 0.143 & 0.145 & 0.105 \\
          & 600   & 0.117 & 0.099 & 0.105 & 0.078 &       & 0.109 & 0.092 & 0.099 & 0.074 &       & 0.102 & 0.080 & 0.096 & 0.063 \\
          \bottomrule 
    \end{tabular}%
           \end{adjustbox}
  \label{tab:oracle}%
\end{table}%
Tables \ref{tab:S_RI} and \ref{tab:m_est} report the performance of the proposed methods relative to the SP method for class-specific linear regression 
($\text{SP}_{\text{L}}$ method), which circumvents the computational complexity of the pairwise-fusion-penalty method \citep{ma2017concave,ma2020exploration}, and the mixture maximum-likelihood method for class-specific linear regression ($\text{MML}$ method) of \citet{jordan1994hierarchical}, with normally distributed errors and logistic class membership probabilities. Estimation performance for the subject-specific conditional means was evaluated using the normalized RSE:
$\sqrt{\sum_{i=1}^{n}\big(\check{m}\big(\check{\beta}^{\top}_{i}Z_i\big) - m\big(\beta^{\text{o}\top}_{i}Z_i\big)\big)^2/n}$,
where $\check{m}(v)$ denotes either a prespecified or an estimated response function, and $\check{\beta}_1, \dots, \check{\beta}_n$ are estimators of subject index coefficient vectors.
For $k=2$, the $\text{SP}_{\text{L}}$ method demonstrates competitive performance in both RI and RSE, suggesting that a linear approximation adequately captures the underlying nonlinear response.
However, its limitations become evident when $k=3$: performance of latent class recovery is substantially 
inferior for $n\geq 400$, and estimation of subject-specific conditional means exhibits larger RSE across the sample sizes considered.
In contrast, the $\text{MML}$ method generally produces lower RI values and higher RSEs, reflecting reduced accuracy in both latent class recovery and estimation.

Table \ref{tab:k} summarizes the performance of three semiparametric information criteria in selecting the number of classes. For $k=2$, all criteria correctly select the true number of classes in almost all replications across all experimental settings. In the more challenging scenario with $k=3$, underestimation occurs at
$n=200$, most notably for $\text{SPIC}_1$. Selection accuracy increases with sample size, with all criteria exhibiting comparable performance by
$n=600$. Overall, the proposed criterion $\text{SPIC}_{0}$, which does not rely on the underlying distributional structure, performs at least as well as the modified criteria $\text{SPIC}_{1}$ and $\text{SPIC}_{2}$.

\begin{table}[htbp]
  \centering
  \caption{Means (proportions of correct selection) of 500 estimated numbers of classes
  from various selection criteria under various model settings.}
  \scriptsize
  \begin{adjustbox}{max width=1\linewidth}
  \begin{tabular}{ccccccccccccc}
    \toprule
          &       & \multicolumn{3}{c}{normal} &       & \multicolumn{3}{c}{logistic} &       & \multicolumn{3}{c}{Gamma}\\
          \cmidrule(rl){3-5} \cmidrule(rl){7-9} \cmidrule(rl){11-13}
          & $n$   & $\text{SPIC}_0$ & $\text{SPIC}_1$ & $\text{SPIC}_2$ &
                  & $\text{SPIC}_0$ & $\text{SPIC}_1$ & $\text{SPIC}_2$ &
                  & $\text{SPIC}_0$ & $\text{SPIC}_1$ & $\text{SPIC}_2$ \\
    \midrule
    $k=2$ & 200 & 2.01    & 2.00    & 2.00    & & 2.00    & 2.00    & 2.00    & & 2.00    & 2.00    & 2.00    \\
          &     & (0.992) & (0.998) & (0.998) & & (1.000) & (1.000) & (1.000) & & (0.996) & (1.000) & (1.000) \\
          & 400 & 2.01    & 2.00    & 2.00    & & 2.00    & 2.00    & 2.00    & & 2.01    & 2.00    & 2.00    \\
          &     & (0.994) & (1.000) & (1.000) & & (0.996) & (1.000) & (1.000) & & (0.994) & (0.998) & (1.000) \\
          & 600 & 2.00    & 2.00    & 2.00    & & 2.00    & 2.00    & 2.00    & & 2.00    & 2.00    & 2.00    \\
          &     & (0.996) & (1.000) & (1.000) & & (0.998) & (0.998) & (0.998) & & (1.000) & (1.000) & (1.000) \\
    \midrule \midrule
    $k=3$ & 200 & 2.73    & 2.01    & 2.34    & & 2.78    & 2.01    & 2.36    & & 2.81    & 2.01    & 2.36    \\
          &     & (0.722) & (0.006) & (0.336) & & (0.774) & (0.002) & (0.356) & & (0.792) & (0.020) & (0.358) \\
          & 400 & 3.01    & 2.87    & 3.00    & & 3.00    & 2.90    & 2.99    & & 3.00    & 2.92    & 2.98    \\
          &     & (0.990) & (0.872) & (0.974) & & (0.988) & (0.898) & (0.958) & & (0.992) & (0.922) & (0.974) \\
          & 600 & 3.00    & 3.00    & 3.00    & & 3.00    & 3.00    & 3.00    & & 3.00    & 3.00    & 2.99    \\
          &     & (0.996) & (1.000) & (1.000) & & (1.000) & (0.996) & (0.998) & & (1.000) & (0.996) & (0.994) \\
    \bottomrule
  \end{tabular}
  \end{adjustbox}
  \label{tab:k}
\end{table}

To evaluate the performance of the criteria associated with the $\text{SP}_{\text{L}}$ and $\text{MML}$ methods for selecting the number of classes, Table \ref{tab:S_k} reports results for the best-performing modified Bayesian information criterion ($\text{BIC}_1$) among those proposed by \citet{ma2017concave} and \citet{ma2020exploration}, as well as the Bayesian-type information criterion ($\text{BIC}_2$) of \citet{khalili2010new}. While the $\text{SP}_{\text{L}}$ method demonstrates robustness to model misspecification, $\text{BIC}_1$ systematically selects the maximum allowable number of classes, indicating that its penalty term inadequately accounts for the structural complexity of the nonlinear response.
In contrast, $\mathrm{BIC}_2$ avoids this severe overestimation but remains biased. For $k=2$, this criterion exhibits a modest upward bias in the selected number of classes, whereas for $k=3$, it underestimates the number of classes for $n \le 400$ and exhibits instability at $n=600$.
The inferior performance of these parametric information criteria underscores the necessity of the proposed semiparametric information criteria when structural or distributional assumptions are violated.

\section{Applications} 
\label{sec:sec6}

We demonstrate the proposed methodology through two illustrative applications in personalized marketing and patient stratification. To ensure comparability across methods, all continuous covariates were standardized to have mean zero and variance one prior to model fitting.  Statistical significance was evaluated at $\alpha=0.05$. 

\subsection{Application in Customer Segmentation}
\label{sec:sec6.1}

Customer segmentation is essential for targeted marketing, allowing firms to identify latent groups with homogeneous purchasing behavior. We apply the proposed methodology to high-frequency transaction data from a major North American supermarket chain collected between 2017 and 2020. The dataset comprises transaction records linked to detailed product information. To ensure temporal stability and avoid COVID-19 distortions, we restrict the analysis to loyal 2018 customers completing at least 13 transactions in both H1 and H2. Following industry practice, multiple transactions occurring within the same week are aggregated into a single purchasing occasion to capture routine shopping behavior better. After this consolidation, the analysis sample comprises 1,906 customers, with 28,376 and 29,111 transactions in H1 and H2, respectively.

We consider an empirical application that estimates the log-transformed purchase amount in H2 ($\textit{pah}2$) using behavioral features measured in H1. The covariates include the log-transformed purchase amount ($\textit{pah}1$), log-transformed average basket size (\textit{abs}), the number of distinct product categories purchased among the top 30 categories (\textit{cate}), an indicator for high-intensity natural food consumption (\textit{nf}; 1 if the purchase amount is at or above the median and 0 otherwise), and an indicator for baby-product purchases (\textit{baby}; 1 if at least one such purchase occurred and 0 otherwise).
The CSIM was specified with class-specific intercepts and class-invariant index coefficients for the covariates, identified during the initialization stage, to account for latent heterogeneity in customer spending. The number of classes was selected using the information criteria $\text{SPIC}_0$, $\text{SPIC}_1$, and $\text{SPIC}_2$, all of which favored a four-class solution. The BR-estimated latent class partition coincided with the SP-estimated latent class partition; Classes 1--4 contained 702, 301, 582, and 321 customers, respectively. The corresponding class-specific mean purchase amounts in H2 (491.20, 714.27, 359.60, and 254.96) indicate that the identified classes capture well-separated levels of customer spending.

\begin{figure}[H]
    \centering
    \includegraphics[width=16cm,height=8cm]{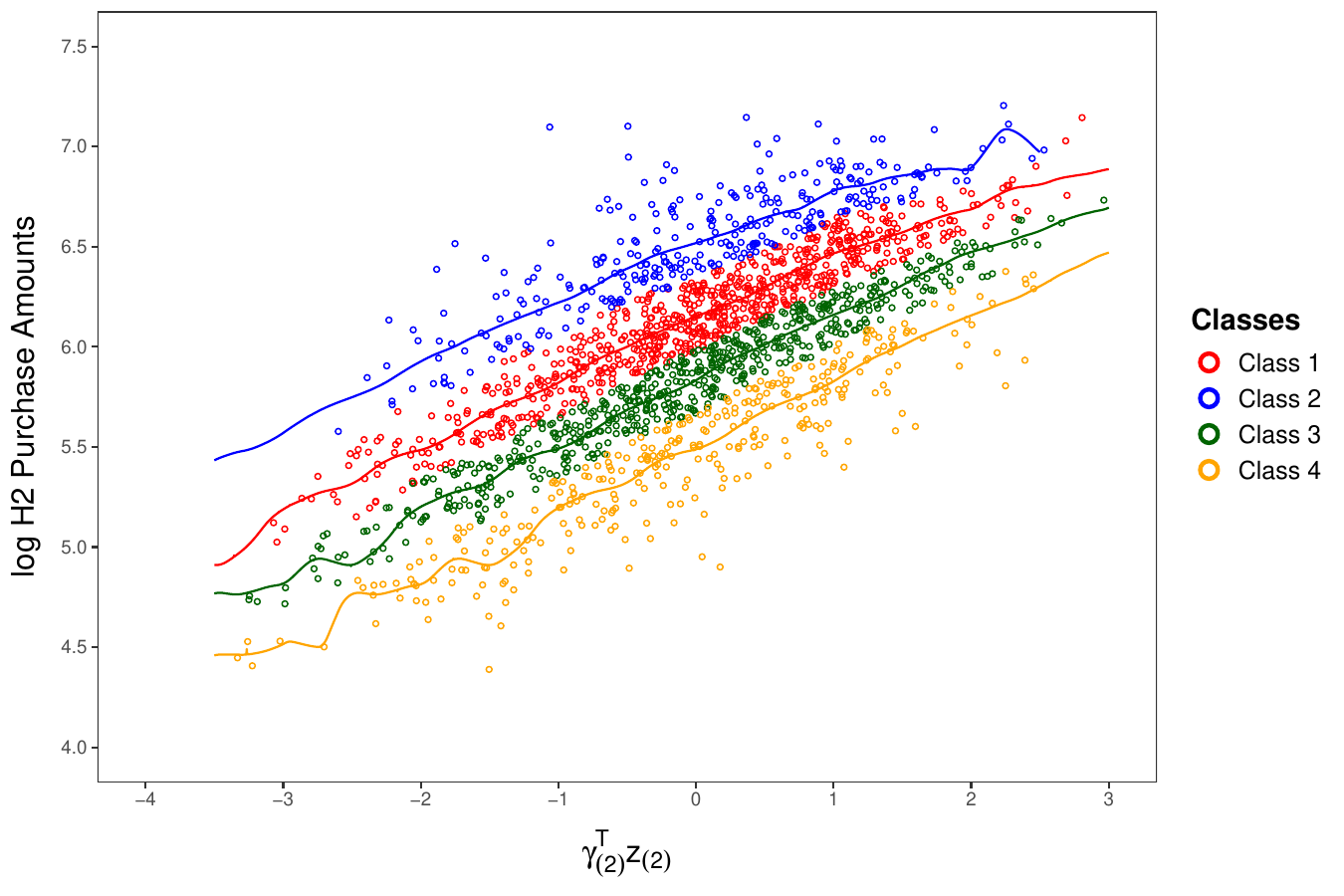}
    \caption{Scatter plot of log-transformed purchase amounts in H2, overlaid with the estimated conditional mean functions given the class-invariant index.}
    \label{fig:ASCE_mean}
\end{figure}

Using the PML method, we obtain the estimates of the CSIM parameters, with standard errors in parentheses. The estimated class-specific intercepts for Classes 1--4 are 0, 1.26 (0.032), -0.97 (0.022), and -1.98 (0.032), respectively. The coefficient of \textit{pah}1 is fixed at 1 for all four classes. The remaining coefficients are common across classes, with estimates of 0.19 (0.013) for \textit{abs}, -0.03 (0.008) for \textit{cate}, -0.05 (0.015) for \textit{nf}, and 0.04 (0.025) for \textit{baby}.
Standard errors were computed from the asymptotic variance matrix estimator $\widetilde{V}^{-1}$ described below Theorem \ref{Thm4.2}. The observed significant differences in the class-specific intercept estimates underscore the necessity of the multi-class structure and support the validity of our class number selection criteria. Among the class-invariant index coefficient estimates, $\textit{pah}1$ and $\textit{abs}$ exhibit significant associations with future spending, whereas $\textit{cate}$ and $\textit{nf}$ are adversely associated. Figure \ref{fig:ASCE_mean} presents the estimated conditional mean functions given the class-invariant index, which exhibit an approximately increasing relationship.

Based on the estimated number of classes and the corresponding estimated latent class partitions, the structural dimension of the SDR-based class membership probabilities was determined to be one, with Class 4 designated as the reference group. The coefficient estimates (standard errors) of the central subspace direction are 1 for \textit{pah}1, 0.59 (0.051) for \textit{abs}, -1.66 (0.077) for \textit{cate}, 3.32 (0.137) for \textit{nf}, and 3.74 (0.131) for \textit{baby}. The effects of $\textit{pah}1$, $\textit{abs}$, $\textit{nf}$, and $\textit{baby}$ are all significant in one direction, while the effect of $\textit{cate}$ is significant in the opposite direction, supporting covariate-dependent class membership probabilities. Figure \ref{fig:ASCE_prob} further shows that the logistic model fails to capture the structure of the latent class membership probabilities.

Based on the near-linear trends observed in the estimated conditional mean functions (Figure \ref{fig:ASCE_mean}) and the limited flexibility of the logistic class membership probabilities (Figure \ref{fig:ASCE_prob}), we fitted a class-specific linear model using the $\text{SP}_{\text{L}}$ method for comparison. Although $\text{BIC}_1$ suggested 15 classes, simulation results indicated that fewer classes were supported. Accordingly, we restricted the model to four classes, aligning it with the fitted CSIM and facilitating an assessment of class agreement. The estimated class-specific intercepts for Classes 1--4 are 0.38 (0.013), 1.08 (0.017), -0.23 (0.012), and -1.05 (0.017), respectively. The coefficient of \textit{pah1} is fixed at 1 for all four classes. The coefficients are common across classes, with estimates (standard errors) of 0.64 (0.013) for \textit{pah}1, 0.11 (0.009) for \textit{abs}, -0.01 (0.008) for \textit{cate}, -0.05 (0.013) for \textit{nf}, and 0.04 (0.020) for \textit{baby}.
The $\text{SP}_{\text{L}}$ estimates have the same interpretations as the PML estimates, and the corresponding linear mean functions (Figure \ref{fig:ASCE_L_mean}) closely match those in Figure \ref{fig:ASCE_mean}. For this dataset, the high RI (0.864) between the $\text{SP}_{\text{L}}$- and BR-estimated latent class partitions provides, at most, evidence of the robustness of class-specific linear regression in the presence of mild model misspecification, highlighting the need for a rigorous evaluation of model goodness-of-fit.

\subsection{Application in ACTG Study 175} 
\label{sec:sec6.2}

The ACTG Study 175 was a landmark randomized clinical trial that had a lasting impact on the development of antiretroviral treatment strategies.
The trial evaluated the comparative efficacy of four antiretroviral regimens in adults infected with HIV-1, all of whom had baseline CD4 T-cell counts between 200 and 500 cells/mm$^{3}$.
Our analysis used the ACTG175 dataset available in R, which comprises 2,139 adults from the trial.
A detailed description of the study design and primary findings is provided in \cite{hammer1996trial}.

Our second application of the proposed methodology analyzes log-transformed CD4 cell counts measured at $20\pm5$ weeks (\textit{cd4}) as the response variable. The covariates include treatment indicator (\textit{tr}; 0 = zidovudine monotherapy, 1 = all other regimens combined), log-transformed baseline CD8 cell counts (\textit{cd8}), age (\textit{age}), weight (\textit{weight}), Karnofsky performance score (\textit{ks}), and gender (\textit{gender}; 0 = female, 1 = male). 
To enhance numerical stability, individuals with extreme values in either the response variable or covariates were excluded, yielding a final sample of 2,054 adult participants. The CSIM was specified with class-specific intercepts and class-specific non-zidovudine treatment effects, identified during the initialization stage, to account for heterogeneity in treatment effects across latent classes.
In this analysis, $\textsc{SPIC}_0$ selected a five-class solution, whereas $\textsc{SPIC}_1$ and $\textsc{SPIC}_2$ favored a four-class configuration. The five-class model exhibited statistically significant differences in the estimated intercepts and treatment coefficients across classes, supporting $k = 5$ as a reliable and plausible representation of latent treatment--response heterogeneity. The BR-estimated latent class partition coincided with the SP-estimated latent class partition. Under both estimated latent class partitions, Classes 1--5 contained 663, 295, 482, 185, and 429 adults, respectively.

\begin{figure}[H]
    \centering
    \includegraphics[width=16cm,height=8cm]{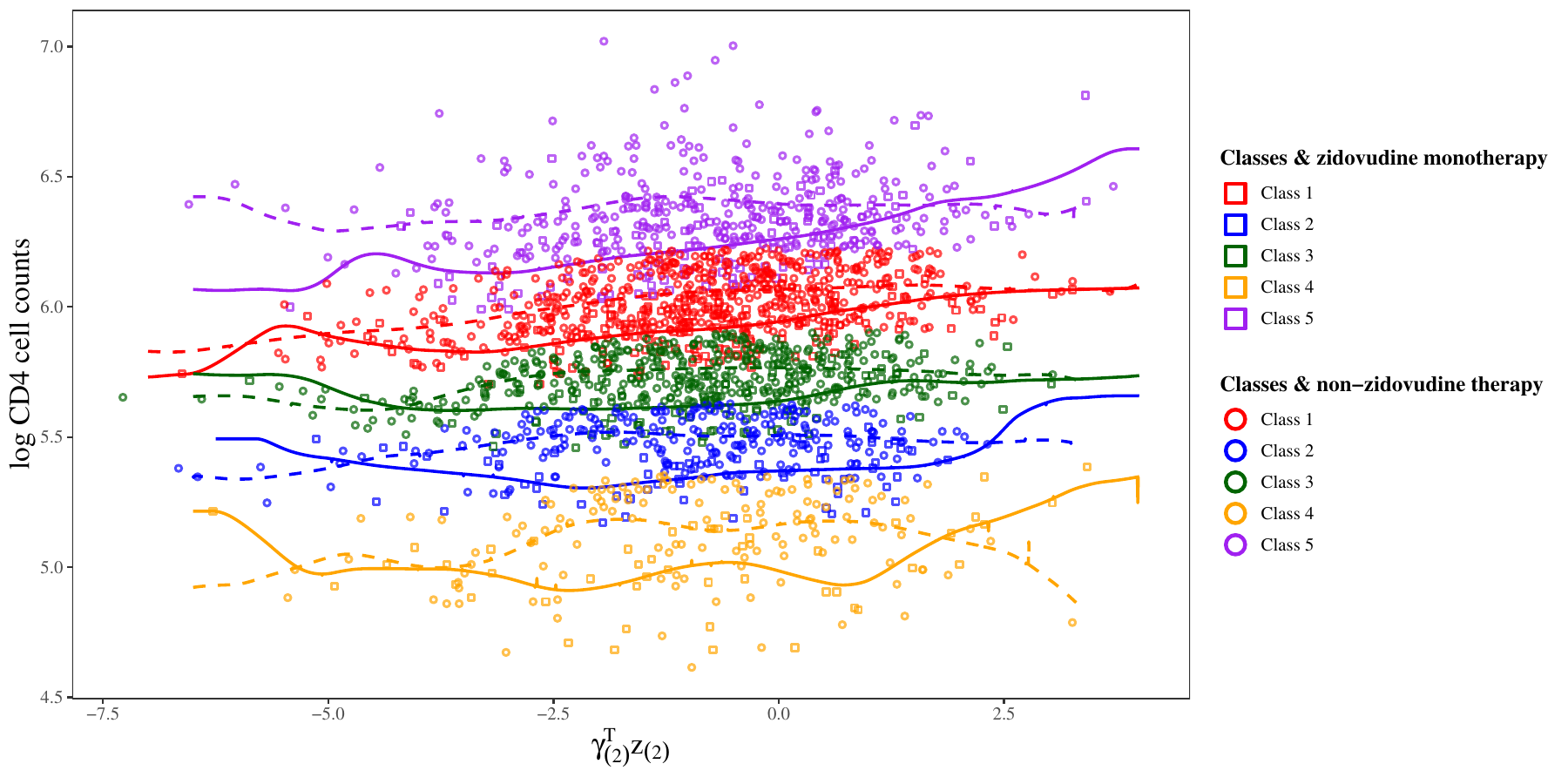} 
    \caption{Scatter plot of log-transformed CD4 cell counts, overlaid with the estimated conditional mean functions given the class-invariant index (solid line: zidovudine monotherapy; dashed line: non-zidovudine therapy).}
    \label{fig:ACTG_mean}
\end{figure}

Using the PML method, we obtain estimates of the CSIM parameters, with standard errors in parentheses. The estimated class-specific intercepts for Classes 1--5 are 0, -29.37 (0.637), -10.80 (0.171), -53.53 (0.125), and 12.52 (0.590), respectively. The coefficient of \textit{cd}8 is fixed at 1 across all five classes. The coefficient of \textit{tr} varies across classes, with estimates of 3.41 (0.227), -10.59 (0.502), -8.52 (0.264), -25.37 (0.272), and 63.03 (0.219) for Classes 1--5, respectively. The remaining coefficients are shared across classes, with estimates of -0.85 (0.056) for \textit{age}, 0.16 (0.053) for \textit{weight}, 0.89 (0.081) for \textit{ks}, and -1.04 (0.161) for \textit{gender}.
The class-specific index coefficient estimates indicate that Class 5 has the highest estimated CD4 cell counts across all treatment conditions, followed in descending order by Classes 1, 3, 2, and 4. The non-zidovudine treatment effects exhibit significant heterogeneity across subgroups, with effects in one direction for Classes 2, 3, and 4 and in the opposite direction for Classes 1 and 5. Among the class-invariant index coefficient estimates, \textit{cd8}, \textit{weight}, and \textit{ks} show significant associations with \textit{cd4}, while \textit{age} and \textit{gender} exhibit significant adverse associations.
Figure \ref{fig:ACTG_mean} presents the estimated conditional mean functions given the class-invariant index, which exhibit a distinct non-monotonic pattern, indicating that class-specific linear regression is inadequate. Additionally, the unreported $\text{SP}_{\text{L}}$ estimates from a class-specific linear model fail to capture the complex treatment--response heterogeneity across subgroups revealed by the fitted CSIM.

With Class 5 designated as the reference group, the structural dimension in the SDR-based class membership probabilities was identified as one. 
The coefficient estimates (standard errors) of the central subspace direction are 1 for \textit{age}, 0.66 (0.164) for \textit{tr}, -0.19 (0.075) for \textit{cd}8, 0.32 (0.082) for \textit{weight}, 0.50 (0.097) for \textit{ks}, and 0.00 (0.135) for \textit{gender}.
The coefficient estimates indicate that \textit{tr}, \textit{age}, and \textit{ks} have significant contributions to the class membership probabilities, whereas \textit{cd8} contributes in the opposite direction. These findings provide compelling evidence of covariate-dependent class membership probabilities. Figure \ref{fig:ACTG_prob} further highlights the limitations of the logistic model in capturing the underlying structure of these probabilities.

\section{Concluding Remarks and Future Challenges}
\label{sec:sec7}

We introduce semiparametric class-specific index regression that characterizes the conditional mean of a response variable and accommodates a general distributional specification. The Bayes classification rule asymptotically maximizes the probability of correct assignment, and the proposed PML estimator achieves semiparametric efficiency.
Building on the subject-specific formulation of the CSIM, we develop a SP estimation procedure implemented via a computationally efficient algorithm that avoids high-dimensional matrix operations and updates auxiliary parameters in explicit closed-form at each ADMM iteration. In our initialization procedure, the index coefficients of the CSIM and the latent class partition of the individual-level index set are estimated by minimizing the pseudo-least-squares criterion in (\ref{fn:pss1}) for each candidate number of classes. This strategy promotes rapid and stable convergence of the SP estimation procedure. Simulation studies and empirical data analyses demonstrate that our method outperforms existing alternatives in estimation precision and latent class recovery accuracy.
 
A central challenge lies in the joint identification and characterization of both the response function and the underlying distributional form within the proposed framework. 
To evaluate the adequacy of the assumed parametric specification for the response function, our estimation procedure---together with its parametric analogue---provides a theoretically grounded and rigorous framework for model assessment. Although the class-specific index distribution model of \citet{teng2025unsupervised} encompasses ours as a special case when the induced response function is nondegenerate, simulation evidence demonstrates that our method achieves superior performance. In particular, the pseudo-maximum likelihood estimators exhibit poor finite-sample properties, largely due to the sparsity inherent in estimating the bivariate probability density function. Nevertheless, the pseudo-least-integrated-squares procedure, combined with our distributional estimator, remains a valuable instrument for evaluating the adequacy of the assumed distributional specification.

Weekly purchase amounts and longitudinally measured CD4 cell counts were analyzed for customer segmentation and HIV/AIDS patient stratification, respectively. The CSIM can be extended to accommodate longitudinal outcomes with class-specific, time-varying covariate effects. A modified pseudo-least-squares criterion, augmented with an $\ell_{1}$ separation penalty, naturally accounts for correlations among repeated measurements within individuals. By remaining distribution-free, the proposed semiparametric information criterion can be adapted to consistently select the number of classes in longitudinal settings. Collectively, the methodology introduced in this work establishes a rigorous framework for inference in complex clustered and longitudinal data structures.

\end{spacing}

\bibliographystyle{biom}
\bibliography{reference}{}

\newpage

\appendix

\section*{Appendix}

\setcounter{thm}{0}
\renewcommand{\thethm}{S\arabic{thm}}
\renewcommand{\theHthm}{S\arabic{thm}}  

\setcounter{rem}{0}
\renewcommand{\therem}{S\arabic{rem}}
\renewcommand{\theHrem}{S\arabic{rem}}  

\setcounter{table}{0}
\renewcommand{\thetable}{S\arabic{table}}
\renewcommand{\theHtable}{S\arabic{table}} 

\setcounter{figure}{0}
\renewcommand{\thefigure}{S\arabic{figure}}
\renewcommand{\theHfigure}{S\arabic{figure}} 

\renewcommand{\theequation}{\thesection.\arabic{equation}}

\section{Summary of Notation} \label{subsec:notation}
\vspace{-0.5in}
\begin{longtable}{@{}p{0.25\textwidth} p{0.7\textwidth}@{}} 
    \caption*{} \label{tab:symbols} \\ 
    \toprule
    \textbf{Symbol} & \textbf{Description} \\
    \midrule
    \endfirsthead 

    \toprule
    \textbf{Symbol} & \textbf{Description} \\
    \midrule
    \endhead 

    \midrule
    \multicolumn{2}{r}{Continued on next page} \\
    \bottomrule
    \endfoot 

    \bottomrule
    \endlastfoot 

$Y$ & Response variable with support $\mathcal{Y}$ (Section \ref{sec:sec2}, beginning)\\
 
$X$ & Covariate vector with support $\mathcal{X}$  (Section \ref{sec:sec2}, beginning)\\

$k$ & Number of classes (Section \ref{sec:sec2}, beginning)\\

$C$ & Latent class variable with support $\mathcal{C} = \{1,\dots,k\}$ (Section \ref{sec:sec2}, beginning)\\

$\gamma_c$ & Index coefficient vector for Class $c \in \mathcal{C}$ in the CSIM (\ref{fn:CIF_1})\\

$Z$ & Augmented covariate vector $Z = (1, X^{\top})^{\top}$ in the CSIM (\ref{fn:CIF_1})\\

$m(v)$ & Response function in the CSIM (\ref{fn:CIF_1})\\

$\beta_i$ & Subject index coefficients in the SSIM (\ref{fn:SIM}) with $\beta_i \in \{ \gamma_{1}, \dots, \gamma_{k}\}$ and $\beta_{i}=\gamma_{c}$ if and only if $C_{i}=c$\\

$U$ & Error term in the SSIM (\ref{fn:SSIM_1})\\

$f(\cdot)$ & Probability density function of $U$\\

$(\widehat{\beta},\widehat{\gamma})$ & SP estimator of $(\beta^{\text{o}}, \gamma^{\text{o}})$ (Section \ref{sec:sec3.1})\\

$\widehat{\mathcal{G}}$ & SP-estimated latent class partition of the realized latent partition $\mathcal{G}^{\text{o}}$ (Section \ref{sec:sec3.1})\\

$\widehat{m}_h(v; \beta)$, $\widehat{m}_h(v; \mathcal{G}, \gamma)$ & Estimator of $m(v)$ (\ref{fn:m_est}) and its alternative representation (\ref{fn:m_n_hat_subject})\\

$A_d$, $d^{\text{o}}$, $A^{\text{o}}$ & Basis matrix, structural dimension, and CS matrix (Section \ref{sec:sec4.1})\\

$\{\pi_{c}(\nu)\}^{k}_{c=1}$ & Class membership functions in the SDR-based model (\ref{fn:SDR})\\

$\widehat{\pi}_{c,h_d}(\nu;A_d)$ & Estimator of $\pi_c(\nu)$ (\ref{fn:SDR})\\

$\widetilde{f}_{\xi}(u;\gamma)$ & Estimator of $f(u)$ (\ref{fn:CSIM_pdf})\\

$\widetilde{\gamma}$ & PML estimator of $\gamma^{\text{o}}$ (Section \ref{sec:sec4.2})\\
\end{longtable}

\section{SP Estimation Procedure}
\subsection{Initialization Strategy} 
\label{sec:IS} 

To implement the SP estimation procedure, we first fit a semiparametric single-index mean model, $m(\gamma_1^{\top} z)$, and a linear mean model, $\gamma_1^{\top}z$. Based on these fits, we compute the residuals
$\big\{Y_i - \widehat{m}^{-i}_{\tilde{h}}(\widehat{\gamma}_1^{\top} Z_i; \widehat{\gamma}_1) \big\}_{i = 1}^n$, where $\widehat{m}_{\tilde{h}}(\widehat{\gamma}_1^{\top} z; \widehat{\gamma}_1)$ is obtained by minimizing the pseudo-least squares criterion in (\ref{fn:pss1}) with $k=1$, and $\big\{Y_i -\widehat{\gamma}_1^{\top} Z_i \big\}_{i = 1}^n$, where $\widehat{\gamma}$ is the ordinary least squares estimator.
We then apply a distribution-free clustering procedure, such as $k$-means or hierarchical clustering, to each set of residuals, yielding initial estimated latent class partitions for a given number of classes $c=2,\dots,\bar{k}$, where $\bar{k}$ may be chosen, for instance, as $\lfloor \sqrt{n} \rfloor$. Substituting these estimators into (\ref{fn:pss1}) yields two pseudo-least squares estimators; we define $\bar{\gamma}^{c}$ as the one achieving the smaller objective value. The corresponding estimated latent class partition and subject coefficient estimators are denoted by $\bar{\mathcal{G}}^{c}=\{\bar{\mathcal{G}}^{c}_{1},\dots,\bar{\mathcal{G}}^{c}_{c}\}$ and $\bar{\beta}^{c}$, respectively.

A new estimated latent class partition $\check{\mathcal{G}}^{c}$ is obtained by assigning each observation $i$ to $\check{\mathcal{G}}^{c}_{c_{0}}$ if 
$\textsc{pss}_{i}(c_{0}) = \min_{\{1 \leq \ell \leq c\}} \textsc{pss}_{i}(\ell)$, where
$\textsc{pss}_{i}(\ell) =  \big(Y_i - \widehat{m}^{-i}_{\tilde{h}}\big(\bar{\gamma}^{c\top}_{\ell}Z_i; \bar{\mathcal{G}}^{c}, \bar{\gamma}^{c}\big) \big)^2$, $i = 1, \dots, n$, $\ell = 1, \dots, c$.
The updated pair $\big(\check{\mathcal{G}}^{c}, \check{\gamma}^{c}\big)$ replaces
$\big(\bar{\mathcal{G}}^{c}, \bar{\gamma}^{c}\big)$ if
$\textsc{pss}\big(\check{\gamma}^{c}, \check{\mathcal{G}}^{c}\big)< \textsc{pss}\big(\bar{\gamma}^{c}, \bar{\mathcal{G}}^{c}\big)$;
otherwise, $\big(\check{\mathcal{G}}^{c},\check{\gamma}^{c}\big)$ is set to $\big(\bar{\mathcal{G}}^{c}, \bar{\gamma}^{c}\big)$, 
where $\check{\gamma}^{c}$ is obtained by minimizing $\textsc{pss}\big(\gamma^{c},\check{\mathcal{G}}^{c}\big)$ with respect to $\gamma^{c}$. 
This procedure is iterated until the objective function attains its minimum or a specified convergence threshold is met. 
For each $c=1,\dots,\bar{k}$, let $\check{\beta}^{c}$ denote the corresponding subject coefficient estimator obtained from initializer $\big(\check{\mathcal{G}}^{c}, \check{\gamma}^{c}\big)$. The reported experiments employ the initialization strategy described above; they do not establish whether alternative initial partitions would lead to equivalent final estimates or partitions.

\subsection{Computational Algorithm}
\label{sec:Algorithm} 
The optimization problem in (\ref{3.1.1}) can be rewritten as
\begin{align}
&\text{Minimize }\frac{1}{2} \sum_{i=1}^n \big(Y_{i}-\widehat{m}^{-i}_{h}\big(\beta_{i}^{\top}Z_i; \beta\big)\big)^2 + \lambda\sum^{n}_{i=1}\min_{1\leq c\leq k} \|\eta_{ic} \|_{1}  \nonumber \\
&\text{subject to } \beta_{(1)i} -\gamma_{(1)c}=\eta_{ic}, ~ i=1,\dots,n, ~ c = 1,\dots, k. \label{A.5}
\end{align}
Because the objective function is separable in $(\beta, \gamma)$ and $\{\eta_{ic}\}$, we minimize the augmented Lagrangian via ADMM:
\begin{align}
 &\frac{1}{2} \sum_{i=1}^n \big(Y_{i}-\widehat{m}^{-i}_{h}\big(\beta_{i}^{\top}Z_i; \beta\big)\big)^2 + \lambda\sum^{n}_{i=1}\min_{1\leq c\leq k} \|\eta_{ic}\|_{1}
 +\sum_{i=1}^{n} \sum_{c=1}^k  \nu_{i c}^{\top} (\beta_{(1)i}-\gamma_{(1)c}-\eta_{i c})\nonumber\\
 &+\frac{1}{2}\sum_{i=1}^{n} \sum_{c=1}^k \| \beta_{(1)i}-\gamma_{(1)c}-\eta_{i c}\|^2, \label{A.6}
\end{align} 
where $\{\nu_{i c}\}$ are the Lagrange multipliers and $\|\cdot \|$ denotes the Frobenius norm. 

Denote by $\widehat{\gamma}^{(t)}_{(1)}$, $\widehat{\beta}^{(t)}$, $\{\widehat{\eta}_{ic}^{(t)}\}$, and $\{\widehat{\nu}_{ic}^{(t)}\}$ the ADMM iterates at iteration $t \geq 0$ for a fixed $\lambda$, with initial values $\widehat{\eta}_{i c}^{(0)} = \check{\beta}_{(1)i} - \check{\gamma}_{(1)c}$ and $\widehat{\nu}_{i c}^{(0)} = 0$ for $i=1,\dots,n$ and $c=1,\dots,k$. Here, $\check{\beta}$ is chosen from $\{\check{\beta}^{c}: c=k,\dots,\lfloor \sqrt{n/\log n} \rfloor\}$ and $\check{\gamma}$ is set to $\check{\gamma}^{k}$. The shrinkage sequence $\lambda_{1} < \dots < \lambda_{J}$ is selected over the interval $[\lambda_{1}, \lambda_{J}]$, 
where $\lambda_{1} = n^{\alpha} \, \min_{\{(i,j,c):(\check{\beta}_{i} - \check{\gamma}_{c})_j \neq 0 \}}|(\check{\beta}_{i}-\check{\gamma}_{c})_j |$ and $\lambda_{J}= n^{\alpha} \, \max_{(i,j,c)} |(\check{\beta}_{i}-\check{\gamma}_{c})_j |$. In our implementation, we set $J=10$ and specify $\alpha$ so that $n^\alpha$ is close to 1.
We compute the SP estimator via the following iterative scheme for $t>0$:
\begin{itemize}
\item \textbf{($\boldsymbol{(\beta,\gamma_{(1)})}$-minimization)} The updates $\widehat{\beta}^{(t+1)}$ and $\widehat{\gamma}^{(t+1)}$ are obtained via the iterations for $s\geq 0$:
\begin{align}
&\hspace{-0.02in}\beta_{(1)i}^{(s+1)}  = \beta_{(1)i}^{(s)} - \big(\text{H}_{1i}\big( \beta_{(1)}^{[i](s)}, \beta_{(2)}^{(s)} \big) + k I_{q} \big)^{-1} 
 \bigg[ \text{S}_{1i}\big( \beta_{(1)}^{[i](s)}, \beta_{(2)}^{(s)} \big) + \bigg( \sum_{c = 1}^k \beta^{(s)}_{(1)i} - \gamma_{(1)c}^{(s)} - \widehat{\eta}^{(t)}_{i c} + {\widehat{\nu}^{(t)}_{i c}}\bigg)\bigg],  \nonumber\\
&\beta_{(2)}^{(s+1)} =\beta_{(2)}^{(s)} - \big(\text{H}_{2}\big(  \beta_{(1)}^{(s+1) }, \beta_{(2)}^{(s)}  \big)\big)^{-1} \text{S}_{2}\big( \beta_{(1)}^{(s+1)}, \beta_{(2)}^{(s) } \big),~
\gamma_{(1)c}^{(s+1)} =  \frac{1}{n} \sum_{i=1}^n \Big( \beta^{(s+1)}_{(1)i} - \widehat{\eta}^{(t)}_{i c} + \widehat{\nu}^{(t)}_{i c} \Big),\nonumber\\
& i = 1, \dots, n, ~c=1,\dots,k,  \label{alg:SP_beta}
\end{align}
where $\beta^{(0)}= \widehat{\beta}^{(t)}$, $\beta^{[i](s)}_{(1)}= \big(\beta_{(1)1}^{(s+1)\top}, \dots, \beta_{(1)i-1}^{(s+1)\top},\beta_{(1)i}^{(s)\top}, \dots, \beta_{(1)n}^{(s)\top}\big)^{\top}$ with $\beta^{[1](s)}_{(1)}=\beta^{(s)}_{(1)}$, $I_{q}$ denotes the $q \times q$ identity matrix,
\begin{align*}
&\hspace{-0.02in}\text{S}_{1i}(\beta)  = - \sum_{j=1}^n \big(Y_j - \widehat{m}_{\hat{h}}^{-j}\big(\beta_j^{\top}Z_j; \beta\big) \big)  \partial_{\beta_{(1)i}} \widehat{m}_{\hat{h}}^{-j}\big(\beta_j^{\top}Z_j; \beta\big),~
\text{H}_{1i}(\beta)  = \sum_{j=1}^n  \big( \partial_{\beta_{(1)i}} \widehat{m}_{\hat{h}}^{-j}\big(\beta_j^{\top}Z_j; \beta\big) \big)^{{\otimes}2}, \\ 
&\text{S}_{2}(\beta)  = - \sum_{i=1}^n \big(Y_i - \widehat{m}_{\hat{h}}^{-i}\big(\beta_i^{\top}Z_i; \beta\big) \big)  \partial_{\beta_{(2)}} \widehat{m}_{\hat{h}}^{-i}\big(\beta_i^{\top}Z_i; \beta\big), ~\text{H}_{2}(\beta)  = \sum_{i=1}^n \big( \partial_{\beta_{(2)}} \widehat{m}_{\hat{h}}^{-i}\big(\beta_i^{\top}Z_i; \beta\big) \big)^{{\otimes}2}.
\end{align*} 
\item \textbf{($\boldsymbol{\eta}$-minimization)} The update $\big\{\widehat{\eta}^{(t+1)}\big\}$ is given by
\begin{align} 
\widehat{\eta}_{ic}^{(t+1)} = \begin{cases} \big( S_\lambda\big(\big(\widehat{\beta}_{(1)i}^{(t+1)} - \widehat{\gamma}_{(1)c}^{(t+1)} + \widehat{\nu}_{ic}^{(t)}\big)_j \big) \big) &c = c_{0},\\
\widehat{\beta}_{(1)i}^{(t+1)} - \widehat{\gamma}_{(1)c}^{(t+1)} + \widehat{\nu}_{ic}^{(t)} &  c \neq c_{0}, 
\end{cases} \label{alg:SP_eta} 
\end{align}
where $S_{\lambda}(z) = \text{sgn}(z)\max(|z|-\lambda, 0)$ and $c_{0}$ is a minimizer of
\begin{align*}
&\lambda\, \Big\| \Big( S_\lambda\Big(\Big(\widehat{\beta}_{(1)i}^{(t+1)} - \widehat{\gamma}_{(1)c}^{(t+1)} + \widehat{\nu}_{ic}^{(t)}\Big)_j \Big) \Big) \Big\|_1 
+ \frac{1}{2} \Big\|  \Big( S_\lambda\big(\big(\widehat{\beta}_{(1)i}^{(t+1)} - \widehat{\gamma}_{(1)c}^{(t+1)} + \widehat{\nu}_{ic}^{(t)}\Big)_j \Big) \Big)\\
\hspace{-1.5in}&~ - \Big(\widehat{\beta}_{(1)i}^{(t+1)} - \widehat{\gamma}_{(1)c}^{(t+1)} + \widehat{\nu}_{ic}^{(t)}\Big) \Big\|^2 .
\end{align*}
\item \textbf{(dual variable update)} The update $\big\{\widehat{\nu}^{(t+1)}_{ic}\big\}$ is given by
\begin{align}
\widehat{\nu}_{i c}^{(t+1)}= \widehat{\nu}_{ic}^{(t)} + \widehat{\beta}_{(1)i}^{(t+1)} - \widehat{\gamma}_{(1)c}^{(t+1)} - \widehat{\eta}_{i c}^{(t+1)}. \label{alg:SP_nu}
\end{align}
\end{itemize}
Following Section 3.3 in \cite{boyd2011distributed}, we terminate the algorithm once the primal residual
$(\sum_{i=1}^n\sum_{c=1}^k\|\widehat\beta^{(t)}_{(1)i}
-\widehat\gamma^{(t)}_{(1)c}-\widehat\eta^{(t)}_{ic}\|^2)^{1/2}$
and the dual residual
$(\sum_{i=1}^n\|\sum_{c=1}^k(\widehat\eta^{(t)}_{ic}-\widehat\eta^{(t-1)}_{ic})\|^2
+\sum_{c=1}^k\|\sum_{i=1}^n(\widehat\eta^{(t)}_{ic}-\widehat\eta^{(t-1)}_{ic})\|^2)^{1/2}$,
fall below prescribed tolerances. Since the $\eta$-update is the exact minimizer of its subproblem and the $(\beta,\gamma)$-update satisfies the first-order optimality condition of its subproblem, the output then satisfies the KKT conditions of (\ref{A.5}) up to these tolerances. This result establishes convergence of the numerical iterates to a stationary point, but does not guarantee that the limiting point is a global minimizer of (\ref{3.1.1}).

\section{Assumptions} 
\label{sec:A.Assumptions}

\subsection{Assumptions for the Oracle Property of the SP Estimator}

Let $\mathcal{W}_n$ denote the collection $\{W_1, \dots, W_n\}$, $f_{\gamma}(v)$ the probability density function of $\gamma^{\top}W$, with support $\mathcal{V}$,  $W_{*}$ the subvector of $W$ corresponding $\gamma_*$, $U_{\gamma}$ the random quantity $Y - m^{[0]}\big(W, \gamma^{\top}W\big)$, with $U^{\text{o}} = Y - m(\gamma^{\text{o}\top}W)$, and $\mathcal{U}_{\gamma n}$ the collection $\{U_{\gamma 1},\dots,U_{\gamma n}\}$. Define
\begin{align*}
&D_{\gamma}=\min_{c \neq \ell } \| \gamma^{\text{o}}_{c} - \gamma^{\text{o}}_{\ell}\|, ~ \phi_{n}= {{\frac{\phi_0}{\log n \cdot  n^{\frac{2}{5}} }}},~N_{sh}(v;\mathcal{G}^{\text{o}}, \gamma)  = \frac{1}{n}\sum_{i=1}^n Y_i^s K_{h}\big(\gamma^{\top}W_i - v\big),\text{ and},\\
&M_{s,t}(w, v)= (-1)^t\E\big[\big(m\big(\gamma^{\text{o} \top}W\big)\big)^s (W_{*} - w_{*})^{\otimes t} \mid \gamma^{\top}W = v\big],~ s=0,1,2,~t=0,1,2,3,4,
\end{align*}
where $\phi_{0}>0$ is a constant.

The following assumptions are imposed to establish the oracle property of the SP estimator: 
\begin{itemize}
\item[A1.] $h = h_0 n^{-\delta}$ for $\delta \in (1/8, 1/5)$ and some constant $h_0>0$.
\item[A2.] $\E\big[Y^4\big]<\infty$ and each component of $X$ is sub-Gaussian.
\item[A3.] $d^{4}_{v}f_{\gamma}(v)$ and $\partial^{t+2}_v M_{s,t}(w, v)$ are bounded for $w \in \mathcal{W}_n$, and Lipschitz continuous in $(v, \gamma_{*})$ for $s=0,1$ and $t = 0, 1, 2$, with Lipschitz constants independent of $w$.
\item[A4.] $B_n=\big\{ \beta: \max_{\{1\leq i\leq n\}} \| \beta_i - \beta^{\text{o}}_i \|_{1} \leq \phi_{n} \big\}$ and $\Gamma_n = \big\{\gamma: \max_{\{c\in\mathcal{C}\}}\|\gamma_{c}-\gamma^{\text{o}}_{c}\|_{1}\leq\phi_{n}\big\}$.
\item[A5.] $\inf_{\{\gamma_{*} \in \Gamma\} }f_{\gamma}(\gamma^{\top} w)$ is far away from zero for $w \in \mathcal{W}_n$.
\item[A6.] $V_{1}$ and $V_{2}$ are nonsingular.
\item[A7.] $D_{\gamma}>4\phi_{n}$ and $\lambda> n^\alpha$ for some sufficiently small fixed $\alpha>0$.
\end{itemize}
Assumption {A1} imposes a restriction on the bandwidth to ensure the $\sqrt{n}$-consistency of the SP estimator.
Assumption {A2} imposes a moment condition on the response variable and a sub-Gaussian tail condition on the covariates for the oracle analysis of a subject-specific formulation of the CSIM. The sub-Gaussian condition can be weakened to a finite-moment condition, $\E[\|X\|^{\iota}\|]<\infty$ with $\iota=21$, provided that $\phi_n= O(n^{-2(\iota+5)/5\iota})$ and $\sqrt{n}\phi_n \rightarrow \infty$.
Assumption {A3} imposes smoothness conditions to guarantee the uniform convergence of the estimated functions to their targets.
Assumptions {A4} and {A7} ensure that the oracle estimator is the unique local minimizer with probability converging to one, while assumptions {A5} and {A6} guarantee the asymptotic normality of the oracle estimator.

\vspace{-0.15in}
\subsection{Assumptions for the Asymptotic Properties of the CS Estimator}

Let $\mathcal{A}_{d}$ denote the compact parameter space of $A_{d*}$, $\mathcal{X}_{n}$ the collection $\{X_{1},\dots,X_{n}\}$, and $f_{A_{d}}(\nu)$ the probability density function of $A^{\top}_{d}X$. For each fixed $d$, define 
\begin{align*}
A^{\text{o}}_{d}\in&~\argmin_{A_{d}}\sum_{c=1}^{k-1} \E\big[\big(I(C = c) - \pi^{[0]}_{c}\big({X}, A^{\top}_{d}X\big) \big)^{2}\big], \\
S_{A_{d}}=&~\sum^{k-1}_{c=1}\big(I(C=c)-\pi^{[0]}_{\ell}\big({X}, A^{\top}_{d}X\big)\big)\pi^{[1]}_{c}\big({X},A^{\top}_{d}X\big), \text{ and}\\
 V_{A_{d}}=&~  \sum^{k-1}_{c=1}E\big[(\pi^{[1]}_{c}\big({X}, A^{\top}_{d}X)\big)^{\otimes 2}-(I(C=c)-\pi^{[0]}_{c}\big({X}, A^{\top}_{d}X)\big)\pi^{[2]}_{c}\big({X}, A^{\top}_{d}X\big)\big]. 
 \end{align*}
By construction, it follows that, for $d = d^{\text{o}}, \dots, p$, we have $E[S_{A^{\text{o}}_{d}}]=0$ and $V_{A^{\text{o}}_{d}} =  \sum^{k-1}_{c=1}E[ (\pi^{[1]}_{c}({X}, $ $A^{\text{o} \top}_{d}X))^{\otimes 2}]$.

The following assumptions are imposed for the asymptotic properties of $\hat{d}$ and $\widehat{A}_*$:
\begin{itemize}
\item[B1.] $h_d = h_0 n^{-\delta}$ for $\delta \in (1/4r, 1/\max\{2d+2,d+4\})$ and some constant $h_0>0$.
\item[B2.] $\partial^{r+2}_{\nu} f_{A_{d}}(\nu)$ and $\partial^{t+r}_{\nu} E\big[ \pi_{c}(A^{\text{o} \top}X) ({X}_{d*} - {x}_{d*})^{\otimes t} \big | A^{\top}_d X = \nu\big]$ are bounded for $x\in\mathcal{X}_{n}$, and Lipschitz continuous in $(\nu, A_{d*})$ for $t = 1, 2$ and  $c=1,\dots,k-1$, with Lipschitz constants independent of $x$.
\item[B3.] $\inf_{\{A_{d*}\in\mathcal{A}_d\}}f_{A_{d}}(A_{d}^{\top} x)$ is far away from zero for $x\in\mathcal{X}_{n}$.
\item[B4.] $\E[S^{\otimes 2}_{A^{\text{o}}}]$ and $V_{A^{\text{o}}}$ are nonsingular.
\end{itemize}
Assumption {B1} imposes a restriction on the bandwidth to ensure the $\sqrt{n}$-consistency of the CS estimator.
Assumption {B2} imposes the smoothness conditions to guarantee the uniform convergence of the estimated functions to their targets.
Assumptions {B3} and {B4} ensure the asymptotic
normality of the CS estimator.

\subsection{Assumptions for the Asymptotic Properties of the PML Estimator}

Define
\begin{align*}
Q_{t_1,t_2}(w,u) = \E\big[\big(- m^{[t_1]}(W, \gamma^{\top}W) + m^{[t_1]}(w, \gamma^{\top}w)\big)^{\otimes t_2} \mid U = u\big], ~t_{1}=1,2,~t_{2}=1,2,4.
\end{align*}
Let $f_0(u;\gamma)$ denote the probability density function of $U_{\gamma}$, and $f(w,u;\gamma)$ the joint probability density function of $(W,U_{\gamma})$. 

The following assumptions are imposed to establish the asymptotic properties of the PML estimator:
\begin{itemize} 
\item[C1.] $\xi/h \longrightarrow \xi_{0}$ as $n \longrightarrow \infty$ for some constant $\xi_{0}>0$.
\item[C2.] $\partial^{4}_{u} f_0(u;\gamma)$, $\partial^{4}_{u} f(w,u;\gamma)$, and $\partial^{t_2 + 2}_{u} Q_{t_1,t_2}(w,u)$ are bounded for $u \in \mathcal{U}_{\gamma n}$ and $w \in \mathcal{W}_n$, and are Lipschitz continuous in $(u, \gamma_*)$ for each $(t_1, t_2) \in \{(1,1), (1,2), (2,1)\}$, with a Lipschitz constant independent of $u$.
\item[C3.] $\inf_{\{\gamma_* \in \Gamma\}} f_0(u;\gamma)$ is far away from zero for $u \in \mathcal{U}_{\gamma n}$.
\item[C4.] $V$ is nonsingular.
\end{itemize}
Assumption {C1} restricts the bandwidth rates to ensure the $\sqrt{n}$-consistency of the PML estimator.
Assumption {C2} imposes the smoothness conditions to guarantee the uniform convergence of the estimated functions to their targets.
Assumptions {C3} and {C4} ensure the asymptotic normality of the PML estimator.
  
\section{Technical Lemmas} 
\label{sec:A.lemma}

\begin{lem} 
Under assumptions {A2} and {A3}, as $n\longrightarrow \infty$ and $h\longrightarrow 0$, with $nh^{2t+1}\longrightarrow \infty$ for $t = 0, 1, 2$,
\begin{align*}
&\sup_{\{(w, \gamma_{*}) \in \mathcal{W}_n \times \Gamma\}} \big\| \partial_{\gamma_{*}}^t N_{sh}(\gamma^{\top}w;\mathcal{G}^{\text{o}},  \gamma) - \partial_{v}^t \big(M_{s,t}\big(w,\gamma^{\top}w \big)f_{\gamma}\big(\gamma^{\top}w\big) \big)\big\| \nonumber \\
&= \sup_{\{(w, \gamma_{*}) \in \mathcal{W}_n \times \Gamma\}} \bigg\| \frac{1}{n} \sum_{i = 1}^n \xi_{is}^{[t]}(w,\gamma) \bigg\|= o_p \bigg( \sqrt{\frac{\log n}{nh^{2t+1}}}\, \bigg) + O(h^2), ~s=0,1,
\end{align*}
where $\xi_{is}^{[t]}(w,\gamma)  =  \partial_{\gamma_{*}}^t ( Y_i^s K_{h} (\gamma^{\top}W_i -\gamma^{\top}w ) ) - \partial_{v}^t (M_{s,t}(w,\gamma^{\top}w )f_{\gamma}(\gamma^{\top}w) )$, $i = 1, \dots, n$.
\label{A:Lemma_1}
\end{lem}

\begin{proof} 
By Lemma 9.6 in \cite{Kosorok2008}, for each $t = 0, 1, 2$, the classes
\begin{align*}
\{ (W_{*}-w_{*})^{\otimes t}: w \in \mathcal{W}_n\} \text{ and } \{ c_1 K^{(t)}(\gamma^{\top}W + c_2): c_1, c_2 \in \mathbb{R}, \gamma_{*} \in \Gamma\}
\end{align*}
have finite VC indices.
It follows from Lemma 2.12 in \cite{pakes1989simulation} that both classes are Euclidean. By Lemma 2.14 therein, their product class
\begin{align*}
\mathcal{F}_{t} = \big\{ c_1 K^{(t)}(\gamma^{\top}W + c_2) (W_{*}-w_{*})^{\otimes t}: w \in \mathcal{W}_n, \gamma_{*} \in \Gamma, c_1, c_2 \in \mathbb{R}\big\}
\end{align*}
is also Euclidean. Under assumptions {A2} and {A3}, an application of H{\"o}lder's inequality gives
\begin{align*}
&\sup_{\{(w, \gamma_{*}) \in \mathcal{W}_n \times \Gamma\}} \bigg\|\E \bigg[\bigg( \frac{1}{h^{t+1}} Y^s K^{(t)} \bigg( \frac{\gamma^{\top}(W-w)}{h} \bigg) \bigg)^2 \big(  W_{*} - w_{*} \big)^{\otimes 2t} \bigg] \bigg\| =O\Big( \frac{1}{h^{2t+1}}\Big)~s=0,1, t=0,1,2.
\end{align*}
Since $\mathcal{F}_{t}$ is Euclidean class, Theorem II.37 in \cite{pollard1984convergence} implies that, for $s= 0, 1$ and $t = 0, 1, 2$,
\begin{align}
&\sup_{\{(w, \gamma_{*}) \in \mathcal{W}_n \times \Gamma\}} \big\| \partial_{\gamma_{*}}^t N_{sh}(\gamma^{\top}w;\mathcal{G}^{\text{o}}, \gamma) - \E\big[\partial_{\gamma_{*}}^t \big( Y^s K_{h}(\gamma^{\top}W - \gamma^{\top}w) \big) \big] \big\| = o_p \bigg( \sqrt{\frac{\log n}{nh^{2t+1}}} \,\bigg). 
\label{eq:lemma1_2}
\end{align} 
By assumption {A3}, the iterated expectation, and Taylor's theorem, 
\begin{align}
&\hspace{0.2in}\E\big[\partial_{\gamma_{*}}^t \big( Y^s K_{h}(\gamma^{\top}W - \gamma^{\top}w) \big) \big] 
=\frac{1}{h^{t+1}} \E \bigg[K^{(t)} \bigg( \frac{\gamma^{\top}(W-w)}{h} \bigg) M_{s,t} \big(w,\gamma^{\top}W \big) \bigg] \nonumber\\
&= \frac{1}{h^{t}} \int K^{(t)} (v) M_{s,t}\big(w,\gamma^{\top}w + v h\big) f_{\gamma}\big(\gamma^{\top}w + v h\big)d v \nonumber\\
&= \partial_{v}^t \big(M_{s,t}\big(w,\gamma^{\top}w \big)f_{\gamma}\big(\gamma^{\top}w\big)\big) + r_{s}^{[t]}\big(w,\gamma \big),~s=0,1,~t=0,1,2,~w \in \mathcal{W}_n,
\label{eq:lemma1_3}
\end{align}
with
$\sup_{\{(w, \gamma_{*}) \in \mathcal{W}_n \times \Gamma\}} \big\|r_{s}^{[t]}\big(w,\gamma \big) \big\| = O(h^2)$.
Substituting (\ref{eq:lemma1_3}) into (\ref{eq:lemma1_2}) completes the proof.
\end{proof}

\begin{lem}
Under assumptions {A2}, {A3} and {A5}, as $n\longrightarrow \infty$ and $h\longrightarrow 0$, with $nh^{2t+1}\longrightarrow \infty$ for $t = 0, 1, 2$,\begin{align*}
\sup_{\{(w, \gamma_{*}) \in \mathcal{W}_n \times \Gamma\}} \bigg\| \partial_{\gamma_{*}}^t\widehat{m}_h(\gamma^{\top}w;\mathcal{G}^{\text{o}}, \gamma) - m^{[t]}(w, \gamma^{\top}w) - \frac{1}{n}\sum_{i = 1}^n \xi_i^{[t]}(w,\gamma) \bigg\| = o_p \bigg({\frac{\log n}{nh^{2t+1}}}\bigg) + O \big(h^4 \big),
\end{align*}
with
\begin{align*}
\sup_{\{(w, \gamma_{*}) \in \mathcal{W}_n \times \Gamma\}} \bigg\|\frac{1}{n} \sum_{i = 1}^n \xi_i^{[t]}(w,\gamma)\bigg\| = & o_p \bigg(\sqrt{\frac{\log n}{nh^{2t+1}}}\,\bigg) + O \big(h^2 \big),
\end{align*}
where $m^{[t]}(w, v)$ and $\xi_i^{[t]}(w,\gamma)$, $t=0,1,2$, $i = 1, \dots, n$, are defined in $(\ref{eq:lemma1_6})$--$(\ref{eq:lemma1_8})$.
\label{A:Lemma_2}
\end{lem}

\begin{proof}
By Taylor's theorem, Lemma \ref{A:Lemma_1}, and assumption {A5}, it follows that, for $t=0,1,2$,
\begin{align}
\frac{\partial_{\gamma_{*}}^t N_{sh}(\gamma^{\top}w;\mathcal{G}^{\text{o}}, \gamma)}{N_{0h}\big(\gamma^{\top}w;\mathcal{G}^{\text{o}}, \gamma\big)} =  \frac{\partial_{v}^t \big(M_{s,t}\big(w,\gamma^{\top}w \big)f_{\gamma}\big(\gamma^{\top}w\big) \big)}{f_{\gamma}(\gamma^{\top}w)} + \frac{1}{n} \sum_{i = 1}^n \xi_{is}^{*[t]}(w,\gamma) + r^{*[t]}_{s}(w,\gamma), \label{eq:lemma1_5}
\end{align}
$s=0,1, w \in \mathcal{W}_n$, with
\begin{align*}
&\xi_{is}^{*[t]}(w,\gamma) = \frac{\xi_{is}^{[t]}(w,\gamma)}{f_{\gamma}(\gamma^{\top}w)} - \frac{\partial_{v}^t \big(M_{s,t}\big(w,\gamma^{\top}w\big)f_{\gamma}\big(\gamma^{\top}w\big) \big) \xi_{i0}^{[0]}(w,\gamma)}{(f_{\gamma}(\gamma^{\top}w))^2},~i=1,\dots,n,\\
&\sup_{\{(w, \gamma_{*}) \in \mathcal{W}_n \times \Gamma\}} \bigg\| \frac{1}{n} \sum_{i = 1}^n \xi_{is}^{*[t]}(w,\gamma) \bigg\| = o_p \Bigg( \sqrt{\frac{\log n}{nh^{2t+1}}}\, \Bigg) + O(h^2), \text{ and}
\end{align*}
\begin{align*}
\hspace{-0.75in}&\sup_{\{(w, \gamma_{*}) \in \mathcal{W}_n \times \Gamma\}} \|r^{*[t]}_{s}(w,\gamma)\| = o_p \bigg( {\frac{\log n}{nh^{2t+1}}} \bigg) + O(h^4).
\end{align*}
For a square matrix $A$, let $\text{sym}(A)\coloneqq (A + A^\top)/2$ denote its symmetric part.
Combining (\ref{eq:lemma1_5}) with the identities
 \begin{align*}
& \widehat{m}_{h}(\gamma^{\top}w;\mathcal{G}^{\text{o}}, \gamma) =  \frac{N_{1h}(\gamma^{\top}w;\mathcal{G}^{\text{o}}, \gamma)}{N_{0h}(\gamma^{\top}w;\mathcal{G}^{\text{o}}, \gamma)}, \partial_{\gamma_{*}} \widehat{m}_{h}(\gamma^{\top}w;\mathcal{G}^{\text{o}}, \gamma) = \sum_{s = 0}^1 \big( - \widehat{m}_h\big(\gamma^{\top}w;\gamma \big)\big)^{1-s} \frac{\partial_{\gamma_{*}}N_{sh}(\gamma^{\top}w;\mathcal{G}^{\text{o}}, \gamma)}{N_{0h}(\gamma^{\top}w;\mathcal{G}^{\text{o}}, \gamma)},\\
& \partial_{\gamma_{*}}^2 \widehat{m}_{h}(\gamma^{\top}w;\mathcal{G}^{\text{o}}, \gamma) 
= \sum_{s = 0}^1 \big( - \widehat{m}_h\big(\gamma^{\top}w;\mathcal{G}^{\text{o}}, \gamma \big)\big)^{1-s} \frac{\partial_{\gamma_{*}}^2N_{sh}(\gamma^{\top}w;\mathcal{G}^{\text{o}}, \gamma)}{N_{0h}(\gamma^{\top}w;\mathcal{G}^{\text{o}}, \gamma)} \\
&\hspace{1.4in} -2 \,\text{sym}\bigg(\frac{\partial_{\gamma_{*}}N_{0h}(\gamma^{\top}w;\mathcal{G}^{\text{o}}, \gamma)}{N_{0h}(\gamma^{\top}w;\mathcal{G}^{\text{o}}, \gamma)} \big(\partial_{\gamma_{*}} \widehat{m}_h \big(\gamma^{\top}w;\mathcal{G}^{\text{o}}, \gamma \big) \big)^{\top}\bigg),
\end{align*} 
yields the following expressions:
\begin{align}
\widehat{m}_h(\gamma^{\top}w;\mathcal{G}^{\text{o}}, \gamma)  =&~ \frac{M_{1,0}\big(w,\gamma^{\top}w \big)f_{\gamma}\big(\gamma^{\top}w\big)}{f_{\gamma}(\gamma^{\top}w)} + \frac{1}{n} \sum_{i = 1}^n \xi_{i1}^{*[0]}(w,\gamma) + r^{*[0]}_{1}(w,\gamma) \nonumber \\
\stackrel{\triangle}{=} &~m^{[0]}(w, \gamma^{\top}w) + \frac{1}{n}\sum_{i = 1}^n \xi^{[0]}_i(w,\gamma) + r^{[0]}(w,\gamma), \label{eq:lemma1_6}\\
\partial_{\gamma_{*}}\widehat{m}_h(\gamma^{\top}w;\mathcal{G}^{\text{o}}, \gamma) =&~ \sum_{s=0}^1 \bigg[ - \bigg( m^{[0]}(w, \gamma^{\top}w) + \frac{1}{n}\sum_{i = 1}^n \xi^{[0]}_i(w,\gamma) + r^{[0]}(w,\gamma) \bigg) \bigg]^{1-s}\hspace{0.5in}  \nonumber\\
&~~ \bigg( \frac{\partial_v \big( M_{s,1}(w,\gamma^{\top}w) f_{\gamma}(\gamma^{\top}w)\big)}{f_{\gamma}(\gamma^{\top}w)} + \frac{1}{n} \sum_{i = 1}^n \xi_{is}^{*[1]}(w,\gamma) + r_{s}^{*[1]}(w,\gamma) \bigg) \nonumber\\
\hspace{0.9in}\stackrel{\triangle}{=}&~ m^{[1]}(w, \gamma^{\top}w) + \frac{1}{n}\sum_{i = 1}^n \xi^{[1]}_i(w,\gamma) + r^{[1]}(w,\gamma),\text{ and} \label{eq:lemma1_7}\\
\partial_{\gamma_{*}}^2\widehat{m}_h(\gamma^{\top}w;\mathcal{G}^{\text{o}}, \gamma) =&~\bigg\{ \sum_{s=0}^1 \bigg[ - \bigg( m^{[0]}(w, \gamma^{\top}w) + \frac{1}{n}\sum_{i = 1}^n \xi^{[0]}_i(w,\gamma) + r^{[0]}(w,\gamma) \bigg) \bigg]^{1-s} \nonumber\\
&~ \bigg( \frac{\partial^{2}_v \big( M_{s,2}(w,\gamma^{\top}w) f_{\gamma}(\gamma^{\top}w)\big)}{f_{\gamma}(\gamma^{\top}w)} + \frac{1}{n} \sum_{i = 1}^n  \xi_{is}^{*[2]}(w,\gamma) + r_{s}^{*[2]}(w,\gamma) \bigg) \bigg\}\nonumber \\
&~ -2 \,\text{sym} \bigg(\bigg[ \frac{\partial_v \big( M_{0,1}(w,\gamma^{\top}w) f_{\gamma}(\gamma^{\top}w)\big)}{f_{\gamma}(\gamma^{\top}w)} + \frac{1}{n}\sum_{i = 1}^n \xi_{i0}^{*[1]}(w,\gamma) + r_{0}^{*[1]}(w,\gamma)  \bigg]\nonumber\\
&~ \bigg(m^{[1]}(w, \gamma^{\top}w) + \frac{1}{n}\sum_{i = 1}^n \xi^{[1]}_i(w,\gamma) + r^{[1]}(w,\gamma)\bigg)^{\top}\bigg) \nonumber\\
\stackrel{\triangle}{=}&~ m^{[2]}(w, \gamma^{\top}w) + \frac{1}{n}\sum_{i = 1}^n \xi^{[2]}_i(w,\gamma) + r^{[2]}(w,\gamma), \label{eq:lemma1_8}
\end{align}
where
\begin{align*}
&m^{[0]}(w,v) = M_{1,0}(w,v), ~m^{[1]}(w,v) = \sum_{s = 0}^1 \big( - m^{[0]}(w,v)\big)^{1-s} \frac{\partial_v(M_{s,1}(w,v)f_{\gamma}(v))}{f_{\gamma}(v)},\text{ and} \\
&m^{[2]}(w,v) =\sum_{s = 0}^1 \big( - m^{[0]}(w,v)\big)^{1-s} \frac{\partial_v(M_{s,2}(w,v)f_{\gamma}(v))}{f_{\gamma}(v)}
- 2\,\text{sym}\bigg(\frac{\partial_v(M_{0,1}(w,v)f_{\gamma}(v))}{f_{\gamma}(v)} \big(m^{[1]}(w,v)\big)^{\top} \bigg),\\
& \text{with } \sup_{\{(w, \gamma_{*}) \in \mathcal{W}_n \times \Gamma\}}  \Big\| r^{[t]}(W_i,\gamma) \Big\|  = o_p \bigg( {\frac{\log n}{nh^{2t+1}}} \bigg) + O(h^4),~ t=0,1,2.
\end{align*}
The lemma follows.
\end{proof}

\begin{lem} 
Under assumptions {A2}, {A3}, {A5}, and {C2}, as $n\longrightarrow \infty$ and $\xi \longrightarrow 0$, with $n\xi^{2t+1}\longrightarrow \infty$ for $t = 0, 1, 2$,
\begin{align*}
&\sup_{\{(w, u, \gamma_*) \in \mathcal{W}_n \times \mathcal{U}_{\gamma n} \times \Gamma\}} \big\| \partial_{\gamma_{*}}^t \bar{f}_{\xi}(u; \gamma) - f^{[t]}(w, u; \gamma) \big\| = \sup_{\{(w, u, \gamma_*) \in \mathcal{W}_n \times \mathcal{U}_{\gamma n} \times \Gamma\}} \bigg\| \frac{1}{n} \sum_{i = 1}^n \zeta^{[t]}_{i}(w, u,\gamma) \bigg\| \\
&= o_p \Bigg( \sqrt{\frac{\log n}{n\xi^{2t+1}}}\, \Bigg) + O(\xi^2)
\end{align*}
and
\begin{eqnarray*}
\sup_{\{(u, \gamma_*) \in \mathcal{U}_{\gamma n} \times \Gamma\}} \Big\|\partial^t_{u} \bar{f}_{{\xi}} (u;\gamma) - \partial^t_{u}  f_0(u; \gamma) \Big\| = \sup_{\{(u, \gamma_*) \in \mathcal{U}_{\gamma n} \times \Gamma\}} \Big\| \frac{1}{n} \sum_{i = 1}^n \varsigma^{[t]}_{i}(u; \gamma)\Big\| = o_p \Bigg( \sqrt{\frac{\log n}{n \xi^{2t+1}}} \, \Bigg) + O(\xi^2), 
\end{eqnarray*}
where $\bar{f}_{{\xi}}(u;\gamma)=\sum^{n}_{i=1}K_{\xi}\big(U_{\gamma i}- u\big) / n$, $f^{[t]}(w, u; \gamma)$, $t=0,1,2$, are defined in 
(\ref{eq:lemma3_3_0})--(\ref{eq:lemma3_3_2}), $\zeta^{[t]}_{i}(w, u,\gamma) = \big( \partial_{\gamma_{*}}^t K_{\xi} \big(U_{\gamma i} - u  \big) \big) - f^{[t]}(w, u; \gamma)$, $\varsigma^{[t]}_{i}(u,\gamma) =\big(  \partial_{u}^t K_{\xi} \big(U_{\gamma i} - u  \big)\big) - \partial^t_{u} f_0(u; \gamma)$, with $f^{[0]}(w, u; \gamma)$ $= f_0(u; \gamma)$ and $\zeta^{[0]}_{i}(w, u,\gamma) = \varsigma^{[0]}_{i}(u,\gamma)$, $i = 1, \dots, n$.
\label{A:Lemma_3}
\end{lem}{}

\begin{proof} 
It suffices to prove the first assertion, as the second assertion follows from \cite{Gine2002rates}.
Applying the chain rule, the summands appearing in $\partial_{\gamma_*} \bar{f}_{\xi}(u;\gamma)$ and $\partial_{\gamma_*}^2 \bar{f}_{\xi}(u;\gamma)$ admit the following representations:
\begin{align}
\partial_{\gamma_{*}} \bigg( \frac{1}{\xi} K\bigg(\frac{U_{\gamma i} - u}{\xi}\bigg) \bigg) = \frac{1}{\xi^2} K^{(1)}\bigg(\frac{U_{\gamma i} - u}{\xi}\bigg) \big( - m^{[1]}(W_i, \gamma^{\top}W_i) + m^{[1]}(w, \gamma^{\top}w) \big) \label{eq:chain_t1}
\end{align}
and
\begin{align}
\partial_{\gamma_{*}}^2 \bigg( \frac{1}{\xi} K\bigg(\frac{U_{\gamma i} - u}{\xi}\bigg) \bigg) = & ~ \frac{1}{\xi^3} K^{(2)}\bigg(\frac{U_{\gamma i} - u}{\xi}\bigg) \big( m^{[1]}(W_i, \gamma^{\top}W_i) - m^{[1]}(w, \gamma^{\top}w) \big)^{\otimes 2} \nonumber 
\end{align}
\begin{align}
&\hspace{2in} + \frac{1}{\xi^2} K^{(1)}\bigg(\frac{U_{\gamma i} - u}{\xi}\bigg) \big( - m^{[2]}(W_i, \gamma^{\top}W_i) + m^{[2]}(w, \gamma^{\top}w) \big), \label{eq:chain_t2}
\end{align}
where $m^{[1]}(w, \gamma^{\top}w)$ and $m^{[2]}(w, \gamma^{\top}w)$ are defined in (\ref{eq:lemma1_7}) and (\ref{eq:lemma1_8}), respectively. 
By Lemma 9.6 in \cite{Kosorok2008}, the class
\begin{align*}
\Big\{ m^{[t]}(W, \gamma^{\top}W) - m^{[t]}(w, \gamma^{\top}w): w \in \mathcal{W}_n, \gamma_* \in \Gamma\Big\}
\end{align*}
is of finite VC index for each $t = 0, 1, 2$. It follows from Lemma 2.12 in \cite{pakes1989simulation} that these classes are Euclidean. By Lemma 2.14 therein, the class
\begin{align*}
\mathcal{F}_{t_1t_2} = \big\{ c_1 K^{(t_1)}(U_{\gamma} + c_2) \big( m^{[t_2]}(W, \gamma^{\top}W) - m^{[t_2]}(w, \gamma^{\top}w) \big)^{\otimes t_1}: w \in \mathcal{W}_n, \gamma_{*} \in \Gamma, c_1, c_2 \in \mathbb{R}\big\}
\end{align*}
is also Euclidean for each $t_{1},t_{2} = 0, 1, 2$.

Under assumptions {A2}, {A3}, {A5}, and {C2}, and using the identities in (\ref{eq:chain_t1}) and (\ref{eq:chain_t2}), we obtain that, for $t = 0, 1, 2$,
\begin{align*}
\sup_{\{(w, u, \gamma_*) \in \mathcal{W}_n \times \mathcal{U}_{\gamma n} \times \Gamma\}} \bigg\|\E \bigg[\bigg(\partial^t_{\gamma_*}\bigg( \frac{1}{\xi} K \bigg( \frac{U_{\gamma} - u}{\xi} \bigg)\bigg)\bigg)^{\otimes 2} \bigg] \bigg\| = O\bigg( \frac{1}{\xi^{2t+1}} \bigg).
\end{align*}
Since $\mathcal{F}_{00}$, $\mathcal{F}_{11}$, and $\mathcal{F}_{12}$ are Euclidean, Theorem II.37 in \cite{pollard1984convergence} implies that, for $t = 0, 1, 2$,
\begin{align}
\sup_{\{(w, u, \gamma_*) \in \mathcal{W}_n \times \mathcal{U}_{\gamma n} \times \Gamma\}} \Big\| \partial_{\gamma_{*}}^t \bar{f}_{\xi}(u; \gamma) - \E\big[\partial_{\gamma_{*}}^t \bar{f}_{\xi}(u; \gamma) \big] \Big\| = o_p \Bigg( \sqrt{\frac{\log n}{n\xi^{2t+1}}} \,\Bigg).
\label{eq:lemma3_2}
\end{align}
Under assumption {C2}, by the iterated expectation and Taylor's theorem, 
\begin{align} 
&\hspace{-0.1in}\E\big[K_{\xi}(U_{\gamma}-u)\big] = \int K(v)f_0(u+ v \xi; \gamma)dv = f_0(u; \gamma) + r_0(u; \gamma) \stackrel{\triangle}{=} f^{[0]}(w,u;\gamma) + r_0(w, u; \gamma), \label{eq:lemma3_3_0}\\
&\hspace{-0.1in}\E\big[\partial_{\gamma_*}K_{\xi}(U_{\gamma}-u)\big] = \frac{1}{\xi^2}\E\bigg[ K^{(1)}\bigg(\frac{U_{\gamma}-u}{\xi}\bigg) Q_{1,1}(w, U_{\gamma})\bigg]= \int \frac{1}{\xi} K^{(1)}(v) Q_{1,1}(w, u + v \xi) f_0(u+ v \xi; \gamma)dv \nonumber\\
&\hspace{-0.1in} = - \partial_u(Q_{1,1}(w,u)f_0(u;\gamma)) + r_1(w, u; \gamma) \stackrel{\triangle}{=} f^{[1]}(w,u;\gamma) + r_1(w, u; \gamma), \text{ and}\label{eq:lemma3_3_1}\\
&\hspace{-0.1in}\E\big[\partial^2_{\gamma_*}K_{\xi}(U_{\gamma}-u)\big] = \frac{1}{\xi^3}\E\bigg[ K^{(2)}\bigg(\frac{U_{\gamma}-u}{\xi}\bigg) Q_{1,2}(w, U_{\gamma})\bigg] + \frac{1}{\xi^2}\E\bigg[ K^{(1)}\bigg(\frac{U_{\gamma}-u}{\xi}\bigg) Q_{2,1}(w,U_{\gamma})\bigg] \nonumber\\
&\hspace{-0.1in}= \int \frac{1}{\xi^2} K^{(2)}(v) Q_{1,2}(w, u + v \xi) f_0(u+ v \xi; \gamma)dv + \int \frac{1}{\xi} K^{(1)}(v) Q_{2,1}(w, u + v \xi) f_0(u+ v \xi; \gamma)dv \nonumber\\
&\hspace{-0.1in}= \partial^2_u(Q_{1,2}(w,u)f_0(u;\gamma)) -   \partial_u(Q_{2,1}(w,u)f_0(u;\gamma)) + r_2(w, u; \gamma) \stackrel{\triangle}{=} f^{[2]}(w,u;\gamma) + r_2(w, u; \gamma), \label{eq:lemma3_3_2} 
\end{align}
with $\sup_{\{(w, u, \gamma_*) \in \mathcal{W}_n \times \mathcal{U}_{\gamma n} \times \Gamma\}} \big\|r_{t}\big(w,u; \gamma \big) \big\| = O(\xi^2)$. Using $\E\big[\partial_{\gamma_{*}}^t \bar{f}_{\xi}(u; \gamma) \big] = \E\big[\partial^t_{\gamma_*}K_{\xi}(U_{\gamma}-u)\big]$ for each $t=0,1,2$, and substituting (\ref{eq:lemma3_3_0})--(\ref{eq:lemma3_3_2}) into (\ref{eq:lemma3_2}), we complete the proof of the first assertion.
\end{proof}

\begin{lem}
Under assumptions {A2}, {A3}, {A5}, and {C2}, as $n\longrightarrow \infty$, $h \longrightarrow 0$, and $\xi \longrightarrow 0$, with $nh^{2t+1}\longrightarrow \infty$ and $n\xi^{2t+1}\longrightarrow \infty$ for $t = 0, 1, 2$,
\begin{align*}
&\sup_{\{i, \gamma_{*}\}} \big\| \partial_{\gamma_{*}}^t \breve{f}^{-i}_{{\xi}} \big(\widehat{U}_{\gamma i};\gamma\big) - f^{[t]}(W_i, U_{\gamma i};\gamma) \big\|  = \sup_{\{i, \gamma_{*}\}} \Big\| \frac{1}{n-1}\sum_{\{j: j \neq i\}} \varphi^{[t]}_j(W_i, U_{\gamma i}; \gamma)\Big\| (1+ o_p(1))\\
&= o_p \Bigg( \sqrt{\frac{\log n}{n \xi_m^{2t+1}}} \, \Bigg) + O(\xi_M^2), 
\end{align*}
where $\xi_m = \min\{h, \xi\}$, $\xi_M = \max\{h, \xi\}$, $\breve{f}^{-i}_{\xi}(u;\gamma)= \sum_{\{j:j\neq i\}} K_{\xi}\big(\widehat{U}_{\gamma j} - u \big) / (n-1)$, $\widehat{U}_{\gamma i} = Y_i - \widehat{m}^{-i}_{\hat{h}}\big(\gamma^{\top}W_i; \gamma\big)$, and $\varphi^{[t]}_i(w, u; \gamma)$ is defined in (\ref{eq:lemma_C_23}), $i = 1,\dots,n$.
\label{A:Lemma_4}
\end{lem}

\begin{proof} 
For each $t=0,1,2$, a direct decomposition gives
\begin{align} 
&\partial_{\gamma_{*}}^t \breve{f}^{-i}_{{\xi}} \big(\widehat{U}_{\gamma i};\gamma \big) - f^{[t]}(W_{i}, U_{\gamma i};\gamma) \nonumber\\
&= \big (\partial_{\gamma_{*}}^t \breve{f}^{-i}_{{\xi}} \big(\widehat{U}_{\gamma i};\gamma \big) - \partial_{\gamma_{*}}^t \bar{f}^{-i}_{{\xi}} (U_{\gamma i};\gamma)\big) +  \big(\partial_{\gamma_{*}}^t \bar{f}^{-i}_{{\xi}} (U_{\gamma i};\gamma) - f^{[t]}(W_{i}, U_{\gamma i};\gamma)\big). \label{eq:lemma_C_1}
\end{align}  
Applying a third-order Taylor expansion of $K\big(\big(\widehat{U}_{\gamma j}-\widehat{U}_{\gamma i}\big)/\xi \big)$ around $\widehat{U}_{\gamma j}-\widehat{U}_{\gamma i} = U_{\gamma j} - U_{\gamma i}$, the first term on the right-hand side of (\ref{eq:lemma_C_1}) with $t=0$ can be written as
\begin{align} 
\breve{f}^{-i}_{{\xi}} \big(\widehat{U}_{\gamma i};\gamma \big) - \bar{f}^{-i}_{{\xi}} (U_{\gamma i};\gamma) = &~ \sum_{\{j:j \neq i\}} \sum_{s=1}^3 \frac{1}{s! (n-1) \,\xi^{s+1}} K^{(s)}\bigg(\frac{U_{\gamma j} - U_{\gamma i}}{\xi}\bigg) \Delta^{s}_{ij} +  \sum_{\{j:j \neq i\}} \frac{1}{6 (n-1) \, \xi^{4}} \nonumber\\
&~ \bigg(K^{(3)}\bigg(\frac{\bar{U}_{\gamma j}-\bar{U}_{\gamma i}}{\xi}\bigg) - K^{(3)}\bigg(\frac{U_{\gamma j} - U_{\gamma i}}{\xi}\bigg) \bigg) \Delta^{3}_{ij}, \label{eq:lemma_C_2}
\end{align}
where 
\begin{align*}
\Delta_{ij} = \big(\widehat{m}^{-i}_h(\gamma^{\top}W_i;\gamma) - \widehat{m}^{-j}_h(\gamma^{\top}W_j;\gamma)\big) - \big( m^{[0]}(W_i, \gamma^{\top}W_i) - m^{[0]}(W_j, \gamma^{\top}W_j)\big)
\end{align*}
and $\bar{U}_{\gamma i} = Y_i - \bar{m}(\gamma^{\top}W_i;\gamma)$, with $\bar{m}(\gamma^{\top}W_i;\gamma)$ lying on the line segment between $\widehat{m}^{-i}_h(\gamma^{\top}W_i;\gamma)$ and $m^{[0]}(W_i, \gamma^{\top}W_i)$ for each $i,j = 1, \dots, n$.
By Lemma \ref{A:Lemma_2} with $t = 0$, the $s=1$ term in the first sum on the right-hand side of (\ref{eq:lemma_C_2}) can be expressed as
\begin{align} 
&\sum_{\{j:j \neq i\}} \frac{1}{(n-1) \xi^2} K^{(1)}\bigg(\frac{U_{\gamma j} - U_{\gamma i}}{\xi}\bigg) \Delta_{ij} \nonumber \\ 
&= - \frac{1}{(n-1)^2 \xi^2} \sum_{\{\ell, j:\ell \neq j, j \neq i\}} K^{(1)}\bigg(\frac{U_{\gamma j} - U_{\gamma i}}{\xi}\bigg) \bar{\xi}^{[0]}_{\ell}(W_j,\gamma) - \frac{1}{(n-1) \xi^2}\sum_{\{j:j \neq i\}} K^{(1)}\bigg(\frac{U_{\gamma j} - U_{\gamma i}}{\xi}\bigg)\nonumber\\
&~~~~\E\big[ {\xi}^{[0]}_{\ell}(W_j,\gamma) \bigm| W_j\big]  - \frac{1}{(n-1)\xi^{2}}\sum_{\{j:j \neq i\}}K^{(1)}\bigg(\frac{U_{\gamma j} - U_{\gamma i}}{\xi}\bigg) r^{[0]}(W_j,\gamma) \nonumber \\
&~~~~ - \partial_{u}  \bar{f}^{-i}_{{\xi}} (U_{i};\gamma) \bigg(  \frac{1}{n-1}\sum_{\{\ell: \ell \neq i\}} \xi^{[0]}_{\ell}(W_i,\gamma) + r^{[0]}(W_i,\gamma) \bigg), \hspace{-0.1in}\label{eq:lemma_C_3}
\end{align}
where $\bar{\xi}^{[0]}_{\ell}(W_j,\gamma) = {\xi}^{[0]}_{\ell}(W_j,\gamma) - \E[ {\xi}^{[0]}_{\ell}(W_j,\gamma) \mid W_j]$, $1 \leq \ell \neq j \leq n$.
The first term in (\ref{eq:lemma_C_3}) is a degenerate second-order $U$-process, as ensured by the identities
\begin{align*}
&\E\bigg[\frac{1}{\xi^2}K^{(1)}\bigg(\frac{U_{\gamma j} - u}{\xi}\bigg)\bar{\xi}^{[0]}_{\ell}(W_j, \gamma)\Bigm | W_{j}, U_{\gamma j}\bigg] = 0 \text{ and }\\
&\E\bigg[\frac{1}{\xi^2}K^{(1)}\bigg(\frac{U_{\gamma j} - u}{\xi}\bigg)\bar{\xi}^{[0]}_{\ell}(W_j, \gamma) \biggm | W_{\ell}, U_{\gamma \ell}\bigg] \\
&= \E\bigg[\frac{1}{\xi^2}K^{(1)}\bigg(\frac{U_{\gamma j} - u}{\xi}\bigg)\E\big[\bar{\xi}^{[0]}_{\ell}(W_j, \gamma) \bigm| U_{\gamma j}, W_{\ell}, U_{\gamma \ell} \big] \biggm | W_{\ell}, U_{\gamma \ell} \bigg] = 0.
\end{align*}
Since the class
\begin{align*}
\bigg\{ K^{(1)}\bigg(\frac{U_{\gamma j} - u}{\xi}\bigg) \bar{\xi}^{[0]}_{\ell}(W_j,\gamma): u \in \mathcal{U}_{\gamma n}, \gamma_* \in \Gamma\bigg\}
\end{align*}
is Euclidean by Lemma 22 in \cite{nolan1987u}, and 
\begin{align*}
\sup_{\{(u, \gamma_*) \in \mathcal{U}_{\gamma n} \times \Gamma\}} \E\Bigg[ \bigg(\frac{1}{\xi^2}K^{(1)}\bigg(\frac{U_{\gamma j} - u}{\xi}\bigg)\bar{\xi}^{[0]}_{\ell}(W_j, \gamma) \bigg)^2\Bigg] = O\bigg( \frac{1}{h\xi^3} \bigg),
\end{align*}
Corollary 4 in \cite{sherman1994maximal} yields
\begin{align}
\sup_{\{(u, \gamma_*) \in \mathcal{U}_{\gamma n} \times \Gamma\}} \bigg|\frac{1}{(n-1)^2 \xi^2} \sum_{\{\ell, j:\ell \neq j \}} K^{(1)}\bigg(\frac{U_{\gamma j} - u}{\xi}\bigg) \bar{\xi}^{[0]}_{\ell}(W_j,\gamma) \bigg| = O_p\bigg( \frac{1}{n\sqrt{h\xi^3}} \bigg). \label{eq:lemma_C_4}
\end{align}
By the definition of ${\xi}^{[0]}_{\ell}(w,\gamma)$ in Lemma \ref{A:Lemma_2}, we have
\begin{align*}
\E\big[{\xi}^{[0]}_{\ell}(W_j,\gamma) \bigm| W_{j} = w \big] = \bigg(\partial^2_v M_{1,0}(w, \gamma^{\top}w) + \frac{\partial_v f_{\gamma}(\gamma^{\top}w)}{f_{\gamma}(\gamma^{\top}w)} \partial_v M_{1,0}(w, \gamma^{\top}w)\bigg)h^2 (1 +o(1)) = O(h^2).
\end{align*}
Together with assumptions {A2}, {A3} and {C2}, the first and second moments of the second term on the right-hand side of (\ref{eq:lemma_C_3}) are given by
\begin{align*}
&\E\bigg[\frac{1}{\xi^2}  K^{(1)}\bigg(\frac{U_{\gamma j} - u}{\xi}\bigg)\E\big[{\xi}^{[0]}_{\ell}(W_j,\gamma) \bigm| W_{j} = w \big]\bigg]  \\
& = h^2 \int  \Big(\partial^2_v M_{1,0}(w, v) + \big({\partial_v \log f_{\gamma}(\gamma^{\top}w)}\big) \partial_v M_{1,0}(w, v)\Big) \partial_u f(w, u; \gamma) dw (1 +o(1)) = O(h^2)\text{ and}\\
&\E\Bigg[\bigg(\frac{1}{\xi^2}  K^{(1)}\bigg(\frac{U_{\gamma j} - u}{\xi}\bigg)\E\big[{\xi}^{[0]}_{\ell}(W_j,\gamma) \bigm| W_{j} = w \big]\bigg)^2\Bigg]  \\
& = \frac{h^4}{\xi^3} \int  \Big(\partial^2_v M_{1,0}(w, v) + \big({\partial_v \log f_{\gamma}(\gamma^{\top}w)}\big) \partial_v M_{1,0}(w, v)\Big)^2  f(w, u; \gamma) dw (1 +o(1)) = O\bigg(\frac{h^4}{\xi^3}\bigg).
\end{align*}
It follows from Corollary 4 in \cite{sherman1994maximal} that
\begin{align}
\sup_{\{i,\gamma_*\}}\bigg|\frac{1}{(n-1) \xi^2} \sum_{\{j: j \neq i\}} K^{(1)}\bigg(\frac{U_{\gamma j} - U_{\gamma i}}{\xi}\bigg)\E\Big[{\xi}^{[0]}_{\ell}(W_j,\gamma) \Bigm| W_{j} = w \Big]\bigg|  =  o_p\bigg( \frac{h^2}{\sqrt{n \xi^3}} \bigg) +  O(h^2). \label{eq:lemma_C_5}
\end{align}
By assumption {C2} and Lemmas \ref{A:Lemma_2} and \ref{A:Lemma_3} with $ t = 0$, the remaining terms on the right-hand side of (\ref{eq:lemma_C_3}) satisfy 
\begin{align}
& \sup_{\{ i, \gamma_* \}}\bigg|\frac{1}{(n-1)\xi^{2}}\sum_{\{j:j \neq i\}} K^{(1)}\bigg(\frac{U_{\gamma j} - U_{\gamma i}}{\xi}\bigg)r^{[0]}(W_j,\gamma) \bigg| = o_p\bigg( {\frac{\log n}{nh}}\bigg) + O\big(h^4\big) \text{ and} \label{eq:lemma_C_7}\\
&(\partial_{u} \bar{f}^{-i}_{{\xi}} (U_{\gamma i};\gamma))\bigg(  \frac{1}{n-1}\sum_{\{\ell: \ell \neq i\}} \xi^{[0]}_{\ell}(W_i,\gamma) + r^{[0]}(W_i,\gamma) \bigg) \nonumber  \\
& = \bigg(  (\partial_{u} f_0(U_{\gamma i}; \gamma) )+ \frac{1}{n-1} \sum_{\{j:j \neq i\}} \varsigma^{[1]}_{j}(U_{\gamma i}; \gamma)\bigg) \bigg(  \frac{1}{n-1}\sum_{\{\ell: \ell \neq i\}} \xi^{[0]}_{\ell}(W_i,\gamma) + r^{[0]}(W_i,\gamma) \bigg) \nonumber\\  
& =  \frac{1}{n-1} \sum_{\{\ell: \ell \neq i\}} (\partial_{u} f_0(U_{\gamma i}; \gamma)) \xi^{[0]}_{\ell}(W_i,\gamma) + \bigg(  o_p \bigg(  {\frac{\log n}{n \xi_m^4}} \, \bigg) + O(\xi_M^4) \bigg), \text{ with}\label{eq:lemma_C_8}\\
&\sup_{\{i, \gamma_*\}} \bigg|\frac{1}{n-1} \sum_{\{j: j \neq i\}} (\partial_{u} f_0(U_{\gamma i}; \gamma) )\xi^{[0]}_{j}(W_i,\gamma) \bigg| = o_p \bigg( \sqrt{\frac{\log n}{n h}} \, \bigg) + O(h^2). \nonumber
\end{align}
Substituting (\ref{eq:lemma_C_4})--(\ref{eq:lemma_C_8}) into (\ref{eq:lemma_C_3}), the $s=1$ term in the first sum on the right-hand side of (\ref{eq:lemma_C_2}) satisfies
\begin{align}
&\sup_{\{i, \gamma_*\}} \bigg| \sum_{\{j:j \neq i\}} \frac{1}{(n-1) \,\xi^2} K^{(1)}\bigg(\frac{U_{\gamma j} - U_{\gamma i}}{\xi}\bigg) \Delta_{ij} \bigg| \nonumber\\
& = \sup_{\{i, \gamma_*\}} \bigg|\sum_{\{j: j \neq i\}}\frac{1}{n-1} \bigg( (\partial_{u} f_0(U_{\gamma i}; \gamma)) \xi^{[0]}_{j}(W_i,\gamma) + \frac{1}{ \xi^2}  K^{(1)}\bigg(\frac{U_{\gamma j} + U_{\gamma i}}{\xi}\bigg)\E\big[{\xi}^{[0]}_{\ell}(W_j,\gamma) \bigm| W_j \big] \bigg)\bigg| \nonumber\\
&= o_p\bigg(\sqrt{\frac{\log n}{nh}}\,\bigg) + O(h^2). \label{eq:lemma_C_9}
\end{align}

Analogous arguments applied to the $s=2, 3$ terms in the first sum on the right-hand side of (\ref{eq:lemma_C_2}) yield
\begin{align}
\sup_{\{i, \gamma_*\}} \bigg|   \sum_{\{j: j \neq i\}} \frac{1}{s! (n-1) \,\xi^{s+1}} K^{(s)}\bigg(\frac{U_{\gamma j} - U_{\gamma i}}{\xi}\bigg) \Delta^{s}_{ij}  \bigg|  = \bigg( o_p\bigg(\sqrt{\frac{\log n}{nh}}\,\bigg) + O(h^2) \bigg)^s. \label{eq:lemma_C_10}
\end{align}
The uniform boundedness of $K^{(3)}$ implies that the last term on the right-hand side of (\ref{eq:lemma_C_2}) satisfies
\begin{align}
&\sup_{\{i, \gamma_*\}} \bigg| \sum_{\{j:j \neq i\}} \frac{1}{6 (n-1) \, \xi^{4}}\bigg(K^{(3)}\bigg(\frac{\bar{U}_{\gamma j} - \bar{U}_{\gamma i}}{\xi}\bigg) - K^{(3)}\bigg(\frac{U_{\gamma j} - U_{\gamma i}}{\xi}\bigg) \bigg) \Delta^{3}_{ij} \bigg|\nonumber\\
& \leq O\bigg( \frac{1}{\xi^4}\bigg) \bigg(\sup_{\{i, j, \gamma_*\}}\big|\Delta_{ij} \big|\bigg)^3= o_p\bigg(\frac{1}{\xi^{4}} \bigg(\frac{\log n}{nh}\bigg)^{\frac{3}{2}} \bigg) + O\bigg(\frac{h^6}{\xi^{4}}\bigg). \label{eq:lemma_C_11}
\end{align}
Substituting (\ref{eq:lemma_C_9})--(\ref{eq:lemma_C_11}) into (\ref{eq:lemma_C_2}) gives
\begin{align}
&\sup_{\{i, \gamma_{*}\}} \Big|\breve{f}^{-i}_{{\xi}} \big(\widehat{U}_{\gamma i};\gamma \big) - \bar{f}^{-i}_{{\xi}}(U_{\gamma i};\gamma)\Big| \hspace{53in}\nonumber\\
&= \sup_{\{i, \gamma_*\}} \bigg|\frac{1}{n-1} \sum_{\{j:j \neq i\}}  (\partial_{u} f_0(U_{\gamma i}; \gamma)) \xi^{[0]}_{j}(W_i,\gamma) - \E\bigg[\frac{1}{ \xi^2}  K^{(1)}\bigg(\frac{U_{\gamma j} - U_{\gamma i}}{\xi}\bigg)\E\big[{\xi}^{[0]}_{\ell}(W_j,\gamma) \bigm| W_j \big]\biggm | U_{\gamma i} \bigg] \bigg| \nonumber\\
&\stackrel{\triangle}{=} \sup_{\{i, \gamma_*\}} \bigg| \frac{1}{n-1}\sum_{\{j: j \neq i\}} (\partial_{u} f_0(U_{\gamma i}; \gamma)) \xi^{[0]}_{j}(W_i,\gamma) - \psi^{[0]}(W_i, U_{\gamma i}; \gamma) \bigg| = o_p \bigg( \sqrt{\frac{\log n}{n h}} \, \bigg) + O(h^2). \label{eq:lemma_C_12}
\end{align}

Applying a third-order Taylor expansion of $\partial^t_{\gamma_*}K\big(\big(\widehat{U}_{\gamma j}-\widehat{U}_{\gamma i} \big)/\xi \big)$ around $\widehat{U}_{\gamma j}-\widehat{U}_{\gamma i}=U_{\gamma j}-U_{\gamma i}$, the first term on the right-hand side of (\ref{eq:lemma_C_1}) with $t=1, 2$ can be written as
\begin{align}  
&\partial_{\gamma_*}\breve{f}^{-i}_{{\xi}} \big(\widehat{U}_{\gamma i};\gamma \big) - \partial_{\gamma_*}\bar{f}^{-i}_{{\xi}} (U_{\gamma i};\gamma) \nonumber\\
&= \frac{1}{n-1} \sum_{\{j:j \neq i\}} \frac{1}{\xi^{2}} K^{(1)}\bigg(\frac{U_{\gamma j} - U_{\gamma i}}{\xi}\bigg) \bigg( \frac{1}{n-1} \sum_{\{\ell:\ell \neq i\}} \xi^{[1]}_{\ell}(W_i, \gamma)  + r^{[1]}(W_i, \gamma)-  \frac{1}{n-1} \sum_{\{\ell:\ell \neq j\}} \xi^{[1]}_{\ell}(W_j, \gamma) \nonumber\\
&~~~~ - r^{[1]}(W_j, \gamma) \bigg) + \frac{1}{n-1}\sum_{\{j:j \neq i\}} \sum_{s=1}^3 \frac{1}{s! \,\xi^{s+2}} K^{(s+1)}\bigg(\frac{U_{\gamma j} - U_{\gamma i}}{\xi}\bigg) \Delta^{s}_{ij}\Big( \partial_{\gamma_*} \widehat{m}^{-i}_h(\gamma^{\top}W_i; \gamma)\nonumber\\
&~~~~ -  \partial_{\gamma_*} \widehat{m}^{-j}_h(\gamma^{\top}W_j; \gamma)  \Big) \text{ and}\label{eq:lemma_C_3_13}
\end{align}
\begin{align}  
&\partial^2_{\gamma_*}\breve{f}^{-i}_{{\xi}} \big(\widehat{U}_{\gamma i};\gamma \big) - \partial^2_{\gamma_*}\bar{f}^{-i}_{{\xi}} (U_{\gamma i};\gamma) \nonumber\\
&=  \frac{1}{n-1} \sum_{\{j:j \neq i\}} \frac{1}{\xi^{3}} K^{(2)}\bigg(\frac{U_{\gamma j} - U_{\gamma i}}{\xi}\bigg) \bigg( \frac{1}{n-1} \sum_{\{\ell:\ell \neq i\}} \xi^{[1]}_{\ell}(W_i, \gamma) + r^{[1]}(W_i, \gamma) - \frac{1}{n-1} \sum_{\{\ell:\ell \neq j\}} \xi^{[1]}_{\ell}(W_j, \gamma) \nonumber\\
& ~~~~  - r^{[1]}(W_j, \gamma) \bigg)^{\otimes 2} + 2\,\text{sym} \bigg(\frac{1}{n-1}\sum_{\{j:j \neq i\}} \frac{1}{\xi^{3}} K^{(2)}\bigg(\frac{U_{\gamma j} - U_{\gamma i}}{\xi}\bigg) \bigg( \frac{1}{n-1} \sum_{\{\ell:\ell \neq i\}} \xi^{[1]}_{\ell}(W_i, \gamma) + r^{[1]}(W_i, \gamma)\nonumber\\
&~~~~ - \frac{1}{n-1} \sum_{\{\ell:\ell \neq j\}} \xi^{[1]}_{\ell}(W_j, \gamma) - r^{[1]}(W_j, \gamma) \bigg) \Big(m^{[1]}(W_i, \gamma^{\top}W_i) - m^{[1]}(W_j, \gamma^{\top}W_j) \Big)^{\top}  \bigg) +\frac{1}{n-1} \sum_{\{j:j \neq i\}} \nonumber\\
& ~~~~   \frac{1}{\xi^{2}} K^{(1)}\bigg(\frac{U_{\gamma j} - U_{\gamma i}}{\xi}\bigg)\bigg(  \frac{1}{n-1} \sum_{\{\ell:\ell \neq i\}} \xi^{[2]}_{\ell}(W_i, \gamma)+ r^{[2]}(W_i, \gamma) - \frac{1}{n-1} \sum_{\{\ell:\ell \neq j\}} \xi^{[2]}_{\ell}(W_j, \gamma)  - r^{[2]}(W_j, \gamma) \bigg)\nonumber\\
& ~~~~ + \frac{1}{n-1}\sum_{\{j:j \neq i\}} \sum_{s=1}^3 \frac{1}{s!  \,\xi^{s+3}} K^{(s+2)}\bigg(\frac{U_{\gamma j} - U_{\gamma i}}{\xi}\bigg) \Delta^{s}_{ij}  \Big( \partial_{\gamma_*} \widehat{m}^{-i}_h(\gamma^{\top}W_i; \gamma) -  \partial_{\gamma_*} \widehat{m}^{-j}_h(\gamma^{\top}W_j; \gamma)  \Big)^{\otimes 2} \nonumber\\
&~~~~ +\frac{1}{n-1} \sum_{\{j:j \neq i\}} \sum_{s=1}^3 \frac{1}{s! \,\xi^{s+2}} K^{(s+1)}\bigg(\frac{U_{\gamma j} - U_{\gamma i}}{\xi}\bigg) \Delta^{s}_{ij} \Big( \partial^2_{\gamma_*} \widehat{m}^{-i}_h(\gamma^{\top}W_i; \gamma) -  \partial^2_{\gamma_*} \widehat{m}^{-j}_h(\gamma^{\top}W_j; \gamma) \Big). \label{eq:lemma_C_14}
\end{align}
Paralleling to the arguments for (\ref{eq:lemma_C_12}), the uniform convergence rates of $\partial_{\gamma_{*}}^t \breve{f}^{-i}_{{\xi}} \big(\widehat{U}_{\gamma i};\gamma \big) - \partial_{\gamma_{*}}^t \bar{f}^{-i}_{{\xi}} (U_{\gamma i};$ $\gamma)$ for $t=1,2$ follow. Specifically,
\begin{align}
&\sup_{\{i, \gamma_{*}\}} \Big\|\partial_{\gamma_*} \breve{f}^{-i}_{{\xi}} \big(\widehat{U}_{\gamma i};\gamma \big) - \partial_{\gamma_*} \bar{f}^{-i}_{{\xi}}(U_{\gamma i};\gamma)\Big\|\hspace{0.3in} \nonumber\\
&=\sup_{\{i, \gamma_*\}} \bigg\|\frac{1}{n-1}\sum_{\{j: j \neq i\}} (\partial_{u} f_0(U_{\gamma i}; \gamma)) \xi^{[1]}_{j}(W_i,\gamma) - \E\bigg[\frac{1}{\xi^2} K^{(1)}\bigg(\frac{U_{\gamma j} - U_{\gamma i}}{\xi}\bigg)\E\big[{\xi}^{[1]}_{\ell}(W_j,\gamma) \bigm| W_j\big]\hspace{0.3in}\nonumber\\
&~~~~ +  (\partial^2_{u} f_0(U_{\gamma i}; \gamma)) \E\big[\xi^{[0]}_{j}(W_i,\gamma)\bigm| W_i\big] m^{[1]}(W_i, \gamma^{\top}W_i) \biggm| U_{\gamma i} \bigg] \nonumber\\
&~~~~   - \E\bigg[\frac{1}{\xi^3} K^{(2)}\bigg(\frac{U_{\gamma j} - U_{\gamma i}}{\xi}\bigg)  \E\big[\xi^{[0]}_{\ell}(W_i,\gamma)\bigm| W_i\big] m^{[1]}(W_j, \gamma^{\top}W_j)\biggm| W_i, U_{\gamma i}\bigg]\nonumber\\
&~~~~ +E\bigg[ \frac{1}{\xi^3} K^{(2)}\bigg(\frac{U_{\gamma j} - U_{\gamma i}}{\xi}\bigg) \E\big[{\xi}^{[0]}_{\ell}(W_j,\gamma) \bigm| W_j\big]\Big(m^{[1]}(W_j, \gamma^{\top}W_j) - m^{[1]}(W_i, \gamma^{\top}W_i)\Big) \biggm | W_i, U_{\gamma i}\bigg] \bigg\| \nonumber\\
& \stackrel{\triangle}{=} \sup_{\{i, \gamma_*\}} \bigg\|\frac{1}{n-1}\sum_{\{j: j \neq i\}} (\partial_{u} f_0(U_{\gamma i}; \gamma)) \xi^{[1]}_{j}(W_i,\gamma)  -  \psi^{[1]}(W_i, U_{\gamma i}; \gamma) \bigg\| = o_p \bigg( \sqrt{\frac{\log n}{n h^{3}}} \, \bigg) + O(h^2) \label{eq:lemma_C_21}
\end{align}
and
\begin{align}
&\hspace{-0.1in}\sup_{\{i, \gamma_{*}\}} \Big\|\partial^2_{\gamma_*} \breve{f}^{-i}_{{\xi}} \big(\widehat{U}_{\gamma i};\gamma \big) - \partial^2_{\gamma_*} \bar{f}^{-i}_{{\xi}}(U_{\gamma i};\gamma)\Big\| \nonumber\\
&\hspace{-0.1in}= \sup_{\{i, \gamma_*\}} \bigg\|\frac{1}{n-1}\sum_{\{j: j \neq i\}} (\partial_{u} f_0(U_{\gamma i}; \gamma)) \xi^{[2]}_j(W_i, \gamma) + (\partial^2_{u} f_0(U_{\gamma i}; \gamma)) \big( \E\big[\xi^{[1]}_j(W_i, \gamma)\bigm| W_i\big] \nonumber\\
& \big(m^{[1]}(W_i, \gamma^{\top}W_i)\big)^{\top} + m^{[1]}(W_i, \gamma^{\top}W_i) \big(\E\big[\xi^{[1]}(W_i, \gamma) \bigm| W_i \big]\big)^{\top}+ \E\big[\xi^{[0]}_j(W_i, \gamma)\bigm| W_i \big] m^{[2]}(W_i, \gamma^{\top}W_i)  \big)\nonumber\\
& + (\partial^3_{u} f_0(U_{\gamma i}; \gamma)) \E\big[\xi^{[0]}(W_i, \gamma) \bigm| W_i \big]\big( m^{[1]}(W_i, \gamma^{\top}W_i) \big)^{\otimes 2} - \E\bigg[\frac{1}{\xi^3} K^{(2)}\bigg(\frac{U_{\gamma j} - U_{\gamma i}}{\xi}\bigg)\Big(  \xi^{[0]}_{\ell}(W_i,\gamma)\nonumber\\
& m^{[2]}(W_j, \gamma^{\top}W_j) +  2 \,\text{sym}\Big( m^{[1]}(W_j, \gamma^{\top}W_j) \big( \xi^{[1]}_{\ell}(W_i,\gamma) \big)^{\top}   \Big) \Big) \biggm| W_i, U_{\gamma i} \bigg] \E\bigg[\frac{1}{\xi^3} K^{(2)}\bigg(\frac{U_{\gamma j} - U_{\gamma i}}{\xi}\bigg) \nonumber\\
& -  \bigg( \big( m^{[1]}(W_j, \gamma^{\top}W_j) - m^{[1]}(W_i, \gamma^{\top}W_i) \big) \Big(\E \big[{\xi}^{[1]}_{\ell}(W_j,\gamma)\bigm| W_j\big]\Big)^{\top}  + \Big(\E \big[{\xi}^{[0]}_{\ell}(W_j,\gamma)\bigm| W_j\big]\Big) \nonumber\\
& \big(  m^{[2]}(W_j, \gamma^{\top}W_j)  - m^{[2]}(W_i, \gamma^{\top}W_i) \big)\bigg) \biggm | W_i, U_{\gamma i}\bigg] - \E\bigg[\frac{1}{\xi^2} K^{(1)}\bigg(\frac{U_{\gamma j} - U_{\gamma i}}{\xi}\bigg) \E \big[{\xi}^{[2]}_{\ell}(W_j,\gamma)\bigm| W_j\big] \bigg| W_i,  \nonumber\\
&  U_{\gamma i} \bigg] - \E\bigg[\frac{1}{\xi^4} K^{(3)}\bigg(\frac{U_{\gamma j} - U_{\gamma i}}{\xi}\bigg)\Big( m^{[1]}(W_i, \gamma^{\top}W_i) - m^{[1]}(W_j, \gamma^{\top}W_j)\Big)^{\otimes 2} \E \big[{\xi}^{[0]}_{\ell}(W_j,\gamma)\bigm| W_j\big] \biggm| W_i, U_{\gamma i} \bigg]\nonumber\\
&   - \E\bigg[\frac{1}{\xi^4} K^{(3)}\bigg(\frac{U_{\gamma j} - U_{\gamma i}}{\xi}\bigg)
\big[ \big( m^{[1]}(W_j, \gamma^{\top}W_j) \big)^{\otimes 2}- 2 \,\text{sym} \big(m^{[1]}(W_i, \gamma^{\top}W_i)\big( m^{[1]}(W_j, \gamma^{\top}W_j) \big)^{\top}\big)    \big] \nonumber\\
& \xi^{[0]}_{\ell}(W_i,\gamma) \biggm| W_i, U_{\gamma i} \bigg] \bigg\| \nonumber\\
&\hspace{-0.1in}\stackrel{\triangle}{=} \sup_{\{i, \gamma_*\}} \bigg\| \frac{1}{n-1}\sum_{\{j: j \neq i\}} (\partial_{u} f_0(U_{\gamma i}; \gamma)) \xi^{[2]}_j(W_i, \gamma)  - \psi^{[2]}(W_i, U_{\gamma i}; \gamma) \bigg\|  = o_p \bigg( \sqrt{\frac{\log n}{n h^{5}}} \, \bigg) + O(h^2). \label{eq:lemma_C_22}
\end{align}
Combining (\ref{eq:lemma_C_12}), (\ref{eq:lemma_C_21}), and (\ref{eq:lemma_C_22}) with Lemma \ref{A:Lemma_3} establishes the assertions of the lemma, with
\begin{align}
\varphi^{[t]}_i(w, u; \gamma) = (\partial_{u} f_0(u; \gamma)) \xi^{[t]}_{i}(w,\gamma) - \psi^{[t]}(w, u; \gamma) + \zeta^{[t]}_i(w, u; \gamma),~t=0,1,2,~i=1,\dots,n.\label{eq:lemma_C_23}
\end{align}

\end{proof}
\vspace{-0.35in}
\section{Proof of Theorem \ref{Thm2.1}}
\label{sec:A.3}

A direct decomposition yields
\begin{align} 
\hspace{-0.1in}\frac{2}{n} \textsc{pss}(\gamma,\mathcal{G}^{\text{o}}) = &~\frac{1}{n} \sum_{i=1}^n  U^2_{\gamma i} + \frac{1}{n} \sum_{i=1}^n \big(m^{[0]}\big(W_i,\gamma^{\top}W_i\big) - \widehat{m}_h^{-i}\big(\gamma^{\top}W_i;\mathcal{G}^{\text{o}},\gamma\big) \big)^2 \nonumber \\
&~ + \frac{2}{n} \sum_{i=1}^n  U_{\gamma i} \big(m^{[0]}\big(W_i, \gamma^{\top}W_i\big) - \widehat{m}_h^{-i}\big(\gamma^{\top}W_i;\mathcal{G}^{\text{o}},\gamma\big) \big)  
\stackrel{\triangle}{=}  I_{1n}(\gamma) + I_{2n}(\gamma) + I_{3n}(\gamma). \hspace{-0.1in} \label{eq:thm1_1}
\end{align}
Under assumptions {A2} and {A3}, the representation $U_{\gamma} = U^{\text{o}}+ (m(\gamma^{\text{o} \top}W) - m^{[0]}(W, \gamma^{\top}W))$, together with $|(\gamma^{\text{o}} - \gamma)^{\top}W| \leq \|(\gamma^{\text{o}} - \gamma)\| \|W\|$ and the compactness of $\Gamma$, implies that $\E[U_{\gamma}^2]<\infty$. 
Since $\{U_{\gamma}: \gamma_* \in \Gamma\}$ is a Euclidean class, which is ensured by Lemma 22 in \cite{nolan1987u}, an application of
Theorem II.24 in \cite{pollard1984convergence} gives
\vspace{-0.015in}
\begin{align}
\sup_{\{\gamma_* \in \Gamma\}} \big| I_{1n}(\gamma) - \E\big[U_{\gamma}^2\big] \big| =  o_p(1). \label{eq:thm1_2}
\end{align}
Under assumptions {A1} and {A2}, and upon setting $t=0$ in Lemma \ref{A:Lemma_2}, we obtain
\begin{align}
\sup_{\{\gamma_* \in \Gamma\}} |I_{2n}(\gamma)| = o_p(1) \text{ and } \sup_{\{\gamma_* \in \Gamma\}}  |I_{3n}(\gamma)| = o_p(1). \label{eq:thm1_3}
\end{align}
Substituting (\ref{eq:thm1_2}) and (\ref{eq:thm1_3}) into (\ref{eq:thm1_1}) and applying the triangle inequality yield
\begin{align}
\sup_{\{\gamma_* \in \Gamma\}}  \bigg| \frac{2}{n} \textsc{pss}(\gamma,\mathcal{G}^{\text{o}}) - \E\big[U_{\gamma}^2\big] \bigg| = o_p (1). \label{eq:thm1_4}
\end{align}
Since $\E[U_{\gamma}^2] \geq \E[(U^{\text{o}})^{2}]$, with strict inequality whenever $\gamma\neq \gamma^{\text{o}}$, it follows that under assumption {A6}, $\gamma^{\text{o}}$ is the unique minimizer of $\E[U_{\gamma}^2]$ over $\Gamma$.
Consequently, by (\ref{eq:thm1_4}) and the law of large numbers, $ 2\textsc{pss}\big(\widehat{\gamma}^{or},\mathcal{G}^{\text{o}}\big) /n - 2\textsc{pss}\big(\gamma^{\text{o}},\mathcal{G}^{\text{o}}\big) / n = \E[U_{\widehat{\gamma}^{or}}^2] - \E[(U^{\text{o}})^{2}] + o_p(1)$, where $\E[U_{\widehat{\gamma}^{or}}^2] - \E[(U^{\text{o}})^{2}] > 0$.
On the other hand, by the definition of $\widehat{\gamma}^{or}$ as a minimizer of $\textsc{pss}\big(\gamma,\mathcal{G}^{\text{o}}\big)$, we have $\textsc{pss}(\gamma^{\text{o}},\mathcal{G}^{\text{o}}) \geq \textsc{pss}(\widehat{\gamma}^{or},\mathcal{G}^{\text{o}})$. Therefore, $\E[U_{\widehat{\gamma}^{or}}^2] - \E[(U^{\text{o}})^{2}] = o_p(1)$.
The consistency of $\widehat{\gamma}^{or}_{*}$ to $\gamma^{\text{o}}_{*}$ then follows from the uniqueness of $\gamma^{\text{o}}_*$ as the minimizer of $\E[U_{\gamma}^{2}]$ over $\Gamma$ and the continuity of $\E[U_{\gamma}^{2}]$ on $\Gamma$.

Define $\textsc{ps}(\gamma)=\partial_{\gamma_{*}} \textsc{pss}({\gamma}, \mathcal{G}^{\text{o}}) / n$ and $\textsc{ph}(\gamma)= \partial_{\gamma_{*}}^2 \textsc{pss}({\gamma}, \mathcal{G}^{\text{o}}) / n$.
For an arbitrary constant vector $\nu\in \mathbb{R}^{k(p+1) - 2}$, a first-order Taylor expansion of $\nu^{\top}\textsc{ps}(\widehat{\gamma}^{or})$ around $\gamma^{\text{o}}_{*}$ yields
\begin{align}
\nu^{\top} (\sqrt{n}\, \textsc{ps}(\gamma^{\text{o}})) + \nu^{\top} \{\textsc{ph}(\bar{\gamma}) [\sqrt{n}\,(\widehat{\gamma}^{or}_{*} - \gamma^{\text{o}}_{*})]\} = 0, \label{eq:thm1_5}
\end{align}
where $\bar{\gamma}$ lies on the line segment between $\widehat{\gamma}^{or}$ and $\gamma^{\text{o}}$.
The pseudo score $\textsc{ps} (\gamma^\text{o})$ in (\ref{eq:thm1_5}) admits the decomposition 
\begin{align}
\textsc{ps} (\gamma^\text{o}) =& \sum_{\ell_1=0}^1 \sum_{\ell_2=0}^1 \frac{1}{n} \sum_{i=1}^n (-1)^{\ell_1}\big( U_i^{\text{o}} \big)^{1-\ell_1} \Big(\widehat{m}^{-i}_h \big(\gamma^{\text{o} \top} W_i;\mathcal{G}^{\text{o}},\gamma^{\text{o}}\big)  -  m\big(\gamma^{\text{o} \top} W_i \big)\Big)^{\ell_1} \Big( m^{[1]} \big(W_i, \gamma^{\text{o} \top} W_i \big) \Big)^{1-\ell_2} \nonumber \\
& \big( \partial_{\gamma^{\text{o}}_{*}} \widehat{m}^{-i}_h \big(\gamma^{\text{o} \top} W_i; \mathcal{G}^{\text{o}},\gamma^{\text{o}}\big) - m^{[1]}\big(W_i, \gamma^{\text{o} \top} W_i \big) \big)^{\ell_2}  
\stackrel{\triangle}{=} \sum_{\ell_1=0}^1 \sum_{\ell_2=0}^1 \textsc{ps}_{\ell_1 \ell_2}(\gamma^{\text{o}}). \label{eq:thm1_6}
\end{align}
Under assumptions {A1} and {A2}, and upon setting $t=0,1$ in Lemma \ref{A:Lemma_2}, it follows that
\begin{align}
\sqrt{n} \textsc{ps}_{01}(\gamma^{\text{o}}) = &~ \sqrt{n} \frac{1}{n} \sum_{i = 1}^n U_i^{\text{o}} \bigg( \frac{1}{n-1} \sum_{\{j: j \neq i\}} \xi^{[1]}_j(W_i, \gamma^{\text{o}}) + r^{[1]}(W_i, \gamma^{\text{o}}) \bigg) \hspace{2in} \nonumber \\
= &~ \frac{ \sqrt{n}}{n(n-1)} \sum_{i \neq j}^n U_i^{\text{o}} \, \E\big[{\xi}^{[1]}_j(W_i, \gamma^{\text{o}}) \bigm| W_i\big] +  \frac{ \sqrt{n}}{n(n-1)} \sum_{i \neq j} U_i^{\text{o}} \, \bar{\xi}^{[1]}_j(W_i, \gamma^{\text{o}})
+ o_p(1),\label{eq:thm1_7} \\
\sqrt{n} \textsc{ps}_{10}(\gamma^{\text{o}}) = &  -\sqrt{n} \frac{1}{n} \sum_{i = 1}^n m^{[1]}\big(W_i, \gamma^{\text{o} \top} W_i\big) \bigg( \frac{1}{n-1} \sum_{\{j: j \neq i\}} \xi^{[0]}_j(W_i, \gamma^{\text{o}}) + r^{[0]}(W_i, \gamma^{\text{o}}) \bigg) \hspace{0.5in} \nonumber \\
= & -\sqrt{n} \frac{1}{n(n-1)} \sum_{i \neq j} m^{[1]}\big(W_i, \gamma^{\text{o} \top} W_i\big) \bar{\xi}^{[0]}_j(W_i, \gamma^{\text{o}})  \nonumber \\
&- \sqrt{n} \frac{1}{n(n-1)} \sum_{i \neq j}^n m^{[1]}\big(W_i, \gamma^{\text{o} \top} W_i\big) \E\big[{\xi}^{[0]}_j(W_i, \gamma^{\text{o}}) \mid W_i\big]  + o_p(1), \text{ and}\label{eq:thm1_7.5}\\
\sqrt{n} \|\textsc{ps}_{11}(\gamma^{\text{o}})\| = &~\sqrt{n}\bigg\| \frac{1}{n} \sum_{i=1}^n \prod_{t=0}^1\big( \partial_{\gamma^{\text{o}}_{*}}^t \widehat{m}^{-i}_h \big(\gamma^{\text{o} \top} W_i; \gamma^{\text{o}}\big) - m^{[t]}\big(W_i, \gamma^{\text{o} \top} W_i \big) \big)  \bigg\| \nonumber \\
\leq &~  \sqrt{n} \prod_{t=0}^1 \sup_{i} \big\| \partial_{\gamma^{\text{o}}_{*}}^t \widehat{m}^{-i}_h \big(\gamma^{\text{o} \top} W_i; \gamma^{\text{o}}\big) - m^{[t]}\big(W_i, \gamma^{\text{o} \top} W_i \big)  \big\| \nonumber\\
= &~ \sqrt{n} \bigg(o_p\bigg(\sqrt{\frac{\log n}{nh}}\,\bigg) + O\big(h^2\big)  \bigg) \bigg( o_p\bigg(\sqrt{\frac{\log n}{nh^3}}\,\bigg) + O\big(h^2\big) \bigg)= o_p(1), \label{eq:thm1_8} 
\end{align}
where $\bar{\xi}^{[t]}_j(W_i, \gamma^{\text{o}}) = {\xi}^{[t]}_j( W_i, \gamma^{\text{o}}) - \E[{\xi}^{[t]}_j( W_i, \gamma^{\text{o}}) \mid W_i]$, $i, j = 1, \dots, n$, $t = 0, 1$.
The identities
\begin{align*}
&\xi^{[0]}_{j}(W_i,\gamma^{\text{o}}) = \frac{\big( Y_j - m\big(\gamma^{\text{o}\top} W_i\big) \big) K_{h}(\gamma^{\text{o}\top} (W_{j} - W_i)) }{f_{\gamma^{\text{o}}}(\gamma^{\text{o}\top}W_i)},\\
&m^{[1]}\big(W_i, \gamma^{\text{o} \top} W_i \big) = \big( \partial_v m \big(\gamma^{\text{o} \top} W_i \big)\big) \big( W_{*i} - H(\gamma^{\text{o} \top} W_i) \big), \text{ with } H(v) = \E[W_{*} \mid \gamma^{\text{o} \top}W = v], \\
&\E\big[ U_i^{\text{o}} \bar{\xi}^{[1]}_j(W_i, \gamma^{\text{o}}) \bigm| W_i, Y_i \big] = 0, ~ \E\big[ m^{[1]}\big(W_i, \gamma^{\text{o} \top} W_i \big) \bar{\xi}^{[0]}_j(W_i, \gamma^{\text{o}}) \bigm| W_i, Y_i \big] = 0,\\
&\E\big[ U_i^{\text{o}} \bar{\xi}^{[1]}_j(W_i, \gamma^{\text{o}}) \bigm| W_j, Y_j \big] = \E\big[ \E\big[ U_i^{\text{o}} \bigm| W_i\big] \bar{\xi}^{[1]}_j(W_i, \gamma^{\text{o}}) \bigm|W_j, Y_j \big] = 0, \text{ and}\\
&\E \big[ m^{[1]}\big(W_i, \gamma^{\text{o} \top} W_i \big) \bar{\xi}^{[0]}_j(W_i, \gamma^{\text{o}}) \bigm| W_j, Y_j \big] \\
&=\E \big[ \big( \partial_v m \big(\gamma^{\text{o} \top} W_i \big)\big) \big( H(\gamma^{\text{o} \top} W_i) - \E[W_{*i} \bigm| \gamma^{\text{o} \top} W_i] \big) \bar{\xi}^{[0]}_j(W_i, \gamma^{\text{o}}) \bigm| W_j, Y_j \big] = 0, ~j = 1, \dots, n,
\end{align*}
further imply that the second-order $U$-statistics in (\ref{eq:thm1_7}) and (\ref{eq:thm1_7.5}) are degenerate.
Since $\E[(U_i^{\text{o}} \bar{\xi}^{[1]}_j$ $(W_i, \gamma^{\text{o}}) )^2]= O(h^{-3})$ and $\E[(m^{[1]}(W_i, \gamma^{\text{o} \top} W_i) \bar{\xi}^{[0]}_j(W_i, \gamma^{\text{o}}) )^2] = O(h^{-1})$, an application of Theorem 8.1 in \cite{hoeffding1948probability} yields
\begin{align}
& \Bigg\| \frac{1}{n(n-1)}\sum_{i \neq j} U_i^{\text{o}} \bar{\xi}^{[1]}_j(W_i, \gamma^{\text{o}}) \Bigg\| = O_p\bigg(\frac{1}{n\sqrt{h^{3}}}\bigg)\text{ and} \label{eq:thm1_9.0}\\
&\Bigg\| \frac{1}{n(n-1)}\sum_{i \neq j} m^{[1]}\big(W_i, \gamma^{\text{o} \top} W_i \big) \bar{\xi}^{[0]}_j(W_i, \gamma^{\text{o}}) \Bigg\| = O_p\bigg(\frac{1}{n\sqrt{h}}\bigg).  \label{eq:thm1_9}
\end{align}
Moreover, Lemma \ref{A:Lemma_2} with $t=0,1$, together with the central limit theorem, implies
\begin{align}
&\bigg\| \frac{1}{n} \sum_{i=1}^n  U_i^{\text{o}} \E\Big[{\xi}^{[1]}_j(W_i, \gamma^{\text{o}}) \Bigm| W_i\Big] \bigg\| = O\bigg( \frac{h^2}{\sqrt{n}} \bigg) \text{ and}\label{eq:thm1_9.5.0}\\
&\bigg\| \frac{1}{n} \sum_{i=1}^n  m^{[1]}\big(W_i, \gamma^{\text{o} \top} W_i \big) \E\Big[{\xi}^{[0]}_j(W_i, \gamma^{\text{o}}) \Bigm| W_i\Big] \bigg\| = O\bigg( \frac{h^2}{\sqrt{n}} \bigg). \label{eq:thm1_9.5}
\end{align}
Substituting  (\ref{eq:thm1_9.0}) and (\ref{eq:thm1_9.5.0}) into (\ref{eq:thm1_7}), and (\ref{eq:thm1_9}) and (\ref{eq:thm1_9.5}) into (\ref{eq:thm1_7.5}), we obtain
\begin{align}
\sqrt{n} \textsc{ps}_{01}(\gamma^{\text{o}}) = o_p(1) \text{ and } \sqrt{n} \textsc{ps}_{10}(\gamma^{\text{o}}) = o_p(1). \label{eq:thm1_10}
\end{align}
Combining (\ref{eq:thm1_8}) and (\ref{eq:thm1_10}) with the central limit theorem result $\sqrt{n} \nu^{\top} \textsc{ps}_{00}(\gamma^{\text{o}}) \stackrel{d}{\longrightarrow} N\big(0, \nu^{\top} V_1 \nu\big)$, and using (\ref{eq:thm1_6}), we obtain
\begin{align}
\sqrt{n} \textsc{ps}(\gamma^{\text{o}})\stackrel{d}{\longrightarrow} N\big(0, V_1\big). \label{eq:thm1_12}
\end{align}

By Lemma \ref{A:Lemma_2} with $t=1, 2$, it follows that
\begin{align}
\textsc{ph}(\gamma) = \frac{1}{n}  \sum_{i=1}^n \big( m^{[1]}\big(W_i, \gamma^{\top}W_i\big) \big)^{\otimes 2} - \frac{1}{n} \sum_{i=1}^n U_i\, m^{[2]} \big(W_i, \gamma^{\top}W_i\big) + o_p\bigg( \sqrt{\frac{\log n}{nh^5}} \,\bigg) + O(h^2). \label{eq:thm1_13}
\end{align}
Under assumption {A1}, applying the law of large numbers to (\ref{eq:thm1_13}) gives
\begin{align}
\textsc{ph}(\gamma)  \stackrel{p}{\longrightarrow} E\big[ \big(m^{[1]}\big(W, \gamma^{\top}W\big)\big)^{\otimes 2} - U\, m^{[2]} \big(W, \gamma^{\top}W\big) \big].  \label{eq:thm1_14}
\end{align}
The continuity of $\textsc{ph}(\gamma)$ in $\gamma$, the consistency of $\widehat{\gamma}^{or}_{*}$ to $\gamma^{\text{o}}_*$, and $\E[U\, m^{[2]} \big(W, \gamma^{\top}W\big)]=0$ for $\gamma = \gamma^{\text{o}}$, yield
\begin{align}
\textsc{ph}(\bar{\gamma}) \stackrel{p}{\longrightarrow} V_2. \label{eq:thm1_15}
\end{align}
Substituting (\ref{eq:thm1_12}) and (\ref{eq:thm1_15}) into (\ref{eq:thm1_5}) and applying Slutsky's theorem, we establish the asymptotic normality of $\widehat{\gamma}^{or}_{*}$ under assumption {A6}.

\section{Proof of Theorem \ref{Thm3.1}}
\label{sec:A.1}
 
The SP estimator $(\widehat{\beta},\widehat{\gamma})$ possesses the oracle property if, with probability converging to one,
\begin{align*}
\textsc{pss}_{\textsc{sp}}\big(\beta,\gamma\big) > \textsc{pss}_{\textsc{sp}}\big(\widehat{\beta}^{or},\widehat{\gamma}^{{or}}\big)
\end{align*}
\noindent
for all $(\beta,\gamma) \in B _n\times \Gamma_n$ such that $(\beta, \gamma)\neq (\widehat{\beta}^{or},\widehat{\gamma}^{or})$. It suffices to show that, with probability converging to one, 
\begin{align*}
\textsc{pss}_{\textsc{sp}}\big(\beta^{*},\gamma^{*}\big)  >\textsc{pss}_{\textsc{sp}}\big(\widehat{\beta}^{or},\widehat{\gamma}^{{or}}\big)\text{ and }\textsc{pss}_{\textsc{sp}}\big(\beta,\gamma\big) \geq \textsc{pss}_{\textsc{sp}}\big(\beta^{*}
,\gamma^{*}\big),
\end{align*} 
where 
\begin{align}
\gamma^{*}_{c}=\frac{1}{|\mathcal{G}_{c}^{\text{o}}|}\sum_{\{i\in\mathcal{G}_{c}^{\text{o}}\}}\beta_{i} \text{ and }
\beta^{*}_{i}=\sum^{k}_{c=1}\gamma^{*}_{c}I(i\in\mathcal{G}_{c}^{\text{o}}), ~c\in\mathcal{C}, ~i=1,\dots, n. \label{eq:A.2.0}
\end{align}

By the definitions of $(\beta^{*},\gamma^{*})$ in (\ref{eq:A.2.0}) and $\textsc{pss}\big(\gamma,\mathcal{G} \big)$ in (\ref{fn:pss1}), the penalty term satisfies $\sum^{n}_{i=1}\min_{c}$ $\big\|\beta^{*}_{(1)i}-\gamma^{*}_{(1)c}\big\|_{1}=0$,
which implies that
\begin{align}
\textsc{pss}_{\textsc{sp}}\big(\beta^{*},\gamma^{*}\big)= \textsc{pss}\big(\gamma^{*},\mathcal{G}^{\text{o}} \big). 
\label{eq:A.2.1}
\end{align}
A second-order Taylor expansion of $\textsc{pss} \big(\gamma^{*}, \mathcal{G}^{\text{o}}\big)$ around $\gamma^{*}_{*}=\widehat{\gamma}^{{or}}_*$, combined with 
$\gamma_*=\gamma^{\text{o}}_{*}+o(1)$, the consistency of $\widehat{\gamma}^{{or}}$ to $\gamma^{\text{o}}_{*}$ in Theorem \ref{Thm2.1}, and (\ref{eq:thm1_15}), yields
\begin{align}
\frac{2}{n} \textsc{pss} \big(\gamma^{*}, \mathcal{G}^{\text{o}}\big) - \frac{2}{n} \textsc{pss} \big(\widehat{\gamma}^{or}, \mathcal{G}^{\text{o}}\big) = \frac{1}{2} \big(\gamma^{*}_{*} - \widehat{\gamma}^{{or}}_*\big)^{\top} V_{2} \big(\gamma^{*}_{*} - \widehat{\gamma}^{{or}}_*\big) (1 + o_p(1)). \label{eq:A.2.2}
\end{align}
It follows from (\ref{eq:A.2.1}) and (\ref{eq:A.2.2}) that under assumption {A6},
\begin{align}
\frac{1}{n} \textsc{pss}_{\textsc{sp}}\big (\beta^{*},\gamma^{*}\big) > \frac{1}{n}\textsc{pss}_{\textsc{sp}} \big(\widehat{\beta}^{or},\widehat{\gamma}^{or}\big) + o_p(1) \text{ for } \beta^{*} \in B \text{ with } \beta^{*} \neq \widehat{\beta}^{or}.
\label{eq:A.2.3}
\end{align}
A first-order Taylor expansion of $\sum_{i=1}^n \big(Y_{i}-\widehat{m}^{-i}_{h}\big(\beta_{i}^{\top}Z_i; \beta\big)\big)^2$ around $\beta = \beta^{*}$ yields
\begin{align}
& \textsc{pss}_{\textsc{sp}}\big(\beta,\gamma\big) - \textsc{pss}_{\textsc{sp}}\big(\beta^{*},\gamma^{*}\big) \nonumber \\
&= - \sum_{i = 1}^n \big(Y_i  - \widehat{m}^{-i}_h \big(\bar{\beta}^{\top}_i Z_i; \bar{\beta}\big)\big) \Bigg[ \sum_{j=1}^n \big(\partial_{\beta_j} \widehat{m}^{-i}_h (\bar{\beta}^{\top}_i Z_i; \bar{\beta})\big)^{\top} \big(\beta_j - \beta^*_j\big)\Bigg] + \lambda \sum^{n}_{i=1}\min_{c}\big\|\beta_{(1)i}-\gamma_{(1)c}\big\|_{1} \nonumber\\
& \stackrel{\triangle}{=}I_{1}+I_{2},  \label{eq:A.2.4}
\end{align}
where $\bar{\beta} = \alpha_0 \beta + (1-\alpha_0) \beta^*$ for some $\alpha_0 \in [0,1]$.

A direct calculation shows that
\begin{align}
I_1 = & -\sum_{i = 1}^n \big(Y_i  - \widehat{m}^{-i}_h \big(\bar{\beta}^{\top}_i Z_i; \bar{\beta}\big)\big)  \big(\partial_{\beta_i} \widehat{m}^{-i}_h (\bar{\beta}^{\top}_i Z_i; \bar{\beta})\big)^{\top} \big(\beta_i - \beta^*_i\big) \nonumber\\
& -\sum_{j = 1}^n  \sum_{\{i: i\neq j\}} \big(Y_i - \widehat{m}^{-i}_h \big(\bar{\beta}^{\top}_i Z_i; \bar{\beta}\big)\big)\,
\big(\partial_{\beta_j} \widehat{m}^{-i}_h (\bar{\beta}^{\top}_i Z_i; \bar{\beta})\big)^{\top} \big(\beta_j - \beta^*_j\big)\nonumber\\
= & \sum_{c = 1}^k \sum_{\{i \in \mathcal{G}_{c}^{\text{o}}\}} \big( \textsc{ps}_{ii} \big(\bar{\beta}\big) \big)^{\top} \sum_{\{j \in \mathcal{G}_{c}^{\text{o}}\}} \frac{\big(\beta_i - \beta_j\big)}{|\mathcal{G}_{c}^{\text{o}}|} + \sum_{j=1}^n  \sum_{\{i:i\neq j\}} \big(\textsc{ps}_{ij} \big(\bar{\beta}\big) \big)^{\top}\big(\beta_j - \beta^*_j\big)  \nonumber\\
= & \sum_{c = 1}^k \sum_{\{i \in \mathcal{G}_{c}^{\text{o}}\}} \big( \textsc{s}_i \big({\beta^{\text{o}}}\big)  \big)^{\top} \sum_{\{j \in \mathcal{G}_{c}^{\text{o}}\}} \frac{\big(\beta_i - \beta_j\big)}{|\mathcal{G}_{c}^{\text{o}}|} + \sum_{c = 1}^k \sum_{\{i \in \mathcal{G}_{c}^{\text{o}}\}} \big( \textsc{ps}_{ii} \big(\bar{\beta}\big) - \textsc{s}_i \big({\beta^{\text{o}}}\big) \big)^{\top} \sum_{\{j \in \mathcal{G}_{c}^{\text{o}}\}} \frac{\big(\beta_i - \beta_j\big)}{|\mathcal{G}_{c}^{\text{o}}|}\nonumber\\
& + \sum_{j=1}^n  \sum_{\{i:i\neq j\}} \big(\textsc{ps}_{ij} \big(\bar{\beta}\big) \big)^{\top}\big(\beta_j - \beta^*_j\big)  \nonumber \\
\stackrel{\triangle}{=} & I_{11} + I_{12} + I_{13},\label{eq:A.2.6}
\end{align}
where $\textsc{ps}_{ij}(\beta)= -\big(Y_i  - \widehat{m}^{-i}_h \big(\beta^{\top}_i Z_i; \beta\big)\big)  \partial_{\beta_j} \widehat{m}^{-i}_h (\beta^{\top}_i Z_i; \beta)$, $\textsc{s}_i \big({\beta^{\text{o}}}\big) = -\sum^{k}_{c=1}I(i\in\mathcal{G}^{\text{o}}_{c})\big(Y_i- {m}^{[0]}_{c} \big(Z_i,$ $\gamma^{\text{o}\top}_{c} Z_i\big)\big)  {m}^{[1]}_{c} \big(Z_i, \gamma^{\text{o}\top}_{c} Z_i\big)$, $i, j = 1, \dots, n$, and $m^{[t]}_{c}(z, v)$ is defined as $m^{[t]}(w, v)$ with $M_{s,t}(w,v)$ replaced by $M_{c,s,t}(z,v) = \E \big[ \big( m\big(\gamma^{\text{o} \top}_{c} Z \big) \big)^s \mid \gamma^{\top}_{c}Z = v \big] (-z_{*})^{\otimes t}$, $s, t = 0, 1$.
A second-order Taylor expansion of $K^{(t)}(\cdot)$ around $(\beta^{\text{o} \top}_j Z_j - \beta^{\text{o} \top}_i Z_i)/h$ yields
\begin{align}
&\frac{1}{(n-1)h^{t+1}} \sum_{\{j:j \neq i\}}  Y_j^s K^{(t)}\bigg(\frac{\bar{\beta}^{\top}_j Z_j - \bar{\beta}^{\top}_i Z_i}{h}\bigg) Z_{*i}^{\otimes t} \hspace{5in}  \nonumber\\
&=  \frac{1}{(n-1)h^{t+1}} \sum_{\{j:j \neq i\}}  Y_j^s  K^{(t)}\bigg(\frac{\beta^{\text{o} \top}_j Z_j - \beta^{\text{o} \top}_i Z_i}{h}\bigg)Z_{*i}^{\otimes t} + \frac{1}{(n-1)h^{t+2}} \sum_{\{j:j \neq i\}} Y_j^s  K^{(t+1)} \bigg(\frac{\beta^{\text{o} \top}_j Z_j - \beta^{\text{o} \top}_i Z_i}{h}\bigg) \nonumber \\
&~~~~\big[ Z^{\top}_j \big(\bar{\beta}_j - \beta^{\text{o}}_j\big) - Z^{\top}_i \big(\bar{\beta}_i - \beta^{\text{o} }_i\big) \big]Z_{*i}^{\otimes t} + \frac{1}{2(n-1)h^{t+3}} \sum_{\{j:j \neq i\}} Y_j^s K^{(t+2)} \bigg(\frac{\bar{\beta}^{* \top}_j Z_j - \bar{\beta}^{* \top}_i Z_i}{h}   \bigg) \nonumber\\
&~~~~ \Big[ \Big( Z^{\top}_j \big(\bar{\beta}_j - \beta^{\text{o}}_j\big) \Big)^2 -2 \big(\bar{\beta}_j - \beta^{\text{o}}_j\big)^{\top} Z_j Z^{\top}_i \big(\bar{\beta}_i - \beta^{\text{o} }_i\big) +\Big( Z^{\top}_i \big(\bar{\beta}_i - \beta^{\text{o}}_i\big) \Big)^2\Big]Z_{*i}^{\otimes t},\label{eq:A.2.8.0}
\end{align} 
where $\bar{\beta}^*_i$ lies on the line segment between $\bar{\beta}_i$ and $\beta^{\text{o}}_i$. 
Under assumptions {A2} and {A3}, arguments parallel to those used in the proof of Lemma \ref{A:Lemma_1} establish 
\begin{align}
&\sup_{\{i, c: i\in\mathcal{G}^{\text{o}}_{c}\}} \bigg\| \frac{1}{(n-1)h^{t+1}} \sum_{\{j: j \neq i\}}  Y_j^s  K^{(t)}\bigg(\frac{\beta^{\text{o} \top}_j Z_j - \beta^{\text{o} \top}_i Z_i}{h}\bigg)Z_{*i}^{\otimes t} - \partial_v^t \big( M_{c,s,t}\big(Z_i, \gamma^{\text{o}\top}_{c}Z_i \big) f_{c}\big( \gamma^{\text{o} \top}_{c}Z_i \big) \big)\bigg\| \nonumber \\
&= o_p \bigg( \sqrt{\frac{\log n}{n h^{2t+1}}}\, \bigg) + O \big( h^2 \big)\text{ and}\label{eq:A.2.7}\\
&\hspace{-0.1in}\sup_{\{i, c: i\in\mathcal{G}^{\text{o}}_{c}\}}  \frac{1}{(n-1)h^{t+2}} \sum_{\{j:j \neq i\}}\biggm| Y_j^s K^{(t+1)} \bigg(\frac{\beta^{\text{o} \top}_j Z_j - \beta^{\text{o} \top}_i Z_i}{h}\bigg) \biggm| = O \bigg( \frac{1}{h^{t+1}} \bigg) + o_p \bigg( \sqrt{\frac{\log n}{n h^{2t+3}}} \, \bigg),  \label{eq:A.2.7.1}
\end{align} 
where $f_c(v)$ denotes the probability density function of $\gamma^{\text{o} \top}_{c}Z_i$, $c \in \mathcal{C}$.
Under assumption {A2}, there exists a constant $c_1 > 0$ such that
\begin{align}
P\big(\sup_{i} \|X_i\| \geq c_1\sqrt{2p\log n}\,\big)
\leq \sum_{i=1}^n \sum_{j=1}^p P\big(|X_{ij}| \geq c_1\sqrt{2\log n}\,\big)
\leq  \frac{2p}{n} \longrightarrow 0 \text{ as } n \longrightarrow \infty. \label{eq:A.2.5.0}
\end{align}
By assumption {A4}, the triangle inequality, and the identity 
$\max_{i} \| \beta^*_{(1)i} - \beta^{\text{o}}_{(1)i} \|^2 = \max_{c} \| \gamma^{*}_{(1)c} - \gamma^{\text{o}}_{(1)c} \|^2$,
we obtain
\begin{align}
\max_{i} \big\| \bar{\beta}_{(1)i} - \beta^{\text{o}}_{(1)i} \big\|  \leq \alpha_1 \max_{i}\big\| \beta_{(1)i} - \beta^{\text{o}}_{(1)i} \big\| + (1-\alpha_1) \max_{i} \big\| \beta^*_{(1)i} - \beta^{\text{o}}_{(1)i} \big\| \leq \phi_n. \label{eq:A.2.5} 
\end{align} 
It follows from (\ref{eq:A.2.7.1})--(\ref{eq:A.2.5}) and the uniform boundedness of $K^{(t+2)} (u)$ that
\begin{align}
&\sup_{\{i, c: i\in\mathcal{G}^{\text{o}}_{c}\}}   \bigg\| \frac{1}{(n-1)h^{t+2}}  \sum_{\{j: j \neq i\}} Y_j^s K^{(t+1)} \bigg(\frac{\beta^{\text{o} \top}_j Z_j - \beta^{\text{o} \top}_i Z_i}{h}\bigg)\big[ Z^{\top}_j \big(\bar{\beta}_j - \beta^{\text{o}}_j\big) - Z^{\top}_i \big(\bar{\beta}_i - \beta^{\text{o} }_i\big) \big]Z_{*i}^{\otimes t} \bigg\| \nonumber \\
&\leq\bigg(\sup_{\{i, c: i\in\mathcal{G}^{\text{o}}_{c}\}}  \frac{1}{(n-1)h^{t+2}} \sum_{\{j: j \neq i\}}\biggm| Y_j^s K^{(t+1)} \bigg(\frac{\beta^{\text{o} \top}_j Z_j - \beta^{\text{o} \top}_i Z_i}{h}\bigg) \biggm|\bigg) \Big( 2 \phi_n \sup_{i} \|Z_i\|^{t+1}\Big) \nonumber \\
&=o_p \bigg( \frac{(\log n)^\frac{{t+1}}{2} \phi_n}{h^{t+1}} \bigg) \text{ and}\label{eq:A.2.c.0} \\
&\sup_{\{i, c: i\in\mathcal{G}^{\text{o}}_{c}\}}   \bigg\| \frac{1}{2(n-1)h^{t+3}}  \sum_{\{j: j \neq i\}} Y_j^s K^{(t+2)} \bigg(\frac{\bar{\beta}^{* \top}_j Z_j - \bar{\beta}^{* \top}_i Z_i}{h}   \bigg) \Big[ \Big( Z^{\top}_j \big(\bar{\beta}_j - \beta^{\text{o}}_j\big) \Big)^2 -2 \big(\bar{\beta}_j - \beta^{\text{o}}_j\big)^{\top} Z_j Z^{\top}_i \nonumber \\
&\hspace{0.55in}\big(\bar{\beta}_i - \beta^{\text{o} }_i\big) +\Big( Z^{\top}_i \big(\bar{\beta}_i - \beta^{\text{o}}_i\big) \Big)^2\Big]Z_{*i}^{\otimes t}\bigg\| \nonumber \\
&\leq     \frac{1}{2(n-1)h^{t+3}} \sum_{\{j: j \neq i\}} \big| Y_j^s \big| \bigg| K^{(t+2)} \bigg(\frac{\bar{\beta}^{* \top}_j Z_j - \bar{\beta}^{* \top}_i Z_i}{h}   \bigg) \bigg|\Big( 4 \phi_n^2 \sup_{i} \|Z_i\|^{t+2}\Big) \nonumber \\
&=o_p \bigg( \frac{(\log n)^\frac{{t+2}}{2} \phi_n^2}{h^{t+3}} \bigg). \label{eq:A.2.c.1}
\end{align}
Applying the triangle inequality to (\ref{eq:A.2.8.0}) and using (\ref{eq:A.2.7}), (\ref{eq:A.2.c.0}), and (\ref{eq:A.2.c.1}), we obtain
\begin{align}
& \sup_{\{i, c: i\in\mathcal{G}^{\text{o}}_{c}\}}   \bigg\|\frac{1}{(n-1)h^{t+1}} \sum_{\{j: j \neq i\}}  Y_j^s K^{(t)}\bigg(\frac{\bar{\beta}^{\top}_j Z_j - \bar{\beta}^{\top}_i Z_i}{h}\bigg) Z_{*i}^{\otimes t} -  \partial_v^t \big( M_{c,s,t}\big(Z_i, \gamma^{\text{o} \top}_c Z_i \big) f_{c}\big( \gamma^{\text{o} \top}_{c}Z_i \big) \big) \bigg\|\nonumber\\
& \leq  o_p \bigg( \sqrt{\frac{\log n}{n h^{2t+1}}}\, \bigg) + O \big( h^2 \big) + o_p \bigg( \frac{(\log n)^\frac{{t+1}}{2} \phi_n}{h^{t+1}} \bigg) + o_p \bigg( \frac{(\log n)^\frac{{t+2}}{2} \phi_n^2}{h^{t+3}} \bigg) = o_p(1), ~t,s=0,1. \label{eq:A.2.8.1}
\end{align}
Combining (\ref{eq:A.2.8.1}) with assumption {A1} and proceeding as in the proof of Lemma \ref{A:Lemma_2} with $t = 0, 1$ yields
\begin{align}
\sup_{\{i, c: i\in\mathcal{G}^{\text{o}}_{c}\}}
\Big\|\partial^{t}_{\beta_i}\widehat{m}^{-i}_h\big(\bar{\beta}^{\top}_iZ_i;\bar{\beta}\big)
      -m^{[t]}_{c}\big(Z_{i},\gamma^{\text{o}\top}_cZ_i\big)\Big\|  =o_p \bigg( \frac{(\log n)^\frac{{t+1}}{2} \phi_n}{h^{t+1}} \bigg)=o_{p}(1)~t=0,1. \label{eq:A.2.9}
\end{align}
Consequently, the triangle inequality, (\ref{eq:A.2.9}), and assumptions {A2}--{A4} imply
\begin{align}
 &|I_{12}| \leq 2 \phi_n \sum_{c=1}^k \sum_{\{i \in \mathcal{G}_{c}^{\text{o}}\}} \big\| \textsc{ps}_{ii} \big(\bar{\beta}\big) - \textsc{s}_i \big({\beta^{\text{o}}}\big) \big\| \nonumber \\
 & \leq 2 \phi_n \sum_{c=1}^k \sum_{\{i \in \mathcal{G}_{c}^{\text{o}}\}} \Big[ \big\| U_i^{\text{o}} \big( \partial_{\beta_i} \widehat{m}^{-i}_h \big(\bar{\beta}^{\top}_iZ_i; \bar{\beta} \big) - m^{[1]}_{c}\big(Z_{i},\gamma^{\text{o} \top}_{c}Z_i\big) \big) \big\| + \big| \widehat{m}^{-i}_h \big(\bar{\beta}^{\top}_iZ_i; \bar{\beta} \big) - m^{[0]}_{c}\big(Z_{i},\gamma^{\text{o} \top}_{c} Z_i\big) \big| \nonumber  \\
 & \hspace{0.2in}  \big( \big\| m^{[1]}_{c}(Z_i, \gamma^{\text{o} \top}_{c} Z_i)\big\| + \big\|\partial_{\beta_i} \widehat{m}^{-i}_h \big(\bar{\beta}^{\top}_iZ_i; \bar{\beta} \big) - m^{[1]}_{c}\big(Z_{i},\gamma^{\text{o} \top}_{c} Z_i\big) \big\|  \big)\Big] = o_p(n \, \phi_{n}). \label{eq:A.2.9.1}
\end{align}
Furthermore, using the alternative representation of $I_{11}$ in (\ref{eq:A.2.6}), $I_{11} = - \sum_{c = 1}^k \sum_{\{i,j \in \mathcal{G}_{c}^{\text{o}}, i < j\}}
\big(\textsc{s}_i\big({\beta^{\text{o}}}\big)$ $- \textsc{s}_j\big({\beta^{\text{o}}}\big)\big)^{\top}\big(\beta_j - \beta_i\big)/|\mathcal{G}_{c}^{\text{o}}|$, together with assumptions {A2} and {A3}, we obtain
\begin{align*}
 & P\bigg(|I_{11}| > n^{\alpha} \sum_{c = 1}^k \sum_{\{i,j \in \mathcal{G}_{c}^{\text{o}}, i < j\}} \frac{\big\|\beta_j - \beta_i\big\|}{|\mathcal{G}_{c}^{\text{o}}|} \bigg) \nonumber  \\
 & \leq P\Bigg(\sum_{c = 1}^k \sum_{\{i,j \in \mathcal{G}_{c}^{\text{o}}, i < j\}} \big\| \textsc{s}_i\big({\beta^{\text{o}}}\big) - \textsc{s}_j\big({\beta^{\text{o}}}\big) \big\| \frac{\big\|\beta_j - \beta_i\big\|}{|\mathcal{G}_{c}^{\text{o}}|}> n^{\alpha} \sum_{c = 1}^k \sum_{\{i,j \in \mathcal{G}_{c}^{\text{o}}, i < j\}} \frac{\big\|\beta_j - \beta_i\big\|}{|\mathcal{G}_{c}^{\text{o}}|} \Bigg)  \nonumber  \\
 & \leq 2 n^{-\alpha} \E\big[ \big\| \textsc{s}_1\big({\beta^{\text{o}}}\big)\big\| \big]  \longrightarrow 0 \text{ as } n  \longrightarrow \infty \text{ for some sufficiently small fixed } \alpha > 0. 
\end{align*}
It follows that, with probability converging to one,
\begin{align}
|I_{11}| \leq  n^{\alpha}\Bigg(\sum_{c = 1}^k \frac{1}{ |\mathcal{G}_{c}^{\text{o}}|}\sum_{\{ i,j \in \mathcal{G}_{c}^{\text{o}},i<j\}} \big\| \beta_{(1)i} - \beta_{(1)j} \big\|\Bigg). \label{eq:A.2.9.2}
\end{align}
By the uniform convergence result in (\ref{eq:A.2.8.1}) with $s = t = 0$, $I_{13}$ in (\ref{eq:A.2.6}) has the representation:
\begin{align} 
I_{13}  = \Bigg[\frac{-1}{(n-1)h^{2}}\sum_{j=1}^n \Bigg(\sum_{c=1}^k \sum_{\{i:i \in \mathcal{G}_c^{\text{o}}, i\neq j\}}
\frac{E_{ii} E_{ji}}
     { f_c(\gamma_c^{\text{o} \top} Z_i)} K^{(1)}\Bigg(\frac{{\beta}_j^{\text{o} \top}Z_j-{\beta}_i^{\text{o} \top}Z_i}{h}\Bigg)\Bigg) Z_{j}^{\top}\big(\beta_j - \beta^*_j\big)\Bigg](1+o_p(1)),
\label{eq:A.2.10}
\end{align}
where $E_{ji} = Y_j-\widehat{m}^{-i}_h \big(\bar{\beta}_i^{\top}Z_i;\bar{\beta}\big)
=U_j^{\text{o}}+\big(m(\beta_j^{\text{o}\top}Z_j)-m(\beta_i^{\text{o}\top}Z_i)\big)
 -\big(\widehat{m}^{-i}_h \big(\bar{\beta}_i^{\top}Z_i;\bar{\beta} \big)-m(\beta_i^{\text{o}\top}Z_i)\big)$, 
$i, j = 1, \dots, n$.
Using the argument for $\mathcal{F}_{t_1t_2}$ in the proof of Lemma {\ref{A:Lemma_3}}, we conclude that under assumptions {A1}--{A3} and {A5}, the following class is Euclidean:
\begin{align*}
\bigg\{\frac{U^{\text{o}}}{ f_{c_2}(\gamma_{c_2}^{\text{o} \top} Z)}
      \big(m(\gamma_{c_1}^{\text{o}\top}z)-m(\gamma_{c_2}^{\text{o}\top}Z)\big)^sK^{(1)}\bigg(\frac{{\gamma}_{c_1}^{\text{o} \top}z-{\gamma}_{c_2}^{\text{o} \top}Z}{h}\bigg): x \in \mathcal{X}_n, c_1, c_2 \in \mathcal{C} \bigg\}, ~s = 0, 1.
\end{align*}
By $\E\big[U^{\text{o}} \mid W\big] = 0$ and
\begin{align*}
\sup_{\{x \in \mathcal{X}_n, c_1, c_2 \in \mathcal{C}\} } \E\bigg[ \bigg(\frac{1}{h^2} U^{\text{o}} \big(m(\gamma_{c_1}^{\text{o}\top}z)-m(\gamma_{c_2}^{\text{o}\top}Z)\big)^{s} {K^{(1)}\bigg(\frac{{\gamma}_{c_1}^{\text{o} \top}z-{\gamma}_{c_2}^{\text{o} \top}Z}{h}\bigg)} \bigg)^2 \bigg] = O\bigg({\frac{1}{h^3}}\bigg),~ s = 0, 1,
\end{align*}
Theorem II.37 in \cite{pollard1984convergence} implies
\begin{align}
&\sup_{\{x \in \mathcal{X}_n, c_1, c_2 \in \mathcal{C}\} } \bigg|\frac{1}{(n-1)h^2} \sum_{\{i: i \in \mathcal{G}^{\text{o}}_{c_2}, i \neq j\}} \frac{U_i^{\text{o}}}{ f_{c_2}(\gamma_{c_2}^{\text{o} \top} Z_i)}\big(m(\gamma_{c_1}^{\text{o}\top}z)-m(\gamma_{c_2}^{\text{o}\top}Z_i)\big)^{s} {K^{(1)}\bigg(\frac{{\gamma}_{c_1}^{\text{o} \top}z-{\gamma}_{c_2}^{\text{o} \top}Z_i}{h}\bigg)}\bigg| \nonumber\\
&= o_p\bigg(\sqrt{\frac{\log n}{nh^3}}\,\bigg)=o_{p}(1), ~ s= 0, 1.
\label{eq:A.2.11}
\end{align}
It follows from (\ref{eq:A.2.9}) with $t=0$ that under assumptions {A1}--{A3} and {A5}, 
\begin{align} 
& \sup_{\{j,c\}}\Bigg|\frac{1}{(n-1)h^2} \sum_{\{i:i \in \mathcal{G}_c^{\text{o}}, i\neq j\}} 
\frac{\big(U^{\text{o}}_i\big)^{s_1}}{ f_c(\gamma_c^{\text{o} \top} Z_i)}
\big(m(\beta_j^{\text{o}\top}Z_j)-m(\beta_i^{\text{o}\top}Z_i)\big)^{s_2} 
\big(\widehat{m}^{-i}_h \big(\bar{\beta}_i^{\top}Z_i;\bar{\beta} \big)-m(\beta_i^{\text{o}\top}Z_i)\big)^{s_3}\nonumber\\
&\hspace{0.35in}{K^{(1)}\bigg(\frac{{\beta}_j^{\text{o} \top}Z_j-{\beta}_i^{\text{o} \top}Z_i}{h}\bigg)} \Bigg|\nonumber\\
&\leq \Bigg(\sup_{\{j,c\}}\Bigg|\frac{1}{(n-1)h^2} \sum_{\{i:i \in \mathcal{G}_c^{\text{o}}, i\neq j\}} 
\frac{\big| U^{\text{o}}_i\big|^{s_1}}{ f_c(\gamma_c^{\text{o} \top} Z_i)}
\big| m(\beta_j^{\text{o}\top}Z_j)-m(\beta_i^{\text{o}\top}Z_i)\big|^{s_2} 
{K^{(1)}\bigg(\frac{{\beta}_j^{\text{o} \top}Z_j-{\beta}_i^{\text{o} \top}Z_i}{h}\bigg)} \Bigg|\Bigg)\nonumber\\
&\hspace{0.25in} \bigg(\sup_{\{i, c: i\in\mathcal{G}^{\text{o}}_{c}\}} \Big|\widehat{m}^{-i}_h\big(\bar{\beta}^{\top}_iZ_i;\bar{\beta}\big)  -m\big({\beta}_i^{\text{o} \top}Z_i\big)\Big|\bigg)^{s_3}\nonumber\\
&=\bigg(O\bigg(\frac{1}{h}\bigg)+ o_p\bigg(\sqrt{\frac{\log n}{nh^3}}\,\bigg)\bigg) 
\bigg(o_p \bigg( \frac{(\log n)^\frac{{1}}{2} \phi_n}{h} \bigg)\bigg)^{s_3}
 = o_p(1), ~s_1,s_2=0,1,~s_3=1,2.
\label{eq:A.2.12}
\end{align}
Applying the triangular inequality to (\ref{eq:A.2.10}) and invoking (\ref{eq:A.2.11}), (\ref{eq:A.2.12}), and assumption {A4}, we obtain
\begin{align}
|I_{13}|& \leq  \Bigg(\sup_{j}\Bigg|\frac{1}{(n-1)h^{2}}\sum_{c=1}^k \sum_{\{i:i \in \mathcal{G}_c^{\text{o}}, i\neq j\}}
\frac{E_{ii} E_{ji}}
     { f_c(\gamma_c^{\text{o} \top} Z_i)} K^{(1)}\Bigg(\frac{{\beta}_j^{\text{o} \top}Z_j-{\beta}_i^{\text{o} \top}Z_i}{h}\Bigg)\Bigg|\Bigg)
 \bigg(\sum_{j=1}^n |U^{\text{o}}_j | \|Z_j\| \bigg)\nonumber\\
 &\hspace{0.2in} \Big(\max_{j}\big\| \beta_{j} - \beta^{*}_{j} \big\|\Big)(1+o_{p}(1))= o_p(n \, \phi_n). \label{eq:A.2.13.3}
\end{align}
Substituting (\ref{eq:A.2.9.1}), (\ref{eq:A.2.9.2}), and (\ref{eq:A.2.13.3}) into $I_{1}$ in (\ref{eq:A.2.6}) yields
\begin{align}
I_1 \geq  - n^{\alpha}\bigg(\sum_{c = 1}^k \frac{1}{ |\mathcal{G}_{c}^{\text{o}}|}\sum_{\{ i,j \in \mathcal{G}_{c}^{\text{o}},i<j\}} \big\| \beta_{(1)i} - \beta_{(1)j} \big\|\bigg) (1+o_p(1)).\label{eq:A.2.15}
\end{align}
Given $(\beta, \gamma) \in B_n \times \Gamma_n$, it follows from the triangle inequality that under assumption {A4}, $i\in\mathcal{G}_{c_{1}}^{\text{o}}$, and $c_{1}\neq c_{2}$, 
\begin{align*}
&\|\beta_{(1)i}-\gamma_{(1)c_{1}}\|_{1}\leq  \|\beta_{(1)i}-\beta^{\text{o}}_{(1)i}\|_{1}+\| \gamma_{(1)c_{1}} - \gamma^{\text{o}}_{(1)c_{1}}\|_{1}\leq 2\phi_{n} \text{ and}\\
&\|\beta_{(1)i}-\gamma_{(1)c_{2}}\|_{1} = \|\beta_{(1)i}-\beta^{\text{o}}_{(1)i}+ \gamma^{\text{o}}_{(1)c_{1}}- \gamma^{\text{o}}_{(1)c_{2}}+ \gamma^{\text{o}}_{(1)c_{2}}- \gamma_{(1)c_{2}}\|_{1}\nonumber\\
&\geq \| \gamma^{\text{o}}_{(1)c_{1}}- \gamma^{\text{o}}_{(1)c_{2}}\|_{1}- \|\beta_{(1)i}-\beta^{\text{o}}_{(1)i}\|_{1}-\| \gamma^{\text{o}}_{(1)c_{2}}- \gamma_{(1)c_{2}}\|_{1}\geq D_{\gamma} -2\phi_{n}>2\phi_{n}.
\end{align*}
Consequently,
$\min_{\{c\in\mathcal{C}\}}\|\beta_{(1)i}-\gamma_{(1)c}\|_{1} =\|\beta_{(1)i}-\gamma_{(1)c_{1}}\|_{1}$.
Using this fact and the norm inequality $\| \cdot \|_1 \geq \| \cdot \|$, we obtain
\begin{align}
I_{2}&=\lambda\sum^{k}_{c=1}\sum_{\{i\in\mathcal{G}_{c}^{\text{o}}\}} \|\beta_{(1)i}-\gamma_{(1)c}\|_{1}
=\lambda\sum^{k}_{c=1}\frac{1}{2 |\mathcal{G}_{c}^{\text{o}}|}\sum_{\{i,j\in\mathcal{G}_{c}^{\text{o}}\}} \big(\|\beta_{(1)i}-\gamma_{(1)c}\|_{1}+\|\beta_{(1)j}-\gamma_{(1)c}\|_{1}\big)\nonumber \\
&\geq  \lambda\sum^{k}_{c=1}\frac{1}{ |\mathcal{G}_{c}^{\text{o}}|}\sum_{\{i,j\in\mathcal{G}_{c}^{\text{o}},i<j\}} \|\beta_{(1)i}-\beta_{(1)j}\|_{1} \geq \lambda \sum^{k}_{c=1}\frac{1}{ |\mathcal{G}_{c}^{\text{o}}|}\sum_{\{i,j\in\mathcal{G}_{c}^{\text{o}},i<j\}} \|\beta_{(1)i}-\beta_{(1)j}\|.  \label{eq:A.2.16}
\end{align}

Substituting (\ref{eq:A.2.15}) and (\ref{eq:A.2.16}) into (\ref{eq:A.2.4}) yields
\begin{align}
&\textsc{pss}_{\textsc{sp}}\big(\beta,\gamma;\lambda\big) - \textsc{pss}_{\textsc{sp}}\big(\beta^{*},\gamma^{*};\lambda\big) \nonumber \\
& \geq  \Big(\lambda - n^{\alpha} \Big) \bigg(\sum_{c = 1}^k\frac{1}{ |\mathcal{G}_{c}^{\text{o}}|}\sum_{\{ i,j \in \mathcal{G}_{c}^{\text{o}},i<j\}} \| \beta_{(1)i} - \beta_{(1)j} \|\bigg) (1+o_p(1)).
 \label{eq:A.2.14}
\end{align}
The local oracle property in Theorem \ref{Thm3.1} follows directly from (\ref{eq:A.2.3}) and (\ref{eq:A.2.14}).

\section{Proof of Theorem \ref{Thm3.2}} 
\label{sec:A.4}

Define 
\begin{align*}
N_{sh} \big(v; \mathcal{G}, \gamma \big)=\frac{1}{n}\sum_{i = 1}^n Y_i^s \sum_{c = 1}^{k} I\big( i \in \mathcal{G}_{c}\big) K_{h}\big( \gamma^{\top}_{c} Z_i - v \big),~s=0,1.
\end{align*} 
Since $P\big(\widehat{\mathcal{G}}=\mathcal{G}^{\text{o}}\big) \longrightarrow 1$ as $n \longrightarrow \infty$, we have
\begin{align}
   n^{r}\sup_{\{i, c\}} \big| I\big(i\in \widehat{\mathcal{G}}_c\big)-I\big(i\in \mathcal{G}^{\text{o}}_c\big) \big|=o_{p}(1) \label{A.4.0}
\end{align}
for each positive integer $r$. By (\ref{A.4.0}), $\tilde{h} = O_p(n^{-1/5})$, the uniform boundedness of $K(u)$, and the triangle inequality, it follows that under assumptions {A2} and {A3},
\begin{align}
\hspace{-3.2in} \sup_{\{(w,\gamma_*) \in \mathcal{W}_n \times \Gamma\}}\big| N_{s \tilde{h}}\big(\gamma^{\top}w; \widehat{\mathcal{G}}, \gamma\big) - N_{s \tilde{h}}\big(\gamma^{\top}w; \mathcal{G}^{\text{o}}, \gamma\big) \big| \nonumber
\end{align}
\begin{align}
&= \sup_{\{(w,\gamma_*) \in \mathcal{W}_n \times \Gamma\}}\bigg | \frac{1}{n} \sum_{i = 1}^n Y_i^s \sum_{c = 1}^{k} \big( I\big(i\in\widehat{\mathcal{G}}_{c}\big)-I\big(i\in \mathcal{G}^{\text{o}}_{c} \big) \big) K_{\tilde{h}} \big( \gamma^{\top}_{c}Z_i - \gamma^{\top}w \big) \bigg|  \nonumber \\
& \leq \sup_{\{ i, c\}} \Big| I\big(i\in\widehat{\mathcal{G}}_{c}\big)-I\big(i\in\mathcal{G}^{\text{o}}_{c}\big)\Big|\,
\sup_{\{(w,\gamma_*) \in \mathcal{W}_n \times \Gamma\}}\bigg( \frac{1}{n}\sum_{i = 1}^n \big( \big|m(\gamma^{*\text{o} \top} Z_i) \big| + \big| U_i^{\text{o}}\big| \big)^s\sum^{k}_{c=1}K_{\tilde{h}}\big(\gamma^{\top}_{c}Z_{i}-\gamma^{\top}w\big) \bigg)\nonumber \\
& \leq o_{p}\Bigg(\sqrt{\frac{{{\log n}}}{n^{\frac{8}{5}}}}\,\Bigg),~s=0,1. \label{eq:A.4.0}
\end{align}
Under assumptions {A2}, {A3}, and {A5}, combining (\ref{eq:A.4.0}) with Lemma \ref{A:Lemma_1} (with $t = 0$) yields
\begin{align}
&\hspace{-0.25in}\sup_{\{(w,\gamma_*) \in \mathcal{W}_n \times \Gamma\}} \big|\widehat{m}_{\tilde{h}}\big(\gamma^{\top} w; \widehat{\mathcal{G}},\gamma\big)-\widehat{m}_{\tilde{h}}\big(\gamma^{\top} w; \mathcal{G}^{\text{o}},\gamma\big) \big |  \nonumber\\
& \hspace{-0.25in}\leq  \frac{ \sup_{\{(w,\gamma_*) \in \mathcal{W}_n \times \Gamma\}}\big| N_{0 \tilde{h}}\big(\gamma^{\top} w; \widehat{\mathcal{G}},\gamma\big) - N_{0 \tilde{h}}\big(\gamma^{\top} w; \mathcal{G}^{\text{o}}, \gamma\big) \big|  \sup_{\{(w,\gamma_*) \in \mathcal{W}_n \times \Gamma\}} \big| N_{1 \tilde{h}}\big(\gamma^{\top} w; \mathcal{G}^{\text{o}}, \gamma\big)\big|}{ \Big( \inf_{\{(w,\gamma_*) \in \mathcal{W}_n \times \Gamma\}} f_{\gamma}(\gamma^{\top} w) + o_p\Big( \sqrt{\frac{{\log n}}{n^{\frac{4}{5}}}} \,\Big) \Big)^2   }\nonumber
\end{align}
\begin{align}
&\hspace{-0.25in}~~+ \frac{ \sup_{\{(w,\gamma_*) \in \mathcal{W}_n \times \Gamma\}}\big| N_{1 \tilde{h}}\big(\gamma^{\top} w; \widehat{\mathcal{G}},\gamma\big) - N_{1 \tilde{h}}\big(\gamma^{\top} w; \mathcal{G}^{\text{o}}, \gamma\big) \big|  \sup_{\{(w,\gamma_*) \in \mathcal{W}_n \times \Gamma\}} \big| N_{0 \tilde{h}}\big(\gamma^{\top} w; \mathcal{G}^{\text{o}}, \gamma\big)\big|}{ \Big( \inf_{\{(w,\gamma_*) \in \mathcal{W}_n \times \Gamma\}} f_{\gamma}(\gamma^{\top} w) + o_p\Big( \sqrt{\frac{{\log n}}{n^{\frac{4}{5}}}} \,\Big) \Big)^2   }\nonumber
\end{align}
\begin{align}
&\hspace{-2.25in} = o_{p}\Bigg( \sqrt{\frac{\log n}{n^{\frac{8}{5}}}} \,\Bigg).\hspace{3in}\label{eq:A.4.2} 
\end{align} 
Arguments parallel to those used in the proof of Theorem \ref{Thm2.1} establish
\begin{align}
\| \widehat{\gamma} - \gamma^{\text{o}} \|=\left\{
 \begin{array}{ll}
 O_{p}\Big( \frac{1}{\sqrt{n}} \Big) & \mathcal{G}^{\text{o}} \in \mathcal{C}_{\mathcal{G}},\\ 
  o_{p}(1) & \text{ otherwise,}
  \end{array}\right.\label{eq:A.4.2.5.1}
\end{align}
where
\vspace{-0.05in}
\begin{align*}
\mathcal{C}_{\mathcal{G}}=\bigg\{ \mathcal{G}^{\text{o}}: \mathcal{G}_{c}^{*\text{o}} = \bigcup_{\{\ell:\mathcal{G}^{\text{o}}_{\ell} \subseteq \mathcal{G}_{c}^{*\text{o}},\gamma^{\text{o}}_{\ell}=\gamma^{*\text{o}}_{c}\}}\mathcal{G}^{\text{o}}_{\ell}~\forall c\in\mathcal{C}^{*} \bigg\}.
\end{align*}

The decomposition of $\textsc{pss}\big(\widehat{\gamma},\widehat{\mathcal{G}}\big)$ admits the following representation:
\begin{align}
\frac{2}{n} \textsc{pss}\big(\widehat{\gamma},\widehat{\mathcal{G}}\big) = &~ \frac{1}{n}\sum^{n}_{i=1}  (U^{\text{o}}_i)^2 + \frac{1}{n}\sum^{n}_{i=1} \Big( \widehat{m}^{-i}_{\tilde{h}}\big(\widehat{\gamma}^{\top}\widehat{W}_{i};\widehat{\mathcal{G}},\widehat{\gamma}\big) -m^{[0]}(W_i, \gamma^{\text{o}\top}W_{i}\big) \Big)^{2}   \nonumber\\
& - \frac{2}{n}\sum^{n}_{i=1} U^{\text{o}}_i \Big(\widehat{m}^{-i}_{\tilde{h}}\big(\widehat{\gamma}^{\top}\widehat{W}_{i};\widehat{\mathcal{G}},\widehat{\gamma}\big) -m^{[0]}(W_i, \gamma^{\text{o}\top}W_{i}\big)\Big) 
\stackrel{\triangle}{=}  I_{1}(k) + I_{2}(k) + I_{3}(k), \label{eq:A.4.5}
\end{align}
where $W_{i}=(I(i\in\mathcal{G}^{\text{o}}_{1}), \dots, I(i\in\mathcal{G}^{\text{o}}_{k}))^{\top} \otimes Z_{i}$ denotes the alternative representation of $W_{i}$ introduced in Section \ref{sec:sec3.1}, and $\widehat{W}_{i}=(I(i\in\widehat{\mathcal{G}}_{1}), \dots, I(i\in\widehat{\mathcal{G}}_{k}))^{\top} \otimes Z_{i}$ for $i=1,\dots, n$.
By $U^{\text{o}}_i = U^{*}_i$ when $\mathcal{G}^{\text{o}} \in \mathcal{C}_{\mathcal{G}}$, and by applying the law of large numbers to $I_{1}(k)$ otherwise, we obtain
\begin{align}
I_{1}(k) - I_{1}(k^{*}) = \left\{
\begin{array}{ll}
0
&\mathcal{G}^{\text{o}} \in \mathcal{C}_{\mathcal{G}},\\ 
b^{2}(k) (1+o_p(1))
& \text{otherwise,}
\end{array}\right. \label{eq:A.4.6}
\end{align}
where $b^{2}(k)= \E\big[ \big(m^{[0]}\big(W,\gamma^{\text{o}\top}W\big)-m\big(\gamma^{*\text{o}\top}W^{*}\big)\big)^{2}\big]>0$ when $\mathcal{G}^{\text{o}} \notin \mathcal{C}_{\mathcal{G}}$.
Straightforward algebra yields 
\vspace{-0.05in}
\begin{align}
\hspace{-1.5in}&\widehat{m}^{-i}_{\tilde{h}}\big(\widehat{\gamma}^{\top}\widehat{W}_{i};\widehat{\mathcal{G}},\widehat{\gamma}\big)-m^{[0]}(W_i, \gamma^{\text{o}\top}W_{i}\big)  \nonumber\\
&= \big( \widehat{m}^{-i}_{\tilde{h}}\big(\widehat{\gamma}^{\top}\widehat{W}_{i};\widehat{\mathcal{G}},\widehat{\gamma}\big)-\widehat{m}^{-i}_{\tilde{h}}\big(\widehat{\gamma}^{\top}W_{i};\mathcal{G}^{\text{o}},\widehat{\gamma}\big) \big) + \big(\widehat{m}^{-i}_{\tilde{h}}\big(\widehat{\gamma}^{\top}W_{i};\mathcal{G}^{\text{o}},\widehat{\gamma}\big) - m^{[0]}(W_i, \widehat{\gamma}^{\top}W_{i}\big) \big) \nonumber\\
& ~~~ + \big(  m^{[0]}(W_i, \widehat{\gamma}^{\top}W_{i}\big) - m^{[0]}(W_i, {\gamma}^{ \text{o} \top}W_{i}\big) \big)
\stackrel{\triangle}{=} I\!I_{1i} + I\!I_{2i} + I\!I_{3i}, ~ i = 1, \dots, n.\label{eq:A.4.7.0}
\end{align}
By (\ref{A.4.0}) and Lemma \ref{A:Lemma_2} with $t = 0$, it follows that
\begin{align}
\sup_i |I\!I_{1i}| \leq & ~ k \bigg( \sup_{\{i, c\}} \big| I\big(i\in \widehat{\mathcal{G}}_c\big)-I\big(i\in \mathcal{G}^{\text{o}}_c\big) \big|  \bigg) \bigg(  \sup_{\{(w, \gamma_*) \in \mathcal{W}_n \times \Gamma\}} \Big| \widehat{m}_{{h}}\big(\gamma^{\top}w;\mathcal{G}^{\text{o}},{\gamma}\big) - m^{[0]}(w,\gamma^{\top}w)\Big| \nonumber\\
&~+ \sup_{\{(w, \gamma_*) \in \mathcal{W}_n \times \Gamma\}} \Big| m^{[0]}(w,\gamma^{\top}w)\Big| \bigg)
=o_p\Big(\frac{1}{n}\Big)\text{ and} \label{eq:A.4.3.0.1}\\
\sup_i |I\!I_{2i}| \leq & ~ \sup_{\{(w, \gamma_*) \in \mathcal{W}_n \times \Gamma\}} \Big| \widehat{m}_{{h}}\big(\gamma^{\top}w;\mathcal{G}^{\text{o}},{\gamma}\big) - m^{[0]}(w,\gamma^{\top}w)\Big| = o_p \Bigg( \sqrt{\frac{\log n}{n^{\frac{4}{5}} }} \,\Bigg). \label{eq:A.4.3.0.2}
\end{align}
For each $i=1,\dots,n$, first- and second-order Taylor expansions of $m^{[0]}(w,\widehat{\gamma}^{\top}w)$ around $\widehat{\gamma}=\gamma^{\text{o}}$ yield
\begin{align}
I\!I_{3i} =&~ \big(\partial_{v} m^{[0]}(W_i,\bar{\gamma}^{\top}_{1}W_i)\big) \big(\widehat{\gamma} - \gamma^{\text{o}}\big)^{\top} W_i \nonumber\\
=&~ \partial_{v} m^{[0]}\big(W_i,\gamma^{\text{o}\top}W_i\big) \big(\widehat{\gamma} - \gamma^{\text{o}}\big)^{\top} W_i + \frac{1}{2}\big(\partial^{2}_{v} m^{[0]}(W_i,\bar{\gamma}^{\top}W_i)\big)\Big(\big(\widehat{\gamma} - \gamma^{\text{o}}\big)^{\top} W_i\Big)^{2}, \label{eq:A.4.3.0}
\end{align}
where $\bar{\gamma}_{t}$ lies on the line segment between $\widehat{\gamma}$ and $\gamma^{\text{o}}$, $t=0,1$.
When $\mathcal{G}^{\text{o}} \in \mathcal{C}_{\mathcal{G}}$, we have $\E[U^{\text{o}} W$ $\partial_{v} m^{[0]}(W,\gamma^{\text{o}\top}W) ] = 0$. Moreover, under assumptions {A2} and {A3}, $\E[\| (U^{\text{o}}\partial_{v} m^{[0]}(W,\gamma^{\text{o}\top} W)^{2}W^{\otimes 2})\| ]<\infty$ and $\E[|U^{\text{o}}|^{s}\|W^{\otimes t}\|]<\infty$, $s=0,1$,  $t=1,2$. 
Consequently, by the central limit theorem and the law of large numbers,
\begin{align}
\frac{1}{n}\sum^{n}_{i=1}U^{\text{o}}_i \big(\partial_{v} m^{[0]}\big(W_i,\gamma^{\text{o}\top}W_i\big)\big)W_{i}=O_{p}\Big(\frac{1}{\sqrt{n}}\Big)\text{ and } 
\frac{1}{n}\sum^{n}_{i=1} |U^{\text{o}}_{i}|^{s}\|W^{\otimes t}_{i}\|=O_{p}(1),~ s=0,1,~t=1,2. \label{eq:A.4.3.1_0}
\end{align}
It follows from (\ref{eq:A.4.2.5.1}), the second-order Taylor expansion in (\ref{eq:A.4.3.0}), and (\ref{eq:A.4.3.1_0}) that 
\begin{align*}
\Bigm|\frac{1}{n}\sum^{n}_{i=1} U^{\text{o}}_i I\!I_{3i} \Bigm|
\leq &~ \big\|\widehat{\gamma} - \gamma^{\text{o}}\big\| \, \bigg\| \frac{1}{n} \sum_{i = 1}^n U^{\text{o}}_i \big(\partial_{v} m^{[0]}\big(W_i,\gamma^{\text{o}\top}W_i\big)\big) W_i \bigg\| \\
&~ + \frac{1}{2} \bigg(\sup_{\{(w, \gamma_*) \in \mathcal{W}_n \times \Gamma\}} \bigm| \partial^{2}_{v} m^{[0]}(w,{\gamma}^{\top}w) \bigm| \bigg) \big\|\widehat{\gamma} - \gamma^{\text{o}}\big\|^{2} \bigg( \frac{1}{n} \sum_{i = 1}^n |U^{\text{o}}_i|\, \|W_i\|^{2} \bigg) \\
=&~ O_p\Big(\frac{1}{\sqrt{n}}\Big) O_p\Big(\frac{1}{\sqrt{n}}\Big) + O_p\Big(\frac{1}{n}\Big) = O_p\Big(\frac{1}{n}\Big).
\end{align*}
When $\mathcal{G}^{\text{o}} \notin \mathcal{C}_{\mathcal{G}}$, the first-order Taylor expansion in (\ref{eq:A.4.3.0}), (\ref{eq:A.4.3.1_0}), and assumption {A3} ensure that
\begin{align*}
\Bigm|\frac{1}{n}\sum^{n}_{i=1} U^{\text{o}}_i I\!I_{3i} \Bigm|
\leq \bigg(\sup_{\{(w, \gamma_*) \in \mathcal{W}_n \times \Gamma\}} \bigm| \partial_{v} m^{[0]}(w,{\gamma}^{\top}w) \bigm| \bigg) \big\|\widehat{\gamma} - \gamma^{\text{o}}\big\| \bigg( \frac{1}{n} \sum_{i = 1}^n|U^{\text{o}}_i| \| W_i \| \bigg)
= O_{p}\big( \| \widehat{\gamma} - \gamma^{\text{o}} \| \big).
\end{align*}
Hence,
\begin{align}
\Bigm|\frac{1}{n}\sum^{n}_{i=1}  U^{\text{o}}_i I\!I_{3i} \Bigm|
=\left\{
 \begin{array}{ll}
 O_{p}\Big( \frac{1}{n} \Big) & \mathcal{G}^{\text{o}} \in \mathcal{C}_{\mathcal{G}},\\
  O_{p}\big( \| \widehat{\gamma} - \gamma^{\text{o}} \| \big) & \text{ otherwise.}
  \end{array}\right. \label{eq:A.4.3.1}
\end{align}
Moreover, by (\ref{eq:A.4.2.5.1}),  the first-order Taylor expansion in (\ref{eq:A.4.3.0}), and (\ref{eq:A.4.3.1_0}),
\begin{align}
 \frac{1}{n} \sum_{i = 1}^n I\!I^{2}_{3i}
& ~ \leq \bigg(\sup_{\{(w, \gamma_*) \in \mathcal{W}_n \times \Gamma\}} \bigm| \partial_{v} m^{[0]}(w,{\gamma}^{\top}w) \bigm| \bigg)^2 \big(\widehat{\gamma} - \gamma^{\text{o}}\big)^{\top} \bigg( \frac{1}{n} \sum_{i = 1}^n W_i^{\otimes 2} \bigg) \big(\widehat{\gamma} - \gamma^{\text{o}}\big) \nonumber\\
&~ =\left\{
 \begin{array}{ll}
 O_{p}\Big( \frac{1}{{n}} \Big) & \mathcal{G}^{\text{o}} \in \mathcal{C}_{\mathcal{G}},\\
  O_{p}\big( \| \widehat{\gamma} - \gamma^{\text{o}} \|^2 \big) & \text{ otherwise.}
  \end{array}\right. \label{eq:A.4.3}
\end{align}

Combining $I_2(k)\leq O(  (\sup_i |I\!I_{1i}| )^{2}+ (\sup_i |I\!I_{2i}|)^{2} + \sum^{n}_{i=1}  I\!I^{2}_{3i}/n)$ with (\ref{eq:A.4.3.0.1}), (\ref{eq:A.4.3.0.2}), and (\ref{eq:A.4.3}) yields
\begin{align}
I_2(k) \leq \left\{
\begin{array}{ll}
o_p\Big( {\frac{{\log n}}{n^{\frac{4}{5}}}} \,\Big)
& \mathcal{G}^{\text{o}} \in \mathcal{C}_{\mathcal{G}}, \\ 
o_p\Big( {\frac{{\log n}}{n^{\frac{4}{5}}}} \,\Big) + O_{p}\big( \| \widehat{\gamma} - \gamma^{\text{o}} \|^2 \big)
& \text{otherwise.}
\end{array}\right.\label{eq:A.4.4}
\end{align}
By $\sum_{i=1}^n |U^{\text{o}}_i| / n \stackrel{p}{\longrightarrow} \E[|U^{\text{o}}|]$, (\ref{eq:A.4.3.0.1}), and assumption {A2},
\begin{align}
\Bigm|\frac{1}{n} \sum_{i=1}^n U^{\text{o}}_i I\!I_{1i}\Bigm| \leq \Big(\frac{1}{n} \sum_{i=1}^n |U^{\text{o}}_i| \Big) \sup_i |I\!I_{1i}| = o_p\Big(\frac{1}{n}\Big). \label{eq:A.4.4.1}
\end{align}
It follows from Lemma \ref{A:Lemma_2} with $t = 0$ that
\begin{align*}
\frac{1}{n} \sum_{i=1}^n U^{\text{o}}_i I\!I_{2i} =&~ \frac{1}{n(n-1)} \sum_{i \neq j} U^{\text{o}}_i \xi^{[0]}_j \big( W_i, \widehat{\gamma}\big) + \frac{1}{n} \sum_{i=1}^n U^{\text{o}}_i r^{[0]}\big( W_i, \widehat{\gamma}\big),
\end{align*}
with the second term satisfying
\begin{align*}
\Bigm|\frac{1}{n}\sum_{i=1}^n U^{\text{o}}_i r^{[0]}\big( W_i, \widehat{\gamma}\big) \Bigm| \leq \bigg(\frac{1}{n}\sum_{i=1}^n |U^{\text{o}}_i| \bigg) \sup_{\{(w, \gamma_*) \in \mathcal{W}_n \times \Gamma\}} \bigm|r^{[0]}( w, {\gamma})\bigm| = o_p\bigg(\frac{\log n }{ n^{\frac{4}{5}}}\bigg).
\end{align*}
Hence,
\begin{align}
\Bigm|\frac{1}{n} \sum_{i=1}^n U^{\text{o}}_i I\!I_{2i}\Bigm| \leq \Bigm|\frac{1}{n(n-1)} \sum_{i \neq j} U^{\text{o}}_i \xi^{[0]}_j \big( W_i, \widehat{\gamma}\big)\Bigm| +o_p\bigg(\frac{\log n }{ n^{\frac{4}{5}}}\bigg). \label{eq:A.4.8}
\end{align}
Using the $\sqrt{n}$-consistency of $\widehat{\gamma}_{*}$ to $\gamma^{\text{o}}_{*}$ in (\ref{eq:A.4.2.5.1}), and applying the arguments in (\ref{eq:thm1_9}) and (\ref{eq:thm1_9.5}) to the first term on the right-hand side of (\ref{eq:A.4.8}) when $\mathcal{G}^{\text{o}} \in \mathcal{C}_{\mathcal{G}}$, we obtain 
\begin{align}
\Bigm|\frac{1}{n} \sum_{i=1}^n U^{\text{o}}_i I\!I_{2i} \Bigm|\leq O_{p}\bigg(\frac{1}{n^{\frac{9}{10}}}\bigg)+o_p\bigg(\frac{\log n }{ n^{\frac{4}{5}}}\bigg)=o_p\bigg(\frac{\log n }{ n^{\frac{4}{5}}}\bigg). \label{eq:A.4.8_1}
\end{align}
Combining $| I_3(k) |\leq O( |\sum^{n}_{i=1}  U^{\text{o}}_i I\!I_{1i} /n|+|\sum^{n}_{i=1}  U^{\text{o}}_i I\!I_{2i} /n| +|\sum^{n}_{i=1}  U^{\text{o}}_i I\!I_{3i} /n|)$ with (\ref{eq:A.4.3.1}), (\ref{eq:A.4.4.1}), and (\ref{eq:A.4.8_1})  yields
\begin{align}
I_{3}(k)= o_p\bigg( \frac{{\log n}}{n^{\frac{4}{5}}} \bigg) \text{ if } \mathcal{G}^{\text{o}} \in \mathcal{C}_{\mathcal{G}}.
\label{eq:A.4.9} 
\end{align}
When $\mathcal{G}^{\text{o}} \notin \mathcal{C}_{\mathcal{G}}$, $U^{\text{o}} = U^{*} - (\widehat{m}_{\tilde{h}}\big(\widehat{\gamma}^{\top}W; \mathcal{G}^{\text{o}},\widehat{\gamma}\big)-m^{[0]}(W,\gamma^{\text{o}\top}W))$. Under assumption {A2}, the law of large number implies $\sum_{i=1}^n |U_i^{*}| / n\stackrel{p}{\longrightarrow} \E[|U^{*}|]$. Combining this result with (\ref{eq:A.4.3.1}) yields
\begin{align}
|I_{3}(k)| = o_p\bigg( \frac{{\log n}}{n^{\frac{4}{5}}} \bigg) + O_{p}\big( \| \widehat{\gamma} - \gamma^{\text{o}} \| \big) \text{ if } \mathcal{G}^{\text{o}} \notin \mathcal{C}_{\mathcal{G}}. \label{eq:A.3.10}
\end{align}
Using $\sum_{i=1}^n (Y - \bar{Y})^2/n \stackrel{p}{\longrightarrow} \var(Y)$ and substituting (\ref{eq:A.4.6}), (\ref{eq:A.4.4}), (\ref{eq:A.4.9}), and (\ref{eq:A.3.10}) into (\ref{eq:A.4.5}), we obtain
\begin{align}
\text{SPIC}_0(k)-\text{SPIC}_0(k^{*}) = \left\{
\begin{array}{ll}
 \frac{1}{\var(Y)} \max\Big\{ b^{2}(k),(k-k^{*})\frac{\log n }{n^{\frac{4}{5}}}\Big\}(1+o_{p}(1))
& k > k^{*}, \\
 \frac{b^{2}(k)}{\var(Y)} (1+o_{p}(1))
&k <k^*.
\end{array}\right. \label{eq:A.3.11}
\end{align}
The consistency of $\widehat{k}^{*}$ to $k^{*}$ follows directly from (\ref{eq:A.3.11}).

\section{Proof of Theorem \ref{Thm4.2}}  
\label{sec:A.5}

By assumption {C3}, the triangle inequality, Taylor's theorem, and Lemma \ref{A:Lemma_4} with $t=0$, 
\begin{align}
&\sup_{\gamma_*} \bigg|\frac{1}{n}p{\ell}(\gamma)-\frac{1}{n}\sum^{n}_{i=1}\log \big(f_0(U_{\gamma i};\gamma)\big)\bigg|\nonumber\\
&\leq \sup_{\gamma_*} \bigg|\frac{1}{n}p{\ell}(\gamma)-\frac{1}{n}\sum^{n}_{i=1}  \log \big(\breve{f}_{{\xi}} \big(\widehat{U}_{\gamma i};\gamma\big) \big)\bigg| +\sup_{\gamma_*}\bigg|\frac{1}{n}\sum^{n}_{i=1} \Big[ \log \big(\breve{f}_{{\xi}} \big(\widehat{U}_{\gamma i};\gamma\big) \big)-\log \big(f_0(U_{\gamma i};\gamma)\big)\Big]\bigg| \nonumber\\
&\leq \sup_{\gamma_*} \bigg|\frac{1}{n}p{\ell}(\gamma)-\frac{1}{n}\sum^{n}_{i=1}  \log \big(\breve{f}_{{\xi}} \big(\widehat{U}_{\gamma i};\gamma\big) \big)\bigg| +\frac{o_{p}(1)}{f_{3}+o_{p}(1)}. \label{eq:A.5.1}
\end{align}
Using arguments analogous to those in (\ref{eq:A.4.0}) and (\ref{eq:A.4.2}), it follows that under assumptions {A1}--{A3} and {A5}, 
\begin{align}
\sup_{\{i, \gamma_*\}} \bigg|\sum^{k}_{c=1}I\big(i \in \widehat{\mathcal{G}}_{c}\big) \partial_{\gamma_*}^t \widehat{m}_{{h}}\big(\gamma^{\top}_{c}Z_i; \widehat{\mathcal{G}}, \gamma\big) - \sum^{k}_{c=1}I\big(i \in \mathcal{G}^{\text{o}}_{c}\big) \partial_{\gamma_*}^t\widehat{m}_{{h}}\big(\gamma^{\top}_{c}Z_i; {\mathcal{G}^{\text{o}}}, \gamma\big) \bigg| = o_p\bigg(\frac{1}{n^2}\bigg)~t=0,1,2. \label{eq:A.5.1.0}
\end{align}
Combining (\ref{eq:A.5.1.0}) with the Lipschitz continuity of the kernel function and assumptions {A1} and {C1}, we obtain
\begin{align}
& \sup_{\{i, \gamma_*\}} \big|\partial_{\gamma_*}^t \widetilde{f}_{\xi}\big(\widetilde{U}_{\gamma i};\gamma\big) - \partial_{\gamma_*}^t \breve{f}_{\xi}\big(\widehat{U}_{\gamma i};\gamma \big)\big|  \leq \frac{1}{\xi} \sup_{\{i,j, \gamma_*\}}\bigg| \partial_{\gamma_*}^t K\bigg(  \frac{\widetilde{U}_{\gamma j} - \widetilde{U}_{\gamma i}}{\xi}\bigg) - \partial_{\gamma_*}^t K\bigg( \frac{\widehat{U}_{\gamma j} - \widehat{U}_{\gamma i}}{\xi}  \bigg) \bigg| \nonumber\\
& = o_p\bigg( \frac{1}{n^2 \xi^{t+2}}\bigg) = o_p(1), ~t=0,1,2, \hspace{1in}\label{eq:A.5.1.5}
\end{align}
where $\widetilde{U}_{\gamma i} = Y_i -\sum^{k}_{c=1}I\big(i \in \widehat{\mathcal{G}}_{c}\big) \widehat{m}_{\hat{h}}\big(\gamma^{\top}_{c}Z_i; \widehat{\mathcal{G}},  \gamma\,\big)$, $i = 1, \dots, n$.
Hence,
\begin{align}
    & \sup_{\{\gamma_* \in \Gamma\}} \bigg|\frac{1}{n}p{\ell}(\gamma)-\frac{1}{n}\sum^{n}_{i=1}  \log \big(\breve{f}_{{\xi}} \big(\widehat{U}_{\gamma i};\gamma \big) \big)\bigg|=o_p(1). \label{eq:A.5.2}   
\end{align}
Substituting (\ref{eq:A.5.2}) into (\ref{eq:A.5.1}) yields
\begin{align}
    \sup_{\{\gamma_* \in \Gamma\}} \bigg|\frac{1}{n}p{\ell}(\gamma)-\frac{1}{n}\sum^{n}_{i=1}\log \big(f_0(U_{\gamma i};\gamma)\big)\bigg|=o_p(1).\label{eq:A.5.4}
\end{align} 
Since $f_0(u;\gamma)$ is of bounded variation on $\mathcal{U}_n\times \Gamma$ by assumption {C2}, Lemma 22 in \cite{nolan1987u} implies that the class
\begin{align*}
    \big\{f_0(u;\gamma): u \in \mathcal{U}_{\gamma n}, \gamma_* \in \Gamma \big\}
\end{align*} 
is Euclidean. It follows from Corollary 4 in \cite{sherman1994maximal} that
\vspace{-0.05in}
\begin{align} 
    \sup_{\{\gamma_* \in \Gamma\}} \bigg|   \frac{1}{n}\sum^{n}_{i=1}\log \big(f_0(U_{\gamma};\gamma)\big)-\E\big[\log( f_0(U_{\gamma};\gamma))\big]  \bigg|=O_p\Big(\frac{1}{\sqrt{n}}\Big).\label{eq:A.5.5}
\end{align}
Combining (\ref{eq:A.5.4}) and (\ref{eq:A.5.5}) yields
\begin{align}
    \sup_{\{\gamma_* \in \Gamma\}} \bigg|   \frac{1}{n} p\ell(\gamma)-\E\big[\log( f_0(U_{\gamma};\gamma))\big]  \bigg|=o_p(1). \label{eq:A.5.6}
\end{align}
Let $f_{W}(w)$ denote the probability density function of $W$. By Jensen's inequality and a change of variables,
\begin{align}
&\E\big[\log(f_0(U_{\gamma};\gamma))\big]-\E\big[\log( f_0(U^{\text{o}};\gamma^{\text{o}}))\big]  \nonumber\\
    &\leq  \log\E\bigg[\frac{f_0(U_{\gamma};\gamma)}{f_0(U^{\text{o}};\gamma^{\text{o}})}\bigg] = \log  \int_{\mathcal{Y}}\int_{\mathcal{W}} f_0\big(y - m^{[0]}(w, \gamma^{\top}w); \gamma\big) f_W(w)dwdy \nonumber\\
 &\leq \log \int^{\infty}_{ -\infty} f_0(u; \gamma)du \int_{\mathcal{W}} f_W(w)dw = \log 1=0, \label{eq:A.5.7}
  \end{align} 
with strict inequality for $\gamma \neq \gamma^{\text{o}}$. Hence, under assumption {C4}, $\gamma^{\text{o}}$ uniquely maximizes $\E[\log( f_0(U_{\gamma};$ $\gamma))]$.
Combining (\ref{eq:A.5.6}) and (\ref{eq:A.5.7}) yields $p\ell(\widetilde{\gamma})/n - p\ell(\gamma^{\text{o}}) /n = \E\big[\log(f_0(U_{\widetilde{\gamma}};\widetilde{\gamma}))\big]-\E\big[\log( f_0(U^{\text{o}};\gamma^{\text{o}}))\big]$ $+ o_p(1)$, where $\E\big[\log(f_0(U_{\widetilde{\gamma}};\widetilde{\gamma}))\big]-\E\big[\log( f_0(U^{\text{o}};\gamma^{\text{o}}))\big] < 0$.
On the other hand, by the definition of $\widetilde{\gamma}$ as a maximizer of $p\ell(\gamma)$, we have $p\ell(\widetilde{\gamma}) \geq p\ell(\gamma^{\text{o}})$. Therefore, $\E\big[\log(f_0(U_{\widetilde{\gamma}};\widetilde{\gamma}))\big]-\E\big[\log( f_0(U^{\text{o}};\gamma^{\text{o}}))\big] = o_p(1)$.
The consistency of $\widetilde{\gamma}$ to $\gamma^{\text{o}}$ then follows from the uniqueness of $\gamma^{\text{o}}$ as the maximizer of $\E\big[\log(f_0(U_{\gamma};$ $\gamma))\big]$ over $\Gamma$ and the continuity of $\E\big[\log(f_0(U_{\gamma};$ $\gamma))\big]$ on $\Gamma$.

Define $pS(\gamma)=\partial_{\gamma_*}p\ell(\gamma)$ and $pI(\gamma)= - \partial^{2}_{\gamma_*}p\ell(\gamma)$.
For an arbitrary fixed vector $v \in \mathbb{R}^{k(p+1)-2}$, a first-order Taylor expansion of $v^{\top}pS(\widetilde{\gamma})/\sqrt{n}$ around $\gamma^{\text{o}}_*$ yields
\begin{align}
\frac{1}{\sqrt{n}}v^{\top}pS(\widetilde{\gamma})=\frac{1}{\sqrt{n}}v^{\top}pS(\gamma^{\text{o}}) - \frac{1}{\sqrt{n}}v^{\top}pI(\Bar{\gamma})(\widetilde{\gamma}_*-\gamma^{\text{o}}_*), \label{eq:A.5.8}
\end{align}
where $\Bar{\gamma}$ lies on the line segment between $\widetilde{\gamma}$ and $\gamma^{\text{o}}$.
By straightforward algebra, $pS(\gamma^{\text{o}})/\sqrt{n}$ in (\ref{eq:A.5.8}) is rewritten as  
\begin{align}
\frac{1}{\sqrt{n}}pS(\gamma^{\text{o}}) = & ~ \frac{1}{\sqrt{n}}\sum^{n}_{i=1} \frac{f^{[1]}(W_i, U^{\text{o}}_{i};\gamma^{\text{o}})}{f_0(U^{\text{o}}_{i};\gamma^{\text{o}})} + \Bigg(\frac{1}{\sqrt{n}}pS(\gamma^{\text{o}}) - \frac{1}{\sqrt{n}}\sum^{n}_{i=1} \frac{\partial_{\gamma_*}\breve{f}_{\xi}\big(\widehat{U}_{\gamma^{\text{o}} i};\gamma^{\text{o}}\big)}{\breve{f}_{\xi}\big(\widehat{U}_{\gamma^{\text{o}} i};\gamma^{\text{o}}\big)} \Bigg) \nonumber\\
& ~+ \frac{1}{\sqrt{n}}\sum_{i=1}^n  \sum_{t=0}^{1}  \frac{(-1)^{1-t}f^{[1-t]}(W_i, U^{\text{o}}_{i};\gamma^{\text{o}})\Big(\partial_{\gamma_*}^{t}\breve{f}_{\xi}\big(\widehat{U}_{\gamma^{\text{o}} i};\gamma^{\text{o}}\big)-f^{[t]}(W_i, U^{\text{o}}_{i};\gamma^{\text{o}})\Big)} {f_0(U^{\text{o}}_{i};\gamma^{\text{o}})\breve{f}_{\xi}\big(\widehat{U}_{\gamma^{\text{o}} i};\gamma^{\text{o}}\big)}\nonumber\\
\stackrel{\triangle}{=}&~ \frac{1}{\sqrt{n}}S(\gamma^{\text{o}})+I\!I+I\!I\!I.\hspace{2.75in} \label{eq:A.5.9.0}
\end{align} 
By the triangle inequality, (\ref{eq:A.5.1.5}), and Lemma \ref{A:Lemma_4} with $t = 0, 1$, it follows that under assumptions {A1}, {C1}, and {C3},
\begin{align}
\bigm| I\!I \bigm|  \leq &~ \frac{\sum_{t=0}^1 \sup_i \bigm|\partial_{\gamma_*}^t \breve{f}_{\xi}\big(\widehat{U}_{\gamma^{\text{o}} i};\gamma^{\text{o}}\big)\bigm| \sup_i \bigm|\partial_{\gamma_*}^{1-t} \widetilde{f}_{\xi}\big(\widetilde{U}_{\gamma^{\text{o}} i};\gamma^{\text{o}}\big) - \partial_{\gamma_*}^{1-t} \breve{f}_{\xi}\big(\widehat{U}_{\gamma^{\text{o}} i};\gamma^{\text{o}}\big)\bigm| } {\inf_i \bigm| \widetilde{f}_{\xi}\big(\widetilde{U}_{\gamma^{\text{o}} i};\gamma^{\text{o}}\big) \bigm| \inf_i \bigm| \breve{f}_{\xi}\big(\widehat{U}_{\gamma^{\text{o}} i};\gamma^{\text{o}}\big) \bigm|} \nonumber \\
\leq &~ \frac{\sum_{t=0}^1 \big( \sup_i \bigm| f^{[t]}(W_i, U^{\text{o}}_{i};\gamma^{\text{o}}) \bigm| +o_p(1) \big) o_p(1)}{(f_3 + o_p(1))^2} = o_p(1)\text{ and}\label{eq:A.5.9.1}\\
I\!I\!I =&~ \frac{1}{\sqrt{n}}\sum_{i=1}^n  \sum_{t=0}^{1}  \frac{(-1)^{1-t}f^{[1-t]}(W_i, U^{\text{o}}_{i};\gamma^{\text{o}})\Big(\partial_{\gamma_*}^{t}\breve{f}_{\xi}\big(\widehat{U}_{\gamma^{\text{o}} i};\gamma^{\text{o}}\big)-f^{[t]}(W_i, U^{\text{o}}_{i};\gamma^{\text{o}})\Big)} {f_0(U^{\text{o}}_{i};\gamma^{\text{o}})\breve{f}_{\xi}\big(\widehat{U}_{\gamma^{\text{o}} i};\gamma^{\text{o}}\big)}. \label{eq:A.5.9}
\end{align}  
Moreover, under assumptions {A1} and {C1},
\begin{align}    
I\!I\!I
 =&~ \sqrt{n}  \sum_{t=0}^{1}(-1)^{1-t}\bigg(\frac{1}{n(n-1)}\sum_{i\neq j}\frac{f^{[1-t]}(W_i, U^{\text{o}}_{i};\gamma^{\text{o}})}{(f_0(U^{\text{o}}_{i};\gamma^{\text{o}}))^2}\varphi^{[t]}_j(W_{i}, U^{\text{o}}_{i};\gamma^{\text{o}})\bigg) \nonumber \\
= &~\sqrt{n}\sum^{1}_{t=0}(-1)^{1-t}\bigg(\frac{1}{n(n-1)}\sum_{i\neq j}U_{ij}^{[t]}+\frac{1}{n}\sum_{i=1}^n \bar{U}_i^{[t]}\bigg), \nonumber 
\end{align}
where  
\begin{align*} 
U_{ij}^{[t]}= \bigg(\frac{f^{[1-t]}(W_i, U^{\text{o}}_{i};\gamma^{\text{o}})}{(f_0(U^{\text{o}}_{i};\gamma^{\text{o}}))^2}\varphi^{[t]}_{j}(W_{i}, U^{\text{o}}_{i};\gamma^{\text{o}})\bigg)-\bar{U}_i^{[t]},~\bar{U}_i^{[t]}=\frac{f^{[1-t]}(W_i, U^{\text{o}}_{i};\gamma^{\text{o}})}{(f_0(U^{\text{o}}_{i};\gamma^{\text{o}}))^2}\E\big[\varphi^{[t]}_{j}(W_{i}, U^{\text{o}}_{i};\gamma^{\text{o}})  \mid i\big].
\end{align*}
The definitions of $f^{[1]}(w, u; \gamma)$ in (\ref{eq:lemma3_3_1}), $\xi^{[1]}_{j}(w,\gamma)$ in (\ref{eq:lemma1_7}),  and $\varphi^{[1]}_{j}(w,u;\gamma)$ in (\ref{eq:lemma_C_23}) yield
\begin{align*}
&\E\big[ f^{[1]}(W_i, U^{\text{o}}_i; \gamma^{\text{o}}) \bigm| U^{\text{o}}_i \big] 
 =   \E\big[-\partial_u \big( Q_{1,1}(W_i, U^{\text{o}}_i \big) f(U^{\text{o}}_i; \gamma^{\text{o}})) \bigm| U^{\text{o}}_i \big] \hspace{1.5in}\\
& =  \E\big[ \big( - \E[m^{[1]}(W,\gamma^{\text{o} \top} W)] + m^{[1]}(W_i,\gamma^{\text{o} \top} W_i) \big) \big( - \partial_u  f(U^{\text{o}}_i; \gamma^{\text{o}}) \big) \bigm| U^{\text{o}}_i \big] \\
&=  \big( - \E\big[m^{[1]}(W,\gamma^{\text{o} \top} W)\big] + \E\big[m^{[1]}(W,\gamma^{\text{o} \top} W) \big] \big)\big( - \partial_u  f(U^{\text{o}}_i; \gamma^{\text{o}}) \big) = 0,\\
&\xi^{[1]}_{j}(W_i,\gamma) - \E\big[\xi^{[1]}_{j}(W_i,\gamma) \bigm| i\big] = a(W_j, W_i) W_i, \text{ and}\\
& \E\big[\zeta^{[1]}_{j}(W_{i}, U^{\text{o}}_{i};\gamma^{\text{o}}) \bigm| W_j, U^{\text{o}}_j, U^{\text{o}}_i\big]
= \E\big[ \big(\partial_{\gamma_*} K_{\xi}(U^{\text{o}}_{j} - U^{\text{o}}_{i})\big) - f^{[1]}(W_i, U^{\text{o}}_i; \gamma^{\text{o}}) \bigm| W_j, U^{\text{o}}_j, U^{\text{o}}_i\big] \\
& = \frac{1}{\xi^2} \E\bigg[ K^{(1)}\bigg(\frac{U^{\text{o}}_{j} - U^{\text{o}}_{i}}{\xi}\bigg) Q_{1,1}(W_i, U^{\text{o}}_{j}) \biggm| W_j, U^{\text{o}}_j, U^{\text{o}}_i \bigg] - \E\big[ f^{[1]}(W_i, U^{\text{o}}_i; \gamma^{\text{o}}) \mid U^{\text{o}}_i \big]  \\
& = \frac{1}{\xi^2}K^{(1)}\bigg(\frac{U^{\text{o}}_{j} - U^{\text{o}}_{i}}{\xi}\bigg)\Big( - \E\big[m^{[1]}(W,\gamma^{\text{o} \top} W)\big] + \E\big[m^{[1]}(W,\gamma^{\text{o} \top} W) \big] \Big) = 0,
\end{align*}
where $a(W_j, W_i)$ is a function of $W_j$ and $W_i$.
Hence,
\begin{align*}
\E\big[ U^{[0]}_{ij} \bigm | W_j, U^{\text{o}}_j \big] = \E\bigg[ \frac{\varphi^{[0]}_{j}(W_{i}, U^{\text{o}}_{i};\gamma^{\text{o}}) - \E\big[\varphi^{[0]}_{j}(W_{i}, U^{\text{o}}_{i};\gamma^{\text{o}})  \mid i\big]}{\big( f_0(U^{\text{o}}_i; \gamma^{\text{o}}) \big)^2} \E\big[f^{[1]}(W_i, U^{\text{o}}_i; \gamma^{\text{o}}) \mid U^{\text{o}}_i \big] \biggm| W_j, U^{\text{o}}_j \bigg] = 0
\end{align*}
and
\begin{align*}
&\E\big[ U^{[1]}_{ij} \bigm | W_j, U^{\text{o}}_j \big] 
=  \E\bigg[ \frac{\varphi^{[0]}_{j}(W_{i}, U^{\text{o}}_{i};\gamma^{\text{o}}) - \E\big[\varphi^{[0]}_{j}(W_{i}, U^{\text{o}}_{i};\gamma^{\text{o}})  \mid i\big]}{f_0(U^{\text{o}}_{i};\gamma^{\text{o}})}\biggm| W_j, U^{\text{o}}_j\bigg]\hspace{1.9in} 
\end{align*}
\begin{align*}
\hspace{0.8in}& =  \E\bigg[ \frac{ (\partial_{u} f_0(U^{\text{o}}_i; \gamma^{\text{o}})) \big( \xi^{[1]}_{j}(W_i,\gamma) - \E \big[ \xi^{[1]}_{j}(W_i,\gamma) \mid i\big] \big) }{f_0(U^{\text{o}}_{i};\gamma^{\text{o}})}\biggm| W_j, U^{\text{o}}_j\bigg]\\
&~ ~~~+ \E\bigg[ \frac{1}{f_0(U^{\text{o}}_{i};\gamma^{\text{o}})}\big( \E\big[\zeta^{[1]}_{j}(W_{i}, U^{\text{o}}_{i};\gamma^{\text{o}}) \bigm| W_j, U^{\text{o}}_j, U^{\text{o}}_i\big] - \E\big[\zeta^{[1]}_{j}(W_{i}, U^{\text{o}}_{i};\gamma^{\text{o}}) \bigm| i\big] \big) \biggm | W_j, U^{\text{o}}_j\bigg]  \\
&= \Big( \int \partial_{\gamma_*} f_0(u; \gamma^{\text{o}}) du \Big) \E[a(W_j, w)|W_j]=  \Big( \partial_{\gamma_*} \int  f_0(u; \gamma^{\text{o}}) du \Big) \E[a(W_j, w)|W_j] = 0.
\end{align*}
Consequently, $\E\big[U^{[t]}_{ij} \mid j \big]=0$.
Combining with $\E\big[U^{[t]}_{ij} \mid i \big]=0$,
${U}_{ij}^{[t]}$ is a degenerate second-order $U$-statistic with variance of order $O(\xi_m^{-2t-1})$ for $t = 0, 1$.
By Theorem 8.1 in \cite{hoeffding1948probability}, 
\begin{align}
   \frac{1}{n(n-1)}\sum_{i\neq j} {U}_{ij}^{[t]}=O_p\bigg(\frac{1}{n\sqrt{\xi_m^{2t+1}}}\bigg), t=0,1.\label{eq:A.5.13}
\end{align}
The central limit theorem implies
\begin{align}
    \frac{1}{n}\sum_{i=1}^n \bar{U}_i^{[t]}= O_p\bigg(\frac{\xi_M^{2}}{\sqrt{n}}\bigg), t=0,1.\label{eq:A.5.14}
     \end{align}
Combining (\ref{eq:A.5.13}) and (\ref{eq:A.5.14}) yields
\begin{align}
   &I\!I\!I=o_p(1).\label{eq:A.5.15}
   \end{align}
Substituting (\ref{eq:A.5.9.1}) and (\ref{eq:A.5.15}) into (\ref{eq:A.5.9.0}) and applying the central limit theorem to $S(\gamma^{\text{o}}) / \sqrt{n}$ yield
\begin{align}
\frac{1}{\sqrt{n}}pS(\gamma^{\text{o}})=\frac{1}{\sqrt{n}}S(\gamma^{\text{o}})+o_p(1) 
  \stackrel{d}{\longrightarrow} N\big(0,V\big). \label{eq:A.5.16}
\end{align}

Using arguments  analogous to those for (\ref{eq:A.5.9.1}), we obtain
\begin{align}
  \frac{1}{n}pI(\gamma) & = \frac{1}{n}\sum_{i=1}^n \Bigg[\Bigg(\frac{\partial_{\gamma_*}\breve{f}_{\xi}\big(\widehat{U}_{\gamma i};\gamma\big)} {\breve{f}_{\xi}\big(\widehat{U}_{\gamma i};\gamma\big)} \Bigg) ^{\otimes 2}-\frac{\partial_{\gamma_*}^{2} \breve{f}_{\xi}\big(\widehat{U}_{\gamma i};\gamma\big)} {\breve{f}_{\xi}\big(\widehat{U}_{\gamma i};\gamma\big)}\Bigg] + o_p(1) \nonumber \\
  & =\frac{1}{n}\sum_{i=1}^n \Bigg[\Bigg(\frac{f^{[1]}(W_i, U_{\gamma i};\gamma)} {f_0(U_{\gamma i};\gamma)} \Bigg) ^{\otimes 2}-\frac{f^{[2]}(W_i,U_{\gamma i};\gamma)} {f_0(U_{\gamma i};\gamma)}\Bigg]+o_p(1). \label{eq:A.5.18}
\end{align}
Applying the law of large numbers to the $i.i.d.$ representation in (\ref{eq:A.5.18}) yields
\begin{align}
    &\frac{1}{n}pI(\gamma)\stackrel{p}{\longrightarrow}\E\Bigg[\Bigg(\frac{f^{[1]}(W, U_{\gamma};\gamma)} {f_0(U_{\gamma};\gamma)} \Bigg) ^{\otimes 2}-\frac{f^{[2]}(W, U_{\gamma};\gamma)} {f_0(U_{\gamma};\gamma)}\Bigg]. \label{eq:A.5.20}
\end{align}
By Lemma \ref{A:Lemma_3} with $t = 2$, the uniform integrability of ${\partial_{\gamma_*}^{2} \bar{f}_{\xi}(U^{\text{o}};\gamma^{\text{o}})} / f(U^{\text{o}};\gamma^{\text{o}})$, and
\begin{align*}
\E\Bigg[\frac{\partial_{\gamma_*}^{2} \bar{f}_{\xi}(U^{\text{o}};\gamma^{\text{o}})}{f_0(U^{\text{o}};\gamma^{\text{o}})}\Bigg] = \partial_{\gamma_*}^{2} \int \bar{f}_{\xi}(u;\gamma) du \biggm|_{\gamma = \gamma^{\text{o}}}= 0,
\end{align*}
it follows that
\begin{align*}
\E\Bigg[\frac{ {f}^{[2]}_{\xi}(W_i, U^{\text{o}};\gamma^{\text{o}})}{f_0(U^{\text{o}};\gamma^{\text{o}})}\Bigg] = 0.
\end{align*}
The continuity of $pI(\gamma)$ on $\Gamma$ and the consistency of $\widetilde{\gamma}_*$ to $\gamma^{\text{o}}_*$ imply
\begin{align}
   \frac{1}{n}pI(\Bar{\gamma}) \stackrel{p}{\longrightarrow} V. \label{eq:A.5.21}
\end{align}
Substituting (\ref{eq:A.5.16}) and (\ref{eq:A.5.21}) into (\ref{eq:A.5.8}) and applying Slutsky's theorem, the asymptotic normality of $\widetilde{\gamma}$ is established under assumption {C4}.

\section{Supplementary Tables and Figures}
\label{table-figure}

\begin{table}[H]
  \centering 
  \caption{Means (standard deviations) of 500 RI values for the $\text{SP}_{\text{L}}$- and MML-estimated latent class partitions under various model settings.}
      \begin{adjustbox}{max width=1\linewidth}
    \begin{tabular}{cccccccccccccccccc}
    \toprule
            & \multicolumn{8}{c}{$k=2$} & & \multicolumn{8}{c}{$k=3$}\\
      \cmidrule(rl){2-9} \cmidrule(rl){11-18}
                & \multicolumn{2}{c}{normal} &       & \multicolumn{2}{c}{logistic} &       & \multicolumn{2}{c}{Gamma} & & \multicolumn{2}{c}{normal} &       & \multicolumn{2}{c}{logistic} &       & \multicolumn{2}{c}{Gamma} \\
          \cmidrule(rl){2-3} \cmidrule(rl){5-6} \cmidrule(rl){8-9} \cmidrule(rl){11-12} \cmidrule(rl){14-15} \cmidrule(rl){17-18}
            $n$    & $\text{SP}_{\text{L}}$    & MML    &       & $\text{SP}_{\text{L}}$    & MML    &       & $\text{SP}_{\text{L}}$    & MML &       & $\text{SP}_{\text{L}}$    & MML    &       & $\text{SP}_{\text{L}}$    & MML    &       & $\text{SP}_{\text{L}}$    & MML \\
            \midrule
    200   & 0.961 & 0.769 && 0.957 & 0.797 && 0.963 & 0.769 &       & 0.901 & 0.748 && 0.903 & 0.745 && 0.907 & 0.749 \\
          &(0.0266) &(0.2284) &&(0.0249) &(0.2193) &&(0.0260) &(0.2175) &       &(0.0451) &(0.1217) &&(0.0442) &(0.1109) &&(0.0461) &(0.1157) \\
    400   & 0.970 & 0.806 && 0.963 & 0.820 && 0.973 & 0.799 &       & 0.931 & 0.813 && 0.930 & 0.816 && 0.935 & 0.809 \\
          &(0.0163) &(0.2197) &&(0.0163) &(0.2133) &&(0.0170) &(0.2063) &       &(0.0274) &(0.1269) &&(0.0267) &(0.1243) &&(0.0297) &(0.1225) \\
    600   & 0.973 & 0.825 && 0.967 & 0.812 && 0.974 & 0.788 &       & 0.938 & 0.835 && 0.939 & 0.828 && 0.946 & 0.827 \\
          &(0.0140) &(0.2145) &&(0.0137) &(0.2153) &&(0.0160) &(0.2066) &       &(0.0237) &(0.1188) &&(0.0205) &(0.1246) &&(0.0231) &(0.1215) \\
          \bottomrule 
    \end{tabular}%
           \end{adjustbox}
  \label{tab:S_RI}%
\end{table}%

\begin{table}[H]
  \centering 
  \caption{Means of 500 RSEs of conditional mean function estimates from various methods under various model settings.}
    \scriptsize
      \begin{adjustbox}{max width=1\linewidth}
    \begin{tabular}{cccccccccccccccc}
    \toprule
              &       & \multicolumn{4}{c}{normal} &       & \multicolumn{4}{c}{logistic} &       & \multicolumn{4}{c}{Gamma}\\ 
          \cmidrule(rl){3-6} \cmidrule(rl){8-11} \cmidrule(rl){13-16}
      Model    & $n$      & SP    & PML   & $\text{SP}_{\text{L}}$    & MML   &       & SP    & PML   & $\text{SP}_{\text{L}}$    & MML   &       & SP    & PML   & $\text{SP}_{\text{L}}$    & MML \\
      \midrule
    $k=2$    & 200   & 0.058 & 0.057 & 0.047 & 0.178 &       & 0.058 & 0.058 & 0.044 & 0.170 &       & 0.057 & 0.056 & 0.044 & 0.180 \\
          & 400   & 0.037 & 0.037 & 0.035 & 0.169 &       & 0.039 & 0.036 & 0.035 & 0.167 &       & 0.038 & 0.036 & 0.033 & 0.180 \\
          & 600   & 0.030 & 0.030 & 0.031 & 0.169 &       & 0.032 & 0.029 & 0.031 & 0.173 &       & 0.032 & 0.030 & 0.031 & 0.183 \\
          \midrule\midrule
    $k=3$    & 200   & 0.072 & 0.071 & 0.132 & 0.121 &       & 0.072 & 0.072 & 0.143 & 0.120 &       & 0.073 & 0.072 & 0.168 & 0.122 \\
          & 400   & 0.044 & 0.043 & 0.101 & 0.080 &       & 0.045 & 0.044 & 0.114 & 0.084 &       & 0.045 & 0.044 & 0.152 & 0.091 \\
          & 600   & 0.035 & 0.034 & 0.095 & 0.070 &       & 0.034 & 0.033 & 0.096 & 0.074 &       & 0.036 & 0.034 & 0.133 & 0.076 \\
          \bottomrule 
    \end{tabular}%
           \end{adjustbox}
  \label{tab:m_est}%
\end{table}%

\begin{table}[H]
  \centering 
  \caption{Means (proportions of correct selection) of 500 estimated number of classes across selection criteria under various model settings.}
      \begin{adjustbox}{max width=1\linewidth}
    \begin{tabular}{cccccccccccccccccc}
    \toprule
            & \multicolumn{8}{c}{$k=2$} & & \multicolumn{8}{c}{$k=3$}\\
      \cmidrule(rl){2-9} \cmidrule(rl){11-18}
                & \multicolumn{2}{c}{normal} &       & \multicolumn{2}{c}{logistic} &       & \multicolumn{2}{c}{Gamma} & & \multicolumn{2}{c}{normal} &       & \multicolumn{2}{c}{logistic} &       & \multicolumn{2}{c}{Gamma} \\
          \cmidrule(rl){2-3} \cmidrule(rl){5-6} \cmidrule(rl){8-9} \cmidrule(rl){11-12} \cmidrule(rl){14-15} \cmidrule(rl){17-18}
            $n$    & $\text{BIC}_{1}$    &$\text{BIC}_{2}$    &       &$\text{BIC}_{1}$    &$\text{BIC}_{2}$   &       & $\text{BIC}_{1}$    &$\text{BIC}_{2}$ &       & $\text{BIC}_{1}$    &$\text{BIC}_{2}$    &       & $\text{BIC}_{1}$    &$\text{BIC}_{2}$   &       & $\text{BIC}_{1}$    &$\text{BIC}_{2}$ \\
            \midrule
    200   & 5.79  & 2.44 & & 5.79  & 2.38 & & 5.72  & 2.48  &       & 6.00  & 2.03  && 5.99  & 2.07  && 5.99  & 2.11 \\
          & (0.000) & (0.578) && (0.000) & (0.636) && (0.000) & (0.572) &       & (0.000) & (0.166) && (0.000) & (0.226) && (0.000) & (0.232) \\
    400   & 8.00  & 2.81  && 8.00  & 2.73  && 8.00  & 2.76  &       & 8.00  & 2.75  && 8.00  & 2.83  && 8.00  & 2.76 \\
          & (0.000) & (0.332) && (0.000) & (0.420) && (0.000) & (0.346) &       & (0.000) & (0.718) && (0.000) & (0.782) && (0.000) & (0.672) \\
    600   & 9.00  & 3.15  && 9.00  & 3.12  && 9.00  & 3.07  &       & 9.00  & 2.95  && 9.00  & 3.02  && 9.00  & 2.99 \\
          & (0.000) & (0.136) && (0.000) & (0.178) && (0.000) & (0.112) &       & (0.000) & (0.868) && (0.000) & (0.890) && (0.000) & (0.866) \\
          \bottomrule 
    \end{tabular}%
           \end{adjustbox}
  \label{tab:S_k}%
\end{table}%

\begin{figure}[H]
\centering
\begin{subfigure}[t]{0.45\linewidth} 
    \centering
    \includegraphics[width=\linewidth]{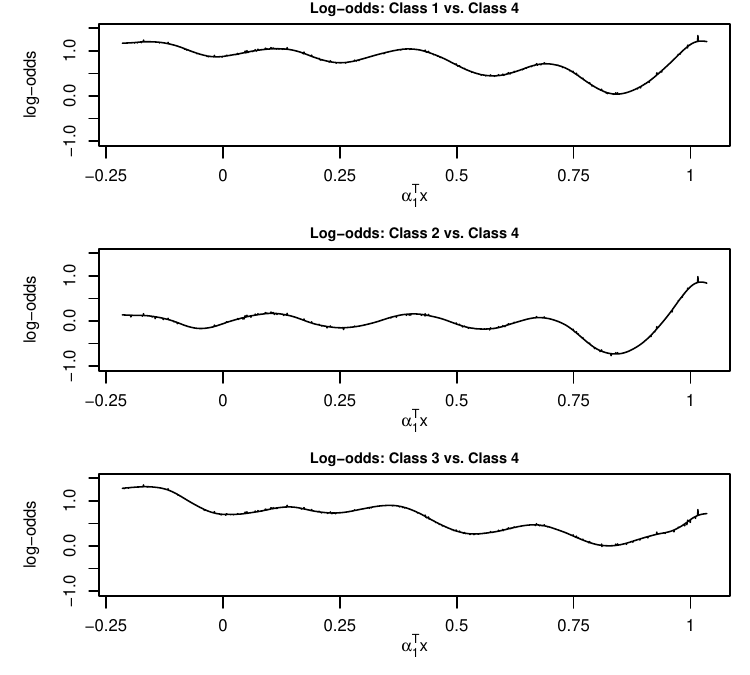} 
    \caption{{\footnotesize SDR-based Class Membership Probabilities}}
\end{subfigure}
\begin{subfigure}[t]{0.45\linewidth} 
    \centering
    \includegraphics[width=\linewidth]{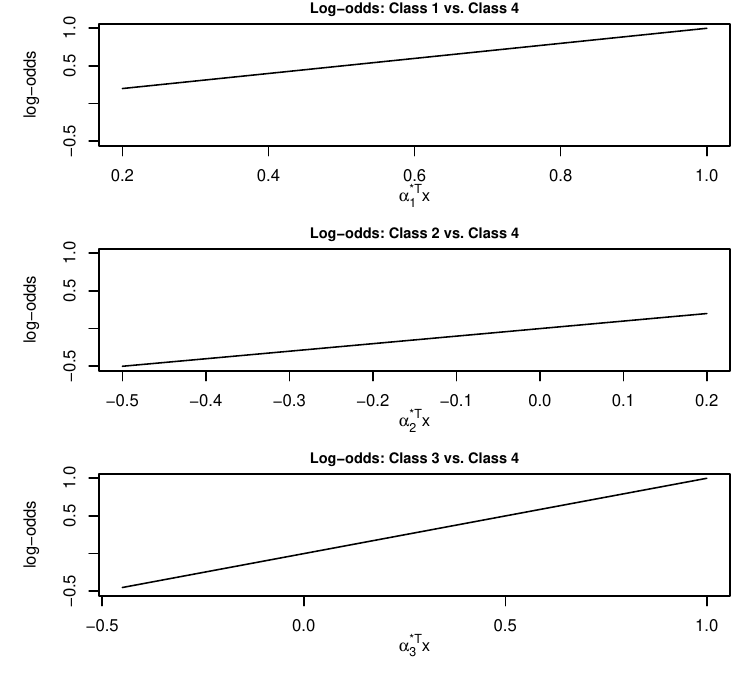} 
    \caption{{\footnotesize Logistic Class Membership Probabilities}}
\end{subfigure}

\caption{{\footnotesize Estimated log-odds of membership in Classes 1--3 relative to Class 4 based on the SDR-based and logistic class membership probabilities.}}
\label{fig:ASCE_prob}
\end{figure}

\begin{figure}[H]
    \centering
    \includegraphics[width=9.5cm,height=5.5cm]{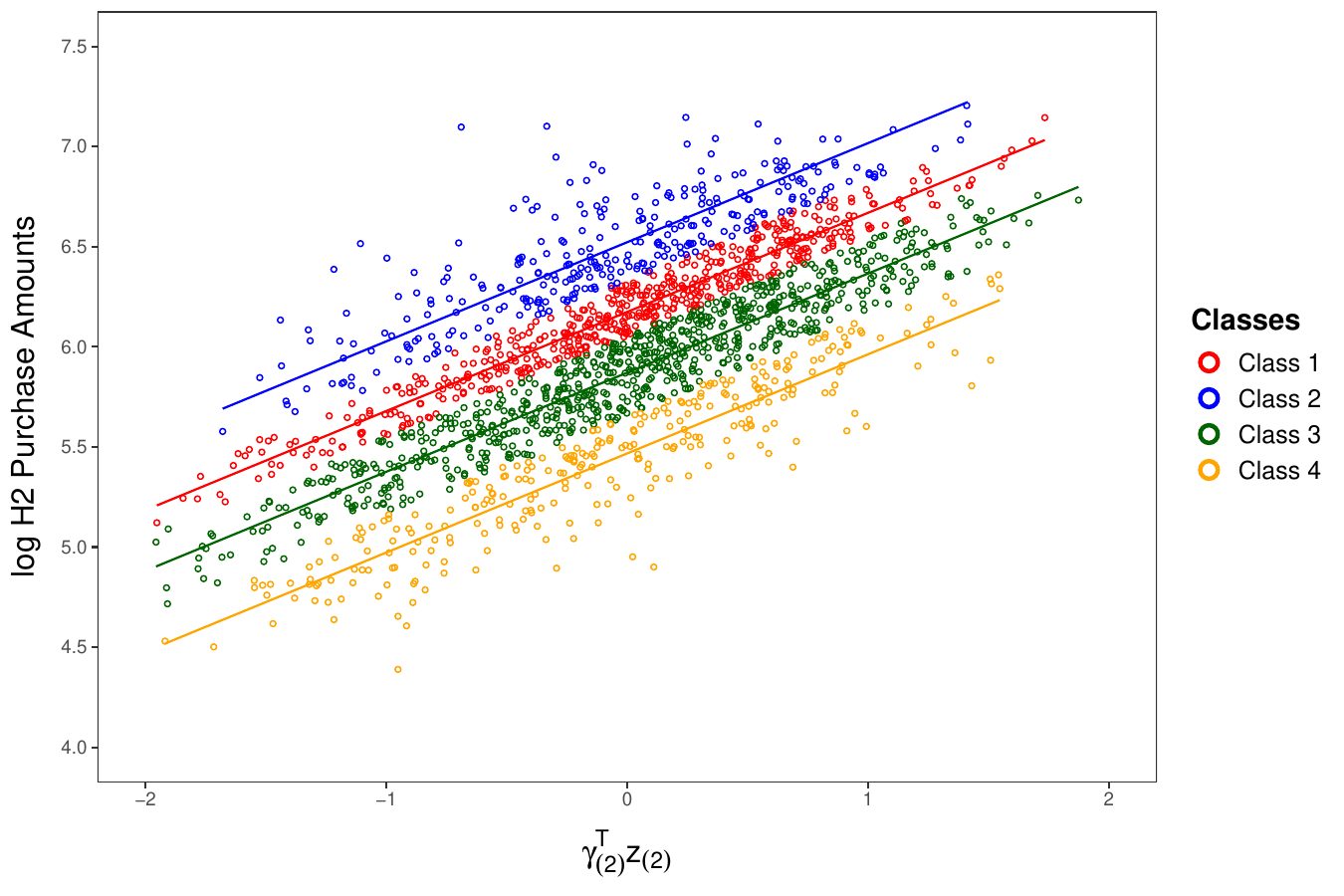}
    \caption{Scatter plot of log-transformed purchase amounts in H2, overlaid with estimated linear mean functions of the class-invariant index.}
    \label{fig:ASCE_L_mean}
\end{figure}

\begin{figure}[H]
\centering
\begin{subfigure}[t]{0.45\linewidth} 
    \centering
    \includegraphics[width=\linewidth]{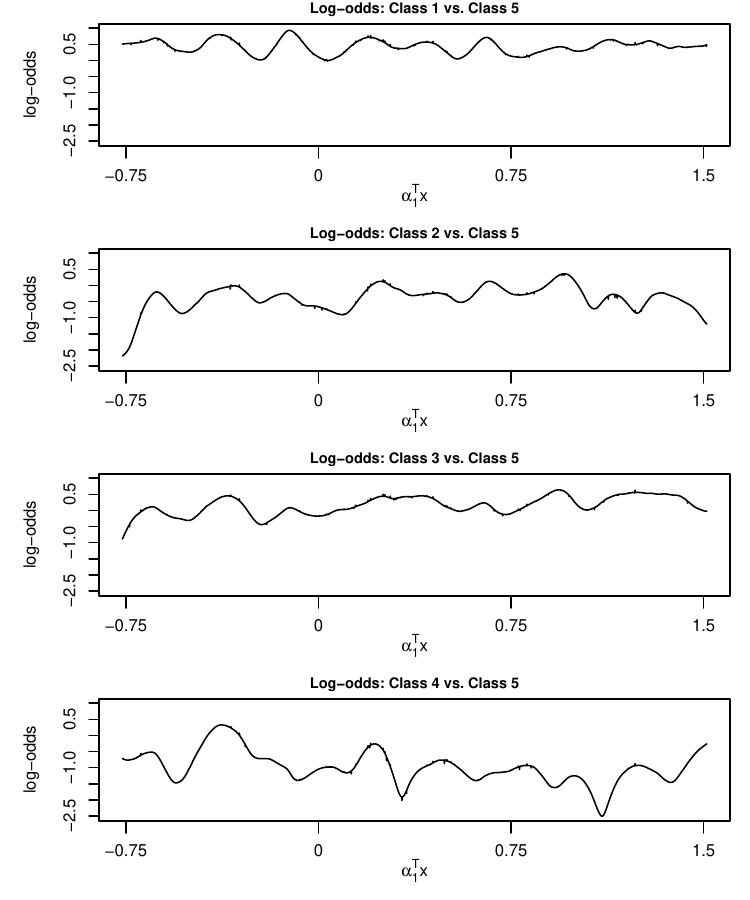} 
    \caption{{\footnotesize SDR-based Class Membership Probabilities}}
\end{subfigure}
\begin{subfigure}[t]{0.45\linewidth} 
    \centering
    \includegraphics[width=\linewidth]{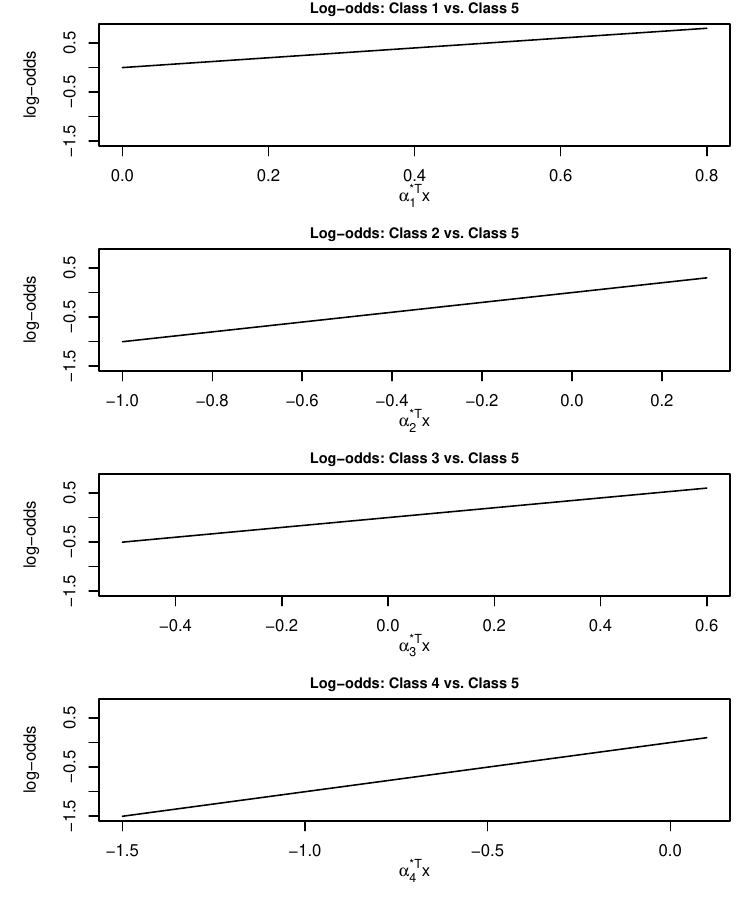} 
    \caption{{\footnotesize Logistic Class Membership Probabilities}}
\end{subfigure}

\caption{{\footnotesize Estimated log-odds of membership in Classes 1--4 relative to Class 5 based on the SDR-based and logistic class membership probabilities.}}
\label{fig:ACTG_prob} 
\end{figure}

\end{document}